\documentclass[12pt]{article}
\usepackage{latexsym,epsfig,amssymb,amsmath,cite,slashed}
\usepackage{authblk}
\newcommand{\be}{\begin{equation}}
\newcommand{\ee}{\end{equation}}
\newcommand{\bea}{\begin{eqnarray}}
\newcommand{\eea}{\end{eqnarray}}

\newcommand{\nn} {\nonumber}
\numberwithin{equation}{section}
\renewcommand\Affilfont{\itshape\small}
\usepackage{hyperref}

\hypersetup{
    colorlinks,
    citecolor=black,
    filecolor=black,
    linkcolor=black,
    urlcolor=black,
linktocpage,
}

\usepackage{array,dcolumn,booktabs}

\begin{document}
\pagenumbering{gobble}
\thispagestyle{empty}
\vspace{1.8cm}

\title{{\bf Anatomy of new SUSY breaking holographic RG flows}} 

\author[1]{Riccardo Argurio\thanks{rargurio@ulb.ac.be}}
\author[2]{Daniele Musso\thanks{dmusso@ictp.it}}
\author[1,3,4]{Diego Redigolo\thanks{dredigol@lpthe.jussieu.fr}}

\affil[1]{Physique Th\'eorique et Math\'ematique and International Solvay Institutes, Universit\'e Libre de Bruxelles,
C.P. 231, 1050 Brussels, Belgium}

\affil[2]{International Center of Theoretical Physics (ICTP), Strada Costiera 11, I 34014 Trieste, Italy}
\affil[3]{Sorbonne Universit\'es, UPMC Univ Paris 06, UMR 7589, LPTHE, F-75005, Paris, France
}
\affil[4]{CNRS, UMR 7589, LPTHE, F-75005, Paris, France
}

\date{}

\maketitle

\begin{abstract}
We find and thoroughly study new supergravity domain wall solutions which are
holographic realizations of supersymmetry breaking strongly coupled
gauge theories. We set ourselves in an $\mathcal{N}=2$ gauged
supergravity with a minimal content in order to reproduce a dual $\mathcal{N}=1$
effective SCFT which has a $U(1)_R$ symmetry, a chiral operator whose
components are responsible for triggering the RG flow, and an
additional $U(1)_F$ symmetry. We present a full three dimensional
parameter space of solutions, which generically break
supersymmetry. Some known solutions are recovered for specific sets
of values of the parameters, with the new solutions interpolating
between them. The generic backgrounds being singular, we provide a stability test of their dual theories by showing that there are no tachyonic resonances in the two point correlators. We compute the latter by holographic renormalization. We also carefully analyze the appearance of
massless modes, such as the dilaton and the R axion, when
the respective symmetries are spontaneously broken, and their lifting
when the breaking is explicit. We further comment on the application of such class of
backgrounds as archetypes of strongly coupled hidden sectors for gauge
mediation of supersymmetry breaking. In particular, we show that it is
possible to model in this way all types of hierarchies between the
visible sector gaugino and sfermion masses.

\end{abstract}

\tableofcontents

\clearpage
\pagenumbering{arabic}

\section{Introduction}
The $AdS$/CFT correspondence \cite{Maldacena:1997re,Gubser:1998bc,Witten:1998qj}
and its generalizations allow us to study strongly coupled gauge
theories in their large $N$ limit. 
In this paper we perform a complete study of a certain class of
strongly coupled renormalization group (RG) flows, characterized by being  $4d$
$\mathcal{N}=1$ SCFTs in the deep UV.  They are therefore 
expected to be well described by asymptotically $AdS$ ($AAdS$) 
domain walls in a $5d$ gauged $\mathcal{N}=2$  supergravity 
(SUGRA). Very much in the spirit of \cite{Heemskerk:2009pn} we assume that in the large $N$ limit the CFT 
dynamics simplifies such that we can focus our attention on the physics of just a subsector of the CFT operators consisting of single trace operators, 
with parametrically smaller dimensions than all other operators. These ``light'' operators organize in $\mathcal{N}=1$ 
multiplets which can be described in terms of $\mathcal{N}=2$
multiplets of local SUGRA fields in the bulk.  

Our basic aim is to study the interplay between the RG flow and
(super)symmetry breaking, both spontaneous and explicit.
With the above mentioned simplification in mind, we will take an
effective approach 
where we assume that no other operator than the ones explicitly
considered can affect the
physics. Translated in gravity, this means that we exclude the
presence of other light fields that could affect
the physical validity of a given a background. This approach has the benefit
of making the issue of the stability of a background (and thus the
validity of the field theory vacuum) a problem that can be solved by directly
analyzing the spectrum without having to resort to generic criteria. On the negative side, one is not
granted to have a well defined and stable uplift to string
theory. However we do not consider this as a drawback, since we are
really interested in the qualitative strong coupling signatures of
symmetry breaking. 

As a prototypical model we consider a simple SUGRA theory which has a
single hypermultiplet (often called universal) whose dynamics is
governed by the coset $SU(2,1)/U(2)\times U(1)$
\cite{Ceresole:2000jd,Ceresole:2001wi,Fitzpatrick:2011hh}.  
Besides the hypermultiplet our theory has the
ubiquitous gravity multiplet, and also one additional
vector multiplet under which the hypermultiplet is not charged in the specific gauging we choose. 
From the CFT point of view we are
keeping in the spectrum of the ``effective'' SCFT just a single
$\mathcal{N}=1$ chiral multiplet, the multiplet
of the stress energy tensor which also contains a $U(1)_R$ symmetry,
and an $\mathcal{N}=1$ linear multiplet
corresponding to an always preserved $U(1)_F$ flavour symmetry. 

The first step of our analysis consists in giving a complete
classification of all the $AAdS$ solutions
which support a non trivial profile for two real scalars in the
universal hypermultiplet:  the dilaton $\phi$ which is dual to
$F_{\mu\nu}^2$ and neutral under R symmetry, 
and the so called squashing mode, that we will call $\eta$ hereafter,
which is dual to the gaugino bilinear $\lambda\lambda$ and is thus R charged.
The space of solutions can be described by a $3d$ parameter space
which depends on three intergration constants determining the scalar profiles.
Moving within the space of solutions in the bulk is equivalent from the boundary perspective to realize
different RG flows which depart from the ``effective'' SCFT we just described, by switching on VEVs
and/or sources for the operators $F_{\mu\nu}^2$ and $\lambda\lambda$. 

In particular, a non trivial profile for the R charged 
scalar corresponds to explicit or
spontaneous 
breaking of the R symmetry in the dual field theory, depending on its
near boundary behavior. Our general class of solutions includes the dilaton domain wall of  \cite{Kehagias:1999tr,Gubser:1999pk}, (half of) the SUSY preserving GPPZ flow \cite{Girardello:1999bd}, the flow featuring a non SUSY IR fixed point of \cite{Distler:1998gb}, plus new numerical backgrounds interpolating between all of the above.

Note that many of the flows under consideration have a naked singularity in the deep interior of the bulk,
and we are not addressing the issues related to their possible UV completions in string theory. Accordingly, we do not consider the usual ``goodness'' criteria for a naked singularity \cite{Gubser:2000nd,Maldacena:2000mw}, but we follow a more bottom up criterion considering as good holographic RG flows the ones  where all the two point correlators in the bosonic sector do not present any tachyonic pole. 

According to this criterion we should compute  holographically the two point
correlators of all the operators inside our effective SCFT, namely the ones of the current supermultiplet and of the supermultiplet containing the stress energy tensor
\cite{Argurio:2012cd,Argurio:2012bi,Argurio:2013uba}. In order to do
that we make use of the standard holographic renormalization
techniques \cite{Mueck:2001cy,Bianchi:2001de,Bianchi:2001kw,Skenderis:2002wp},
slightly generalizing them to $AAdS$ backgrounds which support a
non trivial profile for two scalars (see
\cite{Berg:2005pd,Berg:2006xy} for a similarly involved case, and
\cite{Papadimitriou:2004rz} for a variation of the technique in a
similar context).

The same two point correlators will give much additional information about the dynamics of the RG flows in our parameter space. A general property is that correlators of operators in the same multiplet will satisfy SUSY Ward identities when SUSY is preserved. SUSY breaking will be then manifest as a deviation from the SUSY Ward identities in the correlators. The latter are effectively restored at high external momenta compared to the SUSY breaking parameters since the flows in consideration depart from an $\mathcal{N}=1$ SCFT in the UV.

Given a symmetry of the UV theory, the two point correlator of the
associated conserved current is telling us the fate of that symmetry
along the flow. If the symmetry is spontaneously broken the correlator
of the transverse current has a simple pole associated to the Goldstone
mode. Moreover, the mixed correlator between the current and the
operator whose VEV breaks the symmetry is a pure contact term
proportional to the Goldstone decay constant. Conversely, if the
symmetry is broken explicitly, the current acquires longitudinal
degrees of freedom which should respect Ward identities originating
from the broken symmetry.  Consequently, what was the massless mode in the transverse part of the correlator gets a mass proportional to the parameter of explicit breaking. 

All these features can be reproduced holographically, as it has been
shown in \cite{Bianchi:2001de,Bianchi:2001kw,Skenderis:2002wp}. 
Compared to previous approaches,
our analysis  has the advantage of realizing different dinamical
phases within the same SUGRA model. This allows us to  analyze the parametric dependence of the correlators on the changing of the parameters describing the different SUGRA solutions (i.e. RG flows). In order to attain this task we solved both the differential equations for the background and for the fluctuations numerically.\footnote{An analytical handle on holographic correlators has been developed for very simplified setups in \cite{Hoyos:2013gma,Bajc:2013wha}.}

Let us summarize hereafter the main results of our study:  
\begin{itemize}
\item The pole structure of the two point  functions for the full bosonic spectrum of operators of the model allows us to make precise statements about the \emph{stability} of the backgrounds we are considering, excluding the presence of any tachyonic resonance in the dual QFT. 

\item  Two point correlators in the stress energy tensor multiplet
  highlight the dynamical features of the underlying background such
  as spontaneous/explicit breaking of conformal symmetry, $U(1)_R$
  symmetry and supersymmetry.\footnote{In this paper we will restrict
    our attention on the bosonic sector of the multiplet, leaving the
    discussion on the supercurrent correlators for a forthcoming paper
    \cite{inpreparation}.}
\end{itemize}
To make this last point completely clear, we found it useful to recast
the techniques of  \cite{Bianchi:2001de,Bianchi:2001kw} in a simple
holographic model with just a vector coupled to a charged scalar in
the bulk. In this model the features of the current correlators
discussed above appear nicely without the technical difficulties
associated with the more involved model coming from the SUGRA embedding.

We further compute the correlators of the conserved current multiplet
associated to the preserved $U(1)_F$ on the various backgrounds. These
provide additional information on the strongly coupled theories. Most
notably, 
they are also the building blocks of the general gauge mediation (GGM)
formalism \cite{Meade:2008wd}. Each of our SUGRA solution defines then
a calculable model  for a hidden sector in gauge mediation.

As a phenomenological application,  we study how the ratio of the
gaugino to sfermion masses behaves in our $3d$ parameter space very
much in the spirit of \cite{Buican:2008ws}. We realize  diversified phenomenologies, from gaugino mediation (with either Dirac \cite{Benakli:2008pg} or Majorana masses \cite{Green:2010ww,Sudano:2010vt} for the gaugino) to gaugino mass screening \cite{ArkaniHamed:1998kj,Dumitrescu:2010ha}.

The paper is structured as follows. In Section \ref{sugra} we present
the bulk theory we are going to consider, and motivate holographically
the presence of each multiplet, also relating to previous works. In
Section 3 we derive and classify all the solutions to the system we
consider, with two active scalars and a single warp factor. All the
solutions we find are stable since we later show that there are no tachyonic modes in their spectrum. In Section 4 we present the analysis of representative solutions scanning the parameter space. We display two point correlators of the stress energy tensor multiplet and of the conserved current multiplet highlighting how the physics changes moving around the parameter space. Section 5 contains a thorough compendium of holographic renormalization, the technique we used to derive the correlators of the preceding section. It starts with the toy example of a vector coupled to a charged scalar and then delve in the intricacies of the full supergravity case. In Section 6 we consider using seriously the vector multiplet correlators as an input in the formalism of GGM, and determine how the ratio of gaugino to sfermion masses changes as we move around the parameter space. In Section 7 we give our outlook.

\section{Holography with $\mathcal{N}=2$ gauged supergravity}
\label{sugra}

We begin our discussion by summarizing the SUSY structure of the multiplets which remain light in the large $N$ expansion of our SCFT. 

Considering first the stress energy tensor $T_{\mu\nu}$, we know that in a SCFT it has to be traceless up to improvement transformations. Moreover its multiplet has to contain the supercurrent $S_{\mu}$ and the conserved current $j_{\mu}^{R}$ associated to the superconformal R symmetry. These degrees of freedom can be expressed in terms of a real vector superfield satisfying the constraint 
\begin{equation}
-2\bar{D}\sigma^{\mu}\mathcal{J}_{\mu}=0\ ,
\end{equation}
which ensures the conservation of $T_{\mu\nu}$ and $S_{\mu}$ and also that $T=\bar{\sigma}^{\mu}S_{\mu}=\partial^{\mu}j_{\mu}^{R}=0$. In components we get
\begin{equation}
\mathcal{J}_{\mu}(x,\theta,\bar{\theta})=j_{\mu}^{R}(x)+[i\theta S_{\mu}(x)+ c.c.]+\theta\sigma^{\mu}\bar{\theta}(2T_{\mu\nu}+\frac{1}{2}\epsilon_{\mu\nu\rho\sigma}\partial^{\rho} j^{\sigma})+\dots
\end{equation}

Being interested in RG flows which have a SCFT only as a UV fixed point, we want to add relevant deformations in order to break conformal invariance and possibly supersymmetry. 
Since we are going to consider relevant deformation triggered by an
$\mathcal{N}=1$ chiral multiplet of operators, the additional degrees
of freedom naturally organize in a chiral superfield $X$. The breaking
of conformal symmetry implies that $T\neq0$ and because of
supersymmetry also $\bar{\sigma}^{\mu}S_{\mu}\neq0$ and
$\partial^{\mu}j_{\mu}^{R}\neq0$.  This can be expressed generalizing
the previous superfield to the so called Ferrara-Zumino (FZ) multiplet \cite{Ferrara:1974pz} via the equation	 
\begin{equation} 
-2\bar{D}\sigma^{\mu}\mathcal{J}_{\mu}=DX\ .
\end{equation}
In components we find 
\begin{align}
&\mathcal{J}_{\mu}=j_{\mu}^{R}+\left[\theta(S_{\mu}+\tfrac{1}{3}\sigma_{\mu}\bar{\sigma}^{\nu}S_{\nu})+\theta^2i\partial_{\mu}x^{\ast}+c.c\right]+\theta\sigma^{\nu}\bar{\theta}(2T_{\mu\nu}-\eta_{\mu\nu}\tfrac{2}{3}T+\tfrac{1}{2}\epsilon_{\mu\nu\rho\sigma}\partial^{\rho} j^{\sigma})+\dots\notag\\
&X=x+[\tfrac{1}{3}\theta\bar{\sigma}^{\mu}S_{\mu}+\theta^2(\tfrac{2}{3}T+i\partial^{\mu}j_{\mu}^{R})+c.c.]+\dots
\end{align}
Since $\mathcal{J}_{\mu}$ has the dimension of a current, it follows
that also $X$ has to be of dimension three. 

Let us then analyze the multiplet of a global $U(1)_F$ current $j_\mu$. Its
multiplet contains a fermionic operator $j_{\alpha}$ and a real scalar
operator $J$ and it is often referred to as a linear multiplet because it can be described in terms of a real scalar superfield which satisfies the linear constraint 
\begin{equation}
D^2J=\bar{D}^2J=0
\end{equation}
where 
\begin{equation}
J(x,\theta,\bar{\theta})=J(x)+[i\theta j(x)+ c.c.]+\theta\sigma^{\mu}\bar{\theta}(j_{\mu}-i\partial_{\mu} J)+\dots
\end{equation}
and the linear constraint implies that $\partial^{\mu}j_{\mu}=0$. The
conformal dimension of the conserved current in $4d$ is again fixed to be
three, so that $J$ has dimension two. 

The $AdS$/CFT correspondence maps the $\mathcal{N}=1$ real
vector superfield $\mathcal{J}_{\mu}$ into the $\mathcal{N}=2$ gravity
multiplet, the $\mathcal{N}=1$ chiral multiplet $X$ into an $\mathcal{N}=2$
hypermultiplet, and the
linear multiplet into an $\mathcal{N}=2$ vector multiplet of SUGRA fields with the field masses related to the conformal dimension of the operators in the boundary CFT. The outcome of this mapping is summarized in Table \ref{N = 1hyper}.

\begin{table}[hc]
\begin{center}
 \caption{$\mathcal{N} =1$ multiplets of operators in $4d$ and their dual $\mathcal{N} =2$ multiplets of local supergravity fields in $5d$.\vspace{5pt}}\label{N = 1hyper}
\begin{tabular}{@{}ccccc@{}}
\toprule
 $\mathcal{N}=1$ mult. & $4d$ op. & $\Delta$& $5d$  field  & $AdS$ mass \\[7pt] \midrule 
                            & $j^{R}_{\mu}(x)$                & $\Delta=3$         & $R_{M}(z,x)$  & $m_{R}^2=0$\\[7pt] 
FZ Mult.        &$S_{\mu}(x)$   & $\Delta=7/2$      & $\Psi_{M}(z,x)$& $\vert m_{\Psi}\vert=3/2$\\[7pt] 
                              &$T_{\mu\nu}(x)$       & $\Delta=4$          & $h_{MN}(z,x)$& $m_{h}^2=0$\\[7pt]  \midrule
                              
                                                   &$ x(x) $         & $\Delta=3$         & $\eta(z,x)$  & $m_{\eta}^2=-3$\\[7pt] 
$X$ Mult.        &$S(x)$          & $\Delta=7/2$      & $\psi(z,x)$& $\vert m_{\psi}\vert=3/2$\\[7pt] 
                                                 &$T(x),\partial j^{R}(x)$
& $\Delta=4$          & $\phi(z,x), C_0(z,x)$&
$m_{\phi}^2=m_{C_0}^2=0$\\[7pt] \midrule
                           &   $J(x)$                            & $\Delta=2$         & $D(z,x)$& $m_{D}^2=-4$ \\[7pt] 
Linear Mult.  &$j_{\alpha}(x)$   & $\Delta=5/2$      & $\lambda(z,x)$& $\vert m_{\lambda}\vert=1/2$\\[7pt] 
                            &$j_{\mu}(x)$       & $\Delta=3$          &
$A_M(z,x)$& $m_{A}^2=0$   \\  \bottomrule
    \end{tabular}
  \end{center}
  \end{table}

In the specific backgrounds we are going to present in the next
section, we consider domain walls which support non trivial profiles
for the scalar components of the universal hypermultiplet in the bulk.
The latter corresponds to the chiral multiplet of gauge invariant operators
$X$ in the boundary SCFT. For definiteness, $X$ can be thought
to be proportional to a gaugino bilinear superfield $\mathcal{O}$, so that
we can write the deformation at the boundary directly in superfield notation as a superpotential term, in agreement with \cite{Green:2010da}:
\begin{equation}
\Delta\mathcal{L}\propto \int d^2\theta\ \Phi_0\mathcal{O}+c.c.
= \tfrac{1}{2}\phi_0
  F^{\mu\nu}F_{\mu\nu}+\eta_0\text{Re}(\lambda\lambda)+\dots  \ ,
\end{equation}
where
\begin{subequations}
\begin{align}
 &\Phi_0=\phi_0+\theta\sqrt{2}\psi_0+\theta^2\eta_0+\dots\ ,\\
 &\mathcal{O}=\frac{1}{2}\lambda\lambda+\theta\tfrac{1}{\sqrt{2}}F_{\mu\nu}\sigma^{\mu\nu}\lambda+\theta^2(\tfrac{1}{4}F^{\mu\nu}F_{\mu\nu}-i\epsilon_{\mu\nu\rho\sigma}F^{\mu\nu}F^{\rho\sigma})+\dots \ .
\end{align}
\end{subequations}
Each component in the chiral superfield $\Phi_0$ corresponds to a
source term for a boundary operator and we have suppressed the complex phase in $\phi_0$ and $\eta_0$ for ease of exposition. 
The $AdS$/CFT correspondence relates the sources to the leading modes
of the corresponding bulk fields at the boundary.

Let us notice that a source for the dilaton $\phi_0$ introduces an exactly marginal deformation which redefines the gauge coupling constant $1/g^2$ 
in front of the kinetic term of the boundary gauge degrees of freedom while a source for the squashing mode $\eta_0$ is giving an explicit 
mass to the CFT gauginos.
While a source for the dilaton preserves
supersymmetry a  mass for the gauginos breaks supersymmetry in an
explicit way introducing a soft term. Conversely a non zero VEV for
$F^{\mu\nu}F_{\mu\nu}$ dual to the subleading mode for the dilaton
breaks supersymmetry while the gaugino condensate triggered by the
subleading mode of the squashing field preserves $\mathcal{N}=1$
SUSY. Indeed, these two subleading modes 
are dual to the VEVs for the highest and lowest components of
$\mathcal{O}$, respectively.

We now want to introduce the action for the $5d$ bulk theory. We will
focus on an $\mathcal{N}=2$  SUGRA which has the minimal field content
we are interested in, and which moreover can be thought of as
a truncation to a subsector of the full $\mathcal{N}=8$ SUGRA.

Following the general analysis of \cite{Ceresole:2000jd,Ceresole:2001wi}, the scalars of this theory describe 
a manifold which is a direct product of a very special manifold $\mathcal{S}$ spanned by the scalar inside the vector multiplet 
$D$ and a quaternionic K\"ahler manifold $\mathcal{Q}$, that we take to
be the coset $SU(2,1)/U(2)\times U(1)$, spanned by the four real scalars of  the hypermultiplet $(\phi,C_0,\eta,\alpha)$.
The resulting metric can be written as
\begin{equation}
 \begin{split}
ds^2 = \frac{d D^2 }{2\left(1-\tfrac{D}{2\sqrt{3}}\right)^2} &
               +\frac{1}{2} \cosh^2\eta d \phi^2+2 d \eta^2 \\
             & +\frac{1}{2}\left(e^{\phi}\cosh^2\eta d C_0+2\sinh^2\eta d \alpha\right)^2
               +2\sinh^2\eta d \alpha^2\ . \label{Kmetric}
 \end{split}
\end{equation}
As in our SUGRA model we have only two vectors, we can gauge at most a $U(1)\times U(1)$ 
inside the maximal compact subgroup of the metric isometries. 
Since we are interested in having an unbroken $U(1)$ gauge symmetry in
the bulk,  
we choose to gauge the $U(1)_R$ corresponding to the shift symmetry of the phase $\alpha\to\alpha+c$. 
The gauging gives us a potential for the real scalar $\eta$ and non trivial couplings of the $U(1)_R$ 
gauge field $R_{\mu}$ with the scalars. 

All in all the bosonic Euclidean action for the sector given by the
gravity multiplet and the hypermultiplet  can be written as
\begin{align}
\mathcal{S}_{\text{gravity+hyper}}=\int d^5x&
  \sqrt{G}\left[-\frac{1}{2}R +
\frac{1}{4}\mathcal{R}^{MN}\mathcal{R}_{MN}+\frac{3}{2}\sinh^22\eta R^{M}R_{M}\right. \notag \\ 
&\ \ +
\partial_M\eta\partial^M\eta 
+ \frac{1}{4}\cosh^2\eta\partial_M\phi
\partial^M\phi+\frac{1}{4}e^{2\phi}\cosh^4\eta\partial_{M}C_0\partial^{M}C_0
\notag \\ 
&\ \ +\frac{1}{4} e^{\phi}\sinh^22\eta\partial_{M}C_0\partial^{M}\alpha+\frac{1}{4}\sinh^22\eta\partial^{M}\alpha
\partial_{M}\alpha+\mathcal{V}(\eta) \notag \\
&\ \ \left.-\sqrt{6}e^{\phi}\sinh^2 2\eta\partial^{M}C_0 R_{M}-\frac{\sqrt{6}}{2}\sinh^22\eta\partial^{M}\alpha R_{M}
\right]\ , \label{Sgh}
\end{align}
where the potential is\footnote{ We fix the coupling of the gauging to be $g=\frac{1}{L}=1$.}
\begin{equation}
\mathcal{V}(\eta)=\frac{3}{4}\left(\cosh^22\eta -4 \cosh2\eta-5\right)\ .\label{poteta}
\end{equation}

As for the vector multiplet, 
since we will be eventually interested in two point correlators, we need to know the
action  on the background only at
quadratic order. In a generic background with non trivial $\phi$
and $\eta$, such action reads:
\bea
\mathcal{S}_\mathrm{vector}&=& \int d^5x
\sqrt{G}\left[\frac{1}{4}F^{MN}F_{MN}+\frac{1}{2}(\bar\lambda \slashed{D}
   \lambda +c.c)-\frac{1}{2}(1-\sinh^2\eta)\bar\lambda\lambda\right.\nn\\
&&\qquad \qquad \quad-\frac{i}{2}\big( \sinh
   \eta\bar\lambda\lambda^c-\frac{1}{\cosh\eta}
   \bar\lambda\slashed{\partial}\eta\lambda^c+\frac{1}{2}\sinh\eta 
\bar\lambda\slashed{\partial}\phi\lambda^c+c.c.\big) \nn\\
&&\qquad \qquad \quad \left.+\frac{1}{2}\big(\partial_M
D\partial^M D-(4-2\cosh^2 2\eta+2\cosh2\eta)D^2\big)\right]\ .\label{squad}
\eea
From the action above we can see that the symmetry gauged by the vector $A_M$ does not affect 
the hyperscalars, so that
 non trivial profiles for $\phi$ and $\eta$ do not break the $U(1)_F$ global symmetry
dual to it. Note that the precise form of the actions above depends on
the specific gauging that has been chosen, which conforms with the one
which identifies this $\mathcal{N}=2$ theory as a subsector of $\mathcal{N}=8$ SUGRA.

There are also several other reasons to choose this specific gauging,
leading to the actions above. First, the fact that one
of the two scalars is the dilaton, whose source is dual to the gauge
coupling itself, reduces the space of physically relevant parameters
from four to three, since we should not count the source for $\phi$ as
a parameter of the solutions (rather, it  defines the duality
regime through the ratio between the Planck scale and the $AdS$
scale). 

Further, since our action can be identified as a truncation of maximal
SUGRA has the benefit that its solutions include some solutions which have
already been considered in that context, as
\cite{Kehagias:1999tr,Gubser:1999pk,Girardello:1999bd,Distler:1998gb}. Note
however that stability criteria in maximal SUGRA and in our
``minimal'' context can be, and indeed are, different.

The actions that we consider were also the basis for the work in
\cite{Argurio:2012cd,Argurio:2012bi}, where nevertheless only solutions with
small, non backreacting $\eta$ were considered. Considering here full
solutions with non trivial $\eta$ profiles will allow us on one side
to have more control on the signature of R symmetry breaking, finding
for instance an R axion resonance, while we will at the same time be able
to consider situations in which R symmetry is broken by large VEVs
and/or sources.

We must however also point out two drawbacks of the present set
up. The first is that we cannot turn on a source that breaks
conformality but not SUSY. That would be necessary in order to display
dynamical SUSY breaking. Here on the other hand we have to deal with
backgrounds that seem to have SUSY breaking VEVs and $\langle
T\rangle=0$ at the same time. This can only be motivated by blaming it
on the higher dimensional operators that have been neglected, i.e.~that
acquire large anomalous dimensions in the large $N$ limit (see for instance the
discussion in \cite{Gubser:1999pk}). To revert to a better state of
affairs, having the SUSY breaking dynamics under control in the SUGRA limit, one must consider a different gauging \cite{inpreparation}.

Another problem is related to the fact that though the scalars are not
charged under the gauge field $A_M$, the latter has a non trivial
Chern Simons term. This means that holographically the $U(1)_F$ has a
global anomaly. This does not affect in any way the two point
correlators, however it makes the global symmetry unsuitable to be
gauged. Finding a non anomalous $U(1)$, or in other words a $5d$
vector without CS couplings, would entail having to enlarge the model
to more than one vector multiplet, and then finding the adequate
linear combination. This is beyond the scope of the present work. In
Section 6,  we
will instead take the pragmatic approach of weakly gauging the $U(1)_F$, assuming that the two point functions are qualitatively similar to those of a
non anomalous symmetry.

\section{A full class of RG flow supergravity solutions}\label{allSOL}
In this section we are concerned with finding backgrounds which are solutions of the equations of motion derived from the SUGRA actions \eqref{Sgh} and \eqref{squad}. The reader not interested in the details of our treatment of the background equations of motion can directly go to Section \ref{classification} which summarizes our results.

As already stated, we will be interested in solutions in which only
the scalars $\phi$ and $\eta$ of the hypermultiplet have non trivial
profiles. Indeed, a profile for $C_0$ or $\alpha$ would necessarily
source the graviphoton $R_\mu$. We also demand that all fields of the vector multiplet, including the scalar $D$, are trivial. Requiring Poincar\'e invariance in $4d$, we further restrict the dependence of the scalars, and the metric, to be only along the radial direction.

From the above considerations,  we will take the ``flat domain wall'' ansatz for the metric
\be
 d s^2=\frac{1}{z^2}\left(dz^2+F(z)\eta_{\mu\nu}dx^\mu dx^\nu\right)\ . \label{parametrization}
\ee
This defines $z$ as the radial coordinate, with $z=0$ being the
boundary. We  take $F\to 1$ as $z \to 0$ for the metric to
be asymptotically $AdS$. For the active scalars in the background we write 
\begin{equation}
 \phi(z,x)=\phi(z)\ , \qquad \eta(z,x)=\eta(z)\ .  
\end{equation}

From \eqref{Sgh} we obtain the equations defining the background.  
We have a total of four equations, one of which is redundant:
\begin{subequations}
\begin{align}
&-2\frac{F''}{F}+\frac{{F'}^2}{F^2}+2\frac{F'}{zF}-\frac{4}{z^2} \notag \\
&\ \ \ \ \ \ \ \ \ \ \ \ \ =2{\eta'}^2+\frac{1}{2}\cosh^2\eta {\phi'}^2+\frac{1}{2z^2}\left(\cosh^22\eta
-4 \cosh 2\eta-5\right)\label{Fdprime}\\
&\frac{12}{z^2}\left(1-\frac{zF'}{2F}\right)^2+\frac{3}{2z^2}\left(\cosh^22\eta
-4 \cosh 2\eta-5\right)=2{\eta'}^2+\frac{1}{2}\cosh^2\eta {\phi'}^2
\label{Fprime}\\
& z^2\eta''-3z\eta'+2z^2\frac{F'}{F}\eta'=\frac{1}{8}z^2\sinh 2\eta {\phi'}^2
+\frac{3}{2}\sinh 2\eta (\cosh 2\eta-2) \label{eqeeta}\\
&\partial_z 
\left(\frac{F^2}{z^3}\cosh^2\eta \phi'\right)=0\ ,\label{eqphi}
\end{align}
\end{subequations}
where we have denoted by ${}'$ derivatives with respect to $z$.
In solving for $F$, $\phi$ and $\eta$, we can discard one among the first
two equations. It is usually more efficient to discard the first
one, which is second order in $F$. Similarly, the first integration of
the last equation is trivial and gives $\phi'$ as a function of $F$
and $\eta$,
\begin{equation}
 \phi'(z) = \frac{4\, \tilde{\phi}_4\, z^3}{F(z)^2 \cosh[\eta(z)]^2}\ ,
 \label{exactphi}
\end{equation}
where $\tilde{\phi}_4$ is an integration constant. Having integrated $\phi'(z)$
analytically, the determination of the background amounts to solve a system
of one second order and one first order equations in two variables.  

We now consider the task of finding all the solutions of the above set of
equations for $F$, $\eta$ and $\phi$. One can start solving the equations above using an asymptotic
expansion near the boundary which, for the scalar $\eta$, takes
the following form
\begin{equation}
 \eta\underset{\overset{{z\to0}}{}}{\simeq} z\left(\eta_0+\tilde{\eta_2}\, z^2+ ...\right)\ .
\end{equation}
According to the holographic dictionary, $\eta_0$ and $\tilde{\eta}_2$ are
respectively interpreted as the source and the VEV of the dual boundary operator
associated to $\eta$. Hence, solutions with vanishing $\eta_0$
describe an spontaneous R symmetry breaking; for $\eta_0\neq 0$
the same symmetry breaking is instead explicit.
More precisely, we expand near the $z=0$ (UV) boundary $F$ and $\eta$ as
\begin{subequations}
\bea
F &\underset{\overset{{z\to0}}{}}{\simeq}& 1 + f_2\, z^2  +f_4\, z^4 \log z+ \tilde{f}_4\, z^4+ \dots\label{expaf}\\
 \eta &\underset{\overset{{z\to0}}{}}{\simeq} &z\left(\eta_0+ \eta_2\, z^2 \log z+\tilde{\eta}_2\, z^2+ \dots\right)\ .\label{expaeta}
\eea
\end{subequations}
The expansion of $\phi$ is completely fixed by \eqref{exactphi}. 

Let us count the integration constants for the sake of completeness. 
As we have already stated, there are two equations to solve, one first and one second order. 
We thus need three integration constants. These can be taken to be $\eta_0$ and $\tilde{\eta}_2$, 
while the third integration constant is fixed to be $f_0=1$ by normalization of the metric. 
Note also that we get two additional integration constants from the equation for $\phi$, 
one being $\tilde{\phi}_4$ and the other being $\phi_0$. However,  any solution can be trivially shifted to a new solution with a different value of the constant piece $\phi_0$. Hence the latter can be discarded from the list of parameters defining the solutions and we will fix it to zero from now on. All in all, the parameter space is three dimensional and spanned by
\begin{equation}
\text{SUGRA parameter space}=\lbrace\eta_0, \tilde{\eta}_2, \tilde{\phi}_4\rbrace\ .\label{parspace}
\end{equation}
Plugging  the  expansions \eqref{expaf}--\eqref{expaeta} in the equations of motion, one obtains the expressions for all the other 
coefficients in terms of the integration constants in \eqref{parspace}. For instance:
\begin{equation}\label{backrel}
f_2 = -\frac{1}{3}\eta_0^2\ , \qquad \eta_2 = \frac{8}{3}\eta_0^3\ ,  \qquad f_4 = -\frac{4}{3}\eta_0^4\ ,\qquad 
\tilde{f}_4 = -\frac{1}{2}\eta_0\tilde{\eta}_2 +\frac{5}{18}\eta_0^4\ .
\end{equation}
We see that the terms with logarithmic dependence on $z$ are needed
when $\eta_0\neq 0$.%
\footnote{Such logarithmic terms imply that there
  is an ambiguity in the term with the corresponding power of $z$, as
  for instance $\tilde \eta_2$, that can be fixed through
  holographic renormalization. Note that the scale that must appear in
  the logarithmic terms can be taken to be one of the scales defining
  the background, such as the location of the singularity if there is one, or more simply the $AdS$ radius as we implicitly do here (with $L=1$).}
On the other hand, when $\eta_0=0$ we would need
to go much deeper in the expansion of $F$ in order to see the
distortion created by the scalar profiles (at $z^6$ in the presence of an
$\eta$ profile with $\tilde{\eta}_2\neq 0$ or at $z^8$ for a dilatonic
profile with $\tilde{\phi}_4\neq 0$). 

The trouble with this approach, i.e.~propagating from the boundary, is that
we do not know whether we are building a singular solution or not. It
is thus important to study what kind of singularities are possible, or
under which conditions a non singular solution can be built. To this end,
one has to analyze the equations of motion in the deep bulk, assuming
either that there is no singularity, or that some of the fields
have a specific (singular) behavior. In this way we will be able to
classify all the possible solutions to our system of equations. 

\subsection{Singular backgrounds}
We first turn to backgrounds where the geometry ends at some finite value
of $z$ which determines the position of the naked singularity. 

The geometry will typically end because one of the scalars blows
up. At the same time the warp factor $F$ goes to zero. In fact, backgrounds where $F$ blows up at the singularity 
are unlikely to represent any physical RG flow \cite{Maldacena:2000mw}. Moreover, it can be shown by performing an IR analysis similar to the one we present here below
 that there are no solutions of the equations of motions (\ref{Fdprime}-\ref{eqphi}) having a finite $F$ at the singularity.  
 
We can parametrize the IR behavior near the singularity by a
handful of parameters. First of all, if the singularity is situated at
$z_\mathrm{sing}$, we then choose the new coordinate near the
singularity as $x\equiv z_\mathrm{sing}-z$. The position of the singularity should be thought as a function of all the scales in the problem, namely those defining the parameter space \eqref{parspace}.  The behavior of the fields at the singularity would then be fixed at the leading order by 3 exponents fixing the functional dependence on $x$ plus 3 unknown coefficients which are again function of the UV parameters \eqref{parspace}. 
In what follows, we find two distinct cases depending on which scalar
blows up. We show that in both cases the functional dependence of $F$,
$\eta$ and $\phi$ at the sigularity is completely fixed by the
equations of motion. The unknown coefficients have, instead, different features in the two cases. 

\subsubsection{Backgrouds with $\eta$ blowing up}\label{etablowingup}

We parametrize the way $F$ and $\eta$ approach the singularity by two exponents
\be
F \sim x^\varphi \ , \qquad e^\eta \sim x^{-n}\ ,
\ee
with both $\varphi>0$ and $n>0$.
As usual the behavior of $\phi$ is dictated by \eqref{exactphi}.
The equations of motion for $F$ and $\eta$ are, at leading order near the singularity:
\bea
\frac{n}{x^2}-\frac{2n\varphi}{x^2} & = & c_\eta x^{-4n} + c_\phi x^{2n-4\varphi}\ , \nonumber \\
\frac{3\varphi^2}{x^2}-\frac{2n^2}{x^2} & = & -c_\eta x^{-4n} + 2c_\phi x^{2n-4\varphi} \ ,
\label{eomsing}
\eea
where $c_\eta$ and $c_\phi$ are two positive definite constants that
depend on the various parameters of the solution, and can be read from
the equations of motion. In particular, the vanishing of either of the
constants implies the vanishing of the entire profile for the scalar
in the subscript. 

One can study all the cases, according to whether the terms multiplied by $c_\eta$ and/or $c_\phi$ contribute to determine the solution near  $x=0$. First of all, it is obvious that there is no non trivial solution where one or both of the $c_\eta$ and $c_\phi$ terms dominate over the $x^{-2}$ left hand side.

Then, considering  the $c_\eta$ term to be subdominant, i.e.~$n<1/2$,
one quickly runs into contradictions (both if the $c_\phi$ term is
subdominant or if it goes like $x^{-2}$). We are thus forced to take
$n=1/2$.

We could take all the terms to scale like $x^{-2}$. This fixes also
$\varphi=3/4$ (this would lead to $\phi'\sim x^{-1/2}$). 
However we see that the first of \eqref{eomsing}
becomes $-1/4 = c_\eta +c_\phi$ which is a contradiction because, as
we have already stated, the two constants are positive definite. We thus
conclude that the $c_\phi$ term is necessarily subleading in this
class of solutions.

Finally, we are left with finding solutions with $n=1/2$ and
$\varphi<3/4$. Eliminating the $c_\eta$ term from \eqref{eomsing} we
find $\varphi=1/3$. We can summarize this class of solutions by the
following behavior near the singularity at $x=0$:
\be
F\sim x^{1/3}, \qquad e^\eta \sim x^{-1/2}, \qquad \phi'\sim x^{1/3}
\ .
\ee
Thus we see that the warp factor vanishes, $\eta$ blows up and the
dilaton kinks to a finite value. All functions take
a generic behavior near the singularity (i.e. the three coefficients in front of their leading functional dependence are unspecified), so that no values of the
near boundary parameters is selected. Indeed, numerically, it is easy to realize that for a generic point in the
$3d$ parameter space \eqref{parspace}, we find a solution of this kind. 

\subsubsection*{\it SUSY backgrounds}
Within this class of solutions we find ``half'' of the supersymmetric solution of GPPZ \cite{Girardello:1999bd}. This corresponds to taking a particular limit of the GPPZ background in which the mass deformation goes to zero and just the gaugino condensate is present. In our truncation the solution is obtained for $\eta_0=0$ and $\tilde\phi_4=0$. The remaining parameter $\tilde\eta_2$ fixes the scale of the spontaneous R symmetry breaking (i.e. of the gaugino condensate).

The warp factor and the $\eta$ profile can be written in $z$ coordinates as 
\begin{equation}
F(z)=\left(1-z^6\tilde{\eta}_2^2\right)^{\frac{1}{3}}\ ,\qquad \eta(z)=\frac{1}{2}\ln\left(\frac{1+\tilde{\eta}_2z^3}{1-\tilde{\eta}_2z^3}\right)\ . \label{GPPZback}
\end{equation}
From the analytic solution it is clear that the value of $\tilde{\eta}_2$  determines the position of the singularity which is inversely proportional to $\tilde{\eta}_2^{1/3}$.

Besides the supersymmetric solution, we have a variety of other
solutions, depending on which parameters we turn on. We review some
specific case below. Actually, since the singular behavior of $F$ and $\eta$ in the
whole class of solutions is exactly the same as in the supersymmetric solution, it is quite natural to interpret the non supersymmetric solutions as deformation of the SUSY one. 

\subsubsection*{\it Spontaneous R symmetry breaking backgrounds}
The first deformation is obtained by switching on $\tilde{\phi}_4$. In this class of solutions only parameters which are dual to
VEVs are turned on. In other words, no explicit symmetry breaking is present
and both conformal and $R$ symmetry are spontaneously broken. Supersymmetry is also broken spontaneously because of the non vanishing $\tilde{\phi}_4$. 
\begin{figure}[h]
\centering
\includegraphics[width=50mm]{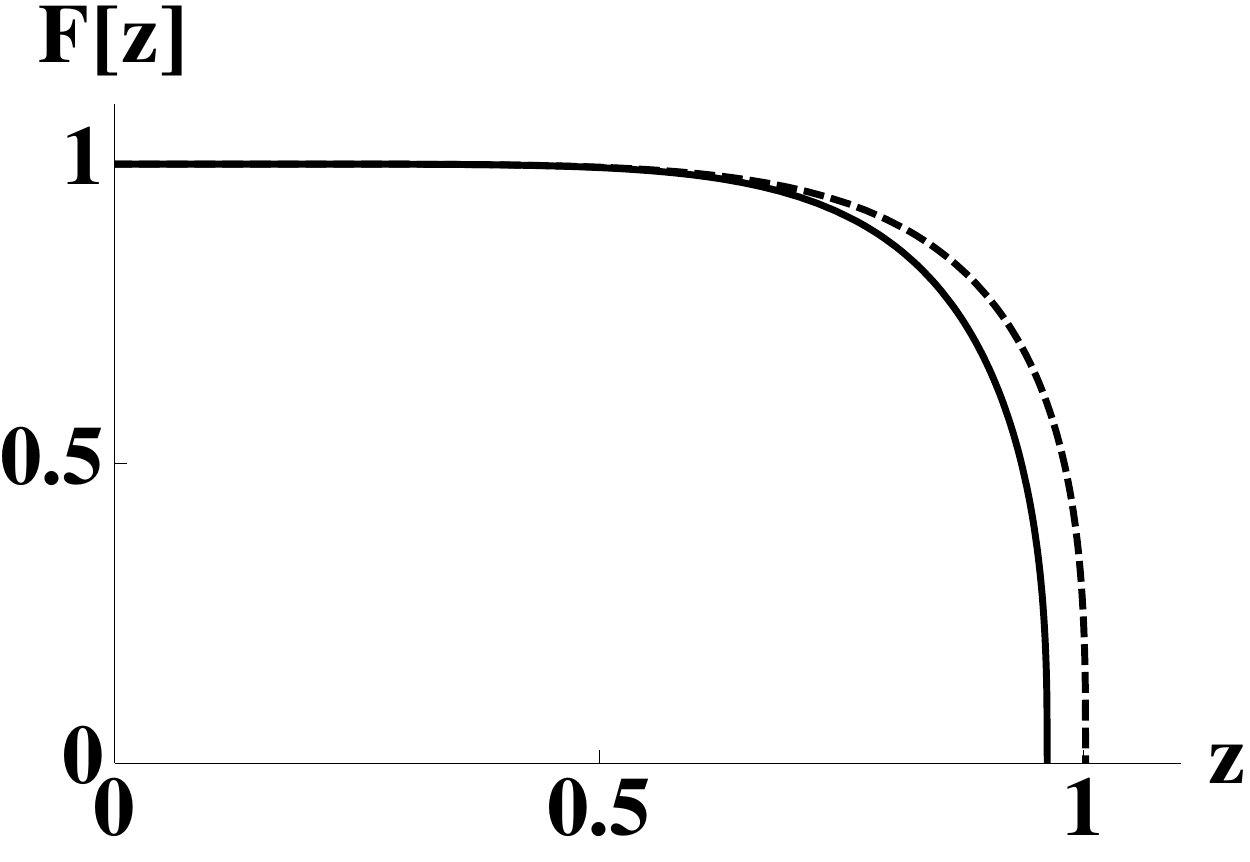} \hspace{0.2cm}
\includegraphics[width=50mm]{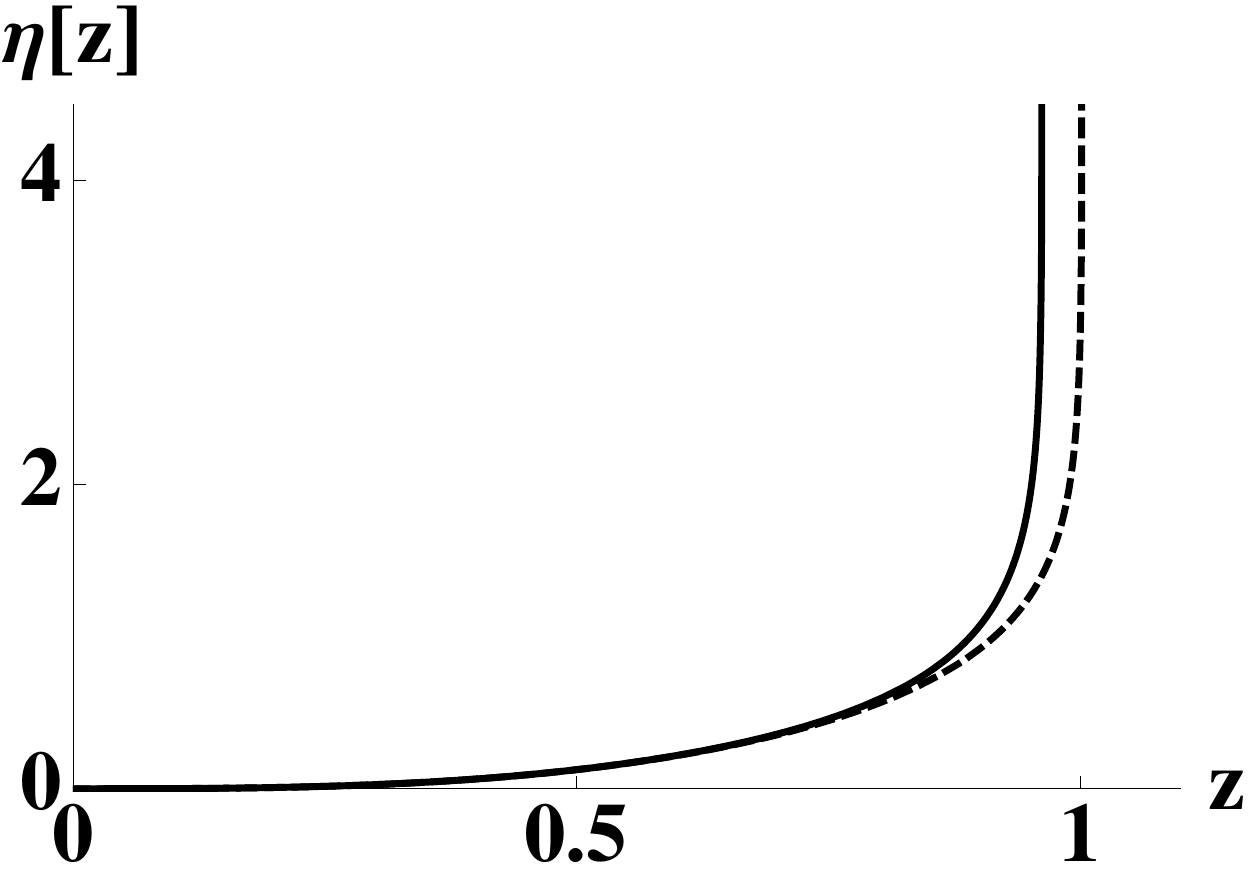}\hspace{0.2cm}
\includegraphics[width=50mm]{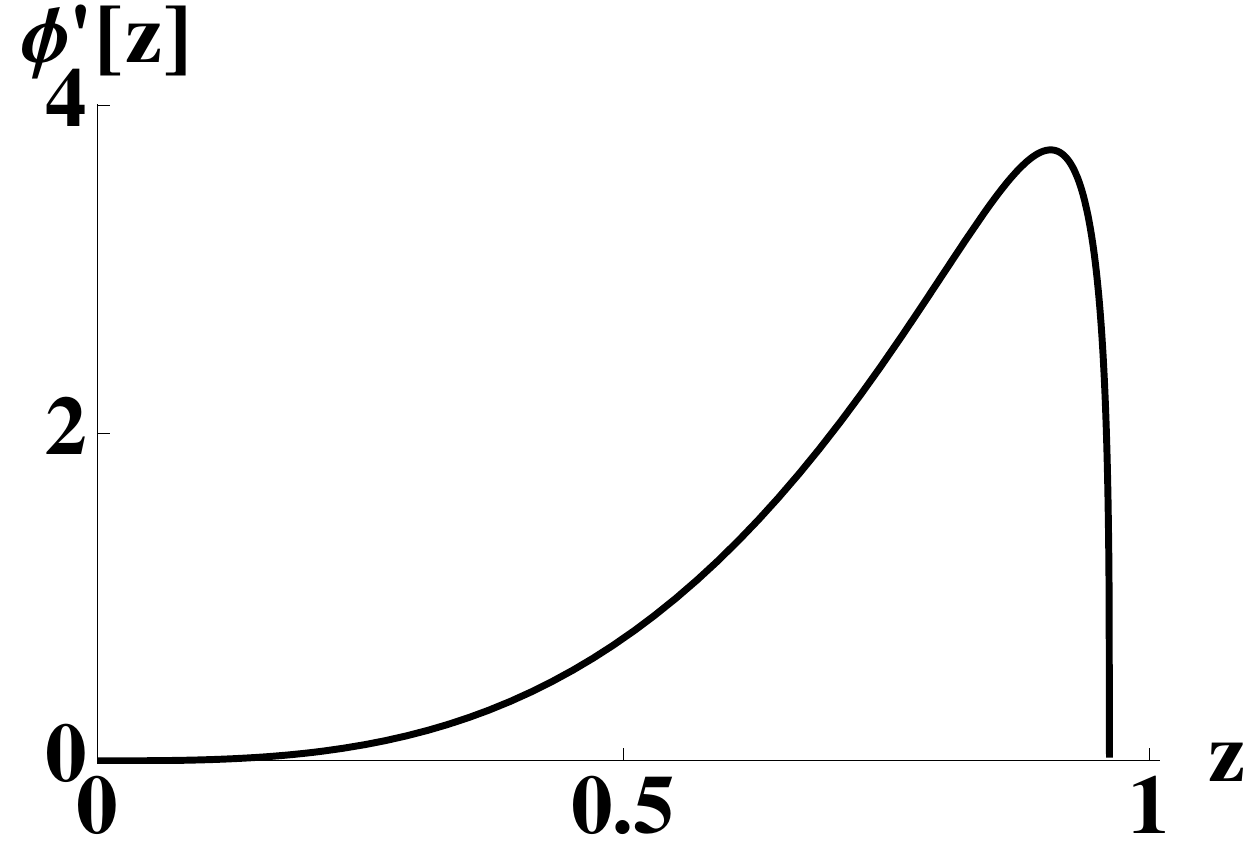}
\caption{$F(z)$, $\eta(z)$ and $\phi'(z)$ profiles in the spontaneous
R symmetry breaking background. The values of the integration constants are specified in the text. The dashed line corresponds to the SUSY background with $\tilde{\eta}_2= 1$ where the $\phi'(z)$ profile vanishes.}
\label{sponta}
\end{figure}

As a representative example, we display the background in Figure \ref{sponta} obtained by numerically integrating the equations of motion with $\{\eta_0, \tilde{\eta}_2, \tilde{\phi}_4\}= \{0,1,1.5\}$. We observe that the geometry becomes singular (i.e. $F(z)$ vanishes)
as the scalar field $\eta$ diverges. The dilaton 
profile instead presents a ``kink'' shape interpolating between two constant
values (note that we plotted $\phi'$). However, the non trivial dilaton profile due to the non vanishing $\tilde{\phi}_4$ contributes to the position of the singularity which gets smaller compared to the SUSY case.

\subsubsection*{\it Explicit R symmetry breaking backgrounds}
\begin{figure}[h]
\centering
\includegraphics[width=50mm]{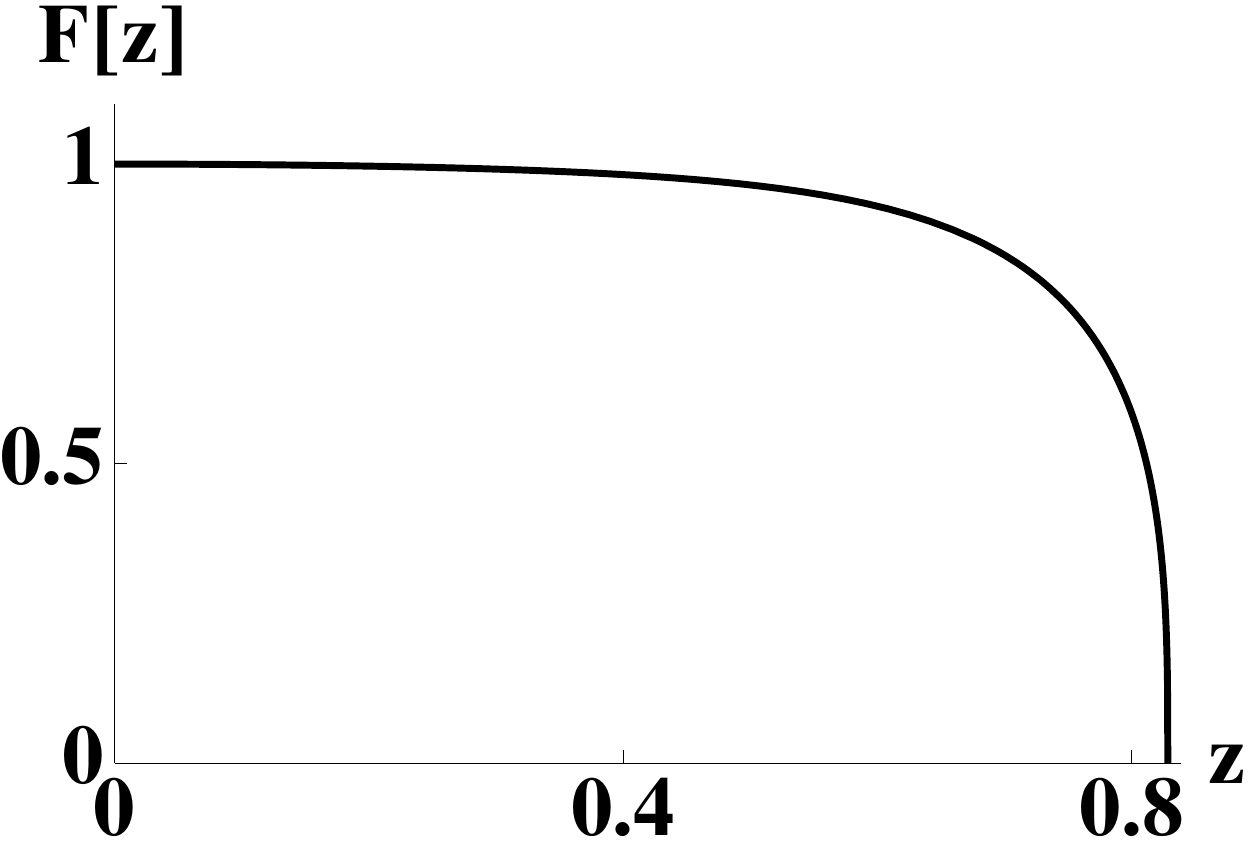}  \hspace{0.2cm}
\includegraphics[width=50mm]{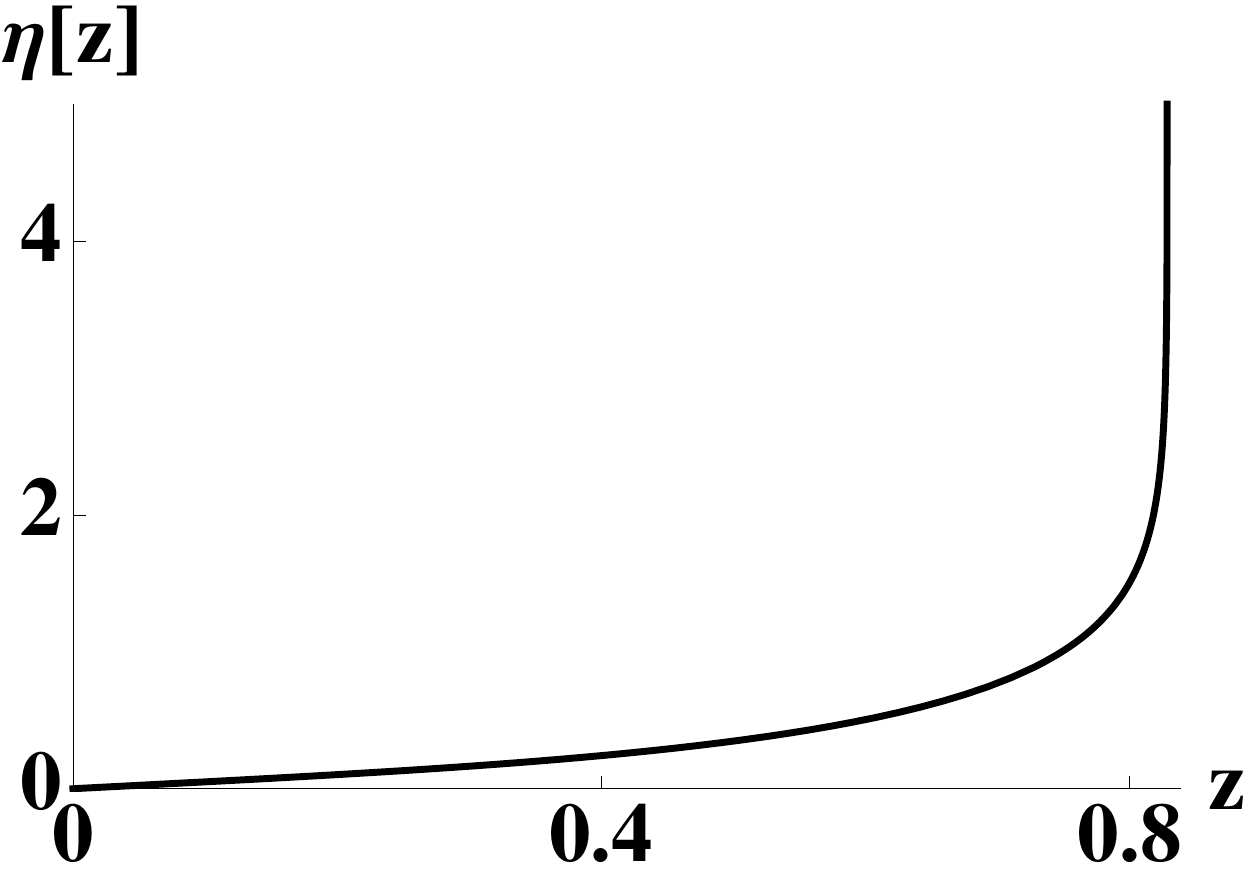} \hspace{0.2cm}
\includegraphics[width=50mm]{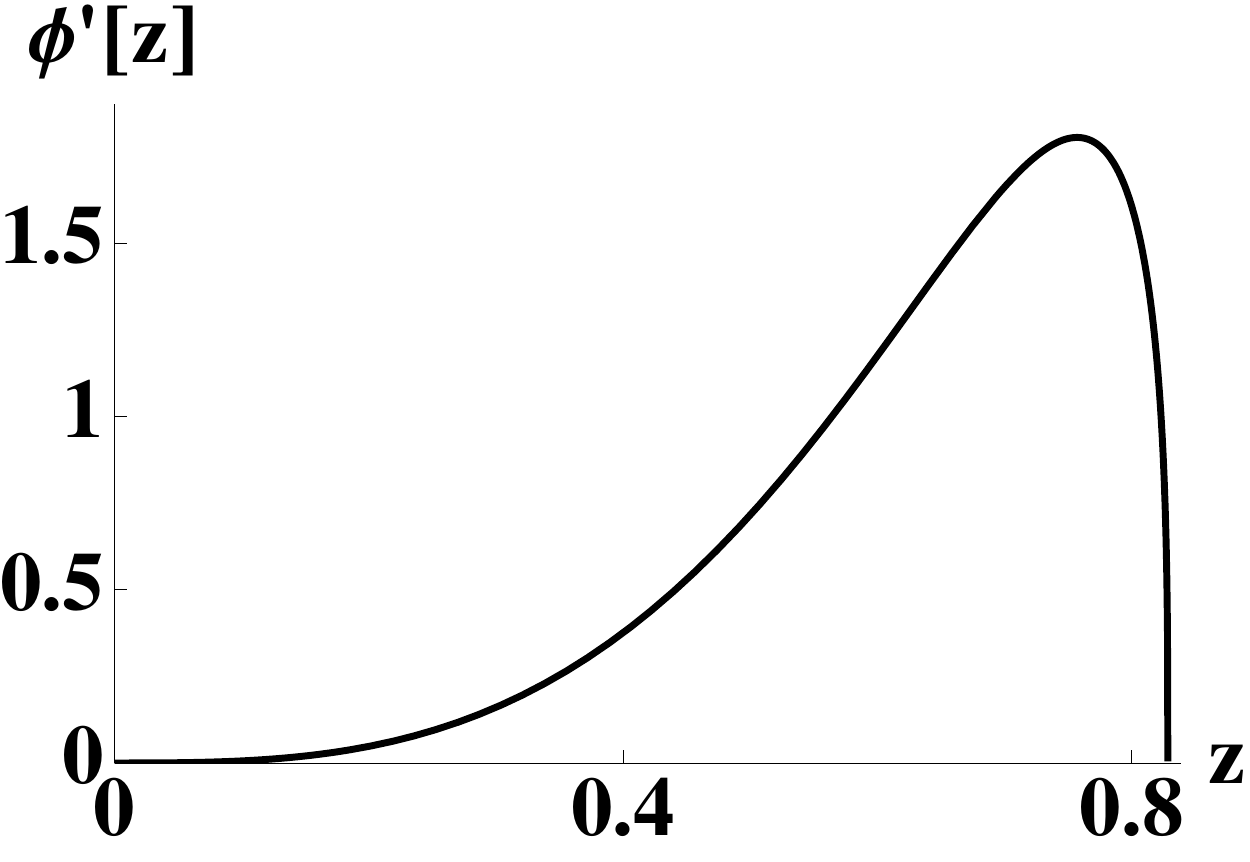}
\caption{$F(z)$,  $\eta(z)$ and $\phi'(z)$ profiles in the explicit
$R$ symmetry breaking background. The values of the integration constants are specified in the text.}
\label{expl}
\end{figure}

We can also allow for a non vanishing source term for the 
scalar field $\eta$ which breaks conformality and $R$ symmetry explicitly. This corresponds to adding a non supersymmetric mass term for the gauginos in the
dual gauge theory. For illustrative purposes, we consider the
background obtained as a solution of the equations of motion with
$\{\eta_0, \tilde{\eta}_2, \tilde{\phi}_4\}= \{0.5,1,1.5\}$.  The
results are plotted in Figure \ref{expl}. Similarly to the spontaneous
R symmetry breaking case, we have that the
geometry becomes singular as the field $\eta$ diverges. Again,
the dilaton interpolates between two constant values and the singularity gets even smaller compared to the previous case because of the presence of a non vanishing $\eta_0$ on top of the other scales.

\subsubsection*{\it Walking backgrounds}

\begin{figure}[h]
\centering
\includegraphics[width=50mm]{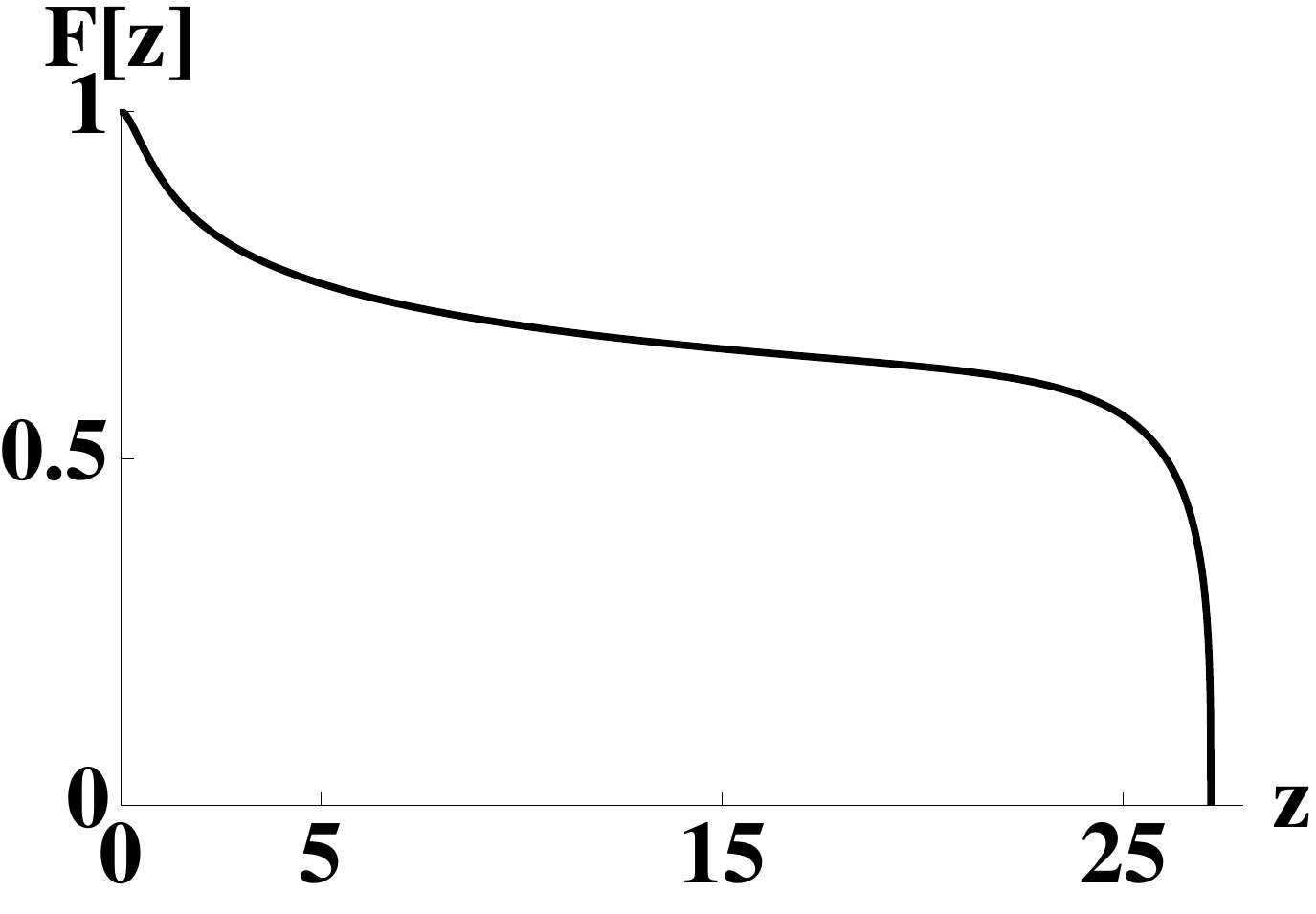} \hspace{0.2cm}
\includegraphics[width=50mm]{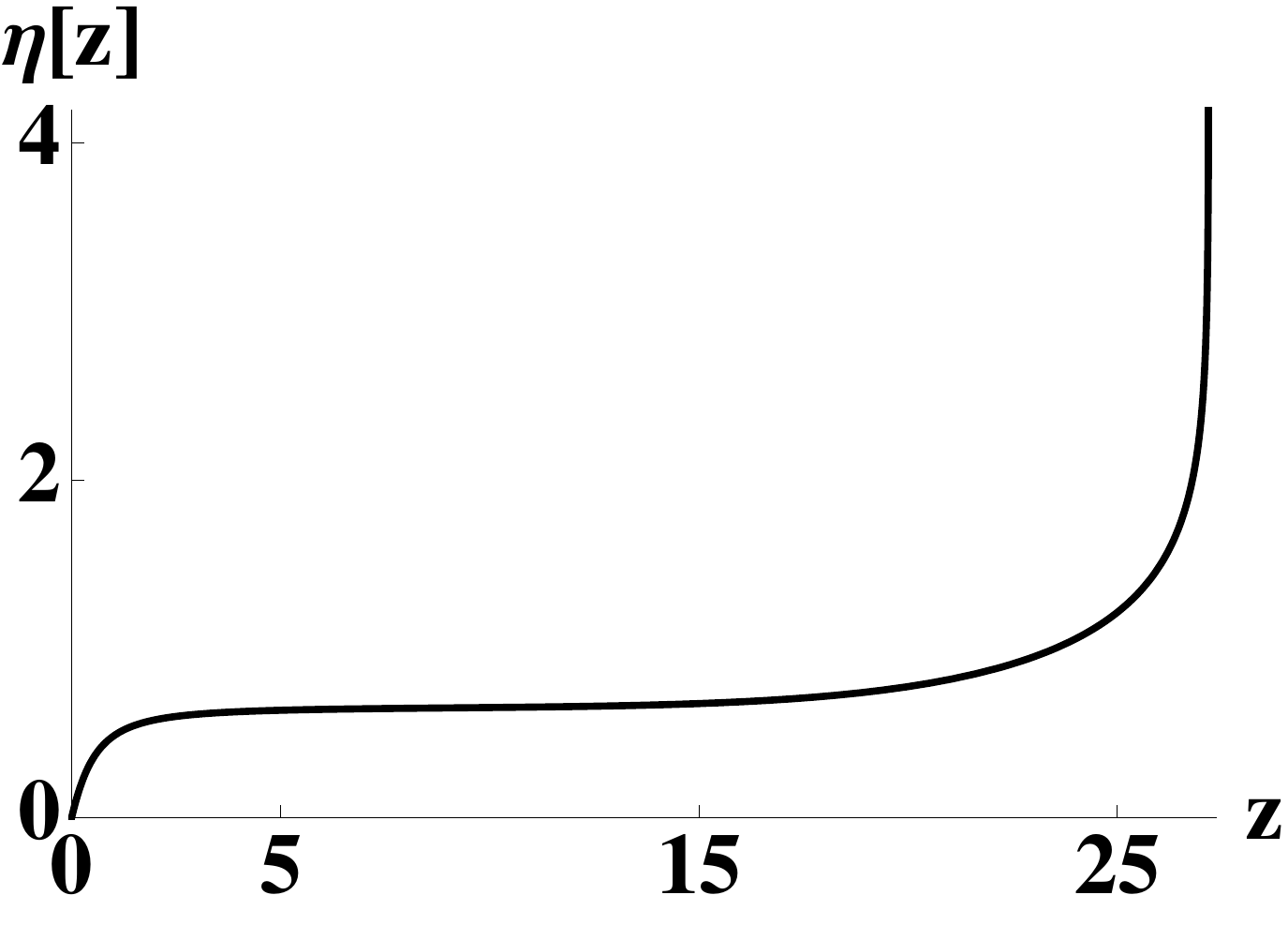} \hspace{0.2cm}
\caption{$F(z)$ and $\eta(z)$ field profiles in the walking solution;
here $\phi'(z)=0$.}
\label{walk}
\end{figure}

Considering still the explicit symmetry breaking solutions but turning
$\tilde{\phi}_4$ off for simplicity,
we can numerically tune the ratio of the $\eta$ source and VEV
(namely $\eta_0/\tilde{\eta}_2$) to be close to the value that we
obtain for the non singular flows which we will present in detail in Subsection \ref{Nonsingular}. As a consequence, 
the singularity is pushed deep towards the infrared region.

We refer to these solutions as ``walking solutions''. They indeed
feature a wide, almost flat regime where $\eta(z)$
is approximately constant, corresponding to the minimum of the potential \eqref{poteta}. Of course,  eventually it
explodes and the geometry ends. In Figure \ref{walk} we plot an instance of such a walking solution.

\subsubsection{Backgrounds with dilaton blowing up}\label{dilatonback}
We now turn to considering solutions where $F\to 0$ but $\eta$ does not
blow up. It is easy to see that $\phi'$ has then to blow up at the
singularity. If we again parametrize $F\sim x^\varphi$ with
$\varphi>0$, we have that $\phi'\sim x^{-2\varphi}$ from \eqref{exactphi}.

From the first order equation \eqref{Fprime} we immediately see that,
for consistency with the fact that $\eta$ does not blow up, the
${\phi'}^2$ term must balance the $(F'/F)^2$ term, so that
$\varphi=1/2$, and the ${\eta'}^2$
term must be subdominant. Note that since $\phi'\sim x^{-1}$ then the
dilaton blows up logarithmically at the singularity. The same equation
also gives a non trivial relation between the parameters of the solutions, which can be written implicitly as  
\begin{equation}
\tilde{\phi}_4=\frac{\sqrt{6}F^2_{\text{sing}}}{8z_{\text{sing}}^4}\ ,\label{relation}
\end{equation}
where  $F_{\text{sing}}$ is
defined by $F\sim F_{\text{sing}}(x/z_{\text{sing}})^{1/2}$, and
$z_{\text{sing}}$ and $F_{\text{sing}}$ should be thought as functions of all the
boundary parameters in \eqref{parspace}.
We now assume that $\eta$ is finite or vanishes at the singularity,
going as $\eta \sim x^{\tilde n}$, with $\tilde n\geq 0$. (Note that it
is now $\eta$ and not $e^\eta$ that scales as a power of $x$ in the vicinity
of the singularity.)

First we observe that the equation of motion for $\eta$ cannot be
satisfied near the singularity if $\eta$ tends to a non vanishing
finite value there. We henceforth assume $\tilde n>0$. Then using the
value of $\tilde \phi_4$ fixed in \eqref{relation}, the equation is
satisfied near the singularity only if $\tilde n=\sqrt{3/8}$. 

This class of solutions is defined by the
following behavior near the singularity (i.e. at $x=0$):
\begin{equation}
F\sim x^{1/2}, \qquad \eta \sim x^{\sqrt{3/8}}, \qquad \phi'\sim x^{-1}\ ,
\end{equation}
plus the relation between the UV coefficients \eqref{relation} which defines a two dimensional subspace of the
3 dimensional parameter space. 

\subsubsection*{\it Dilaton domain wall background}
For a vanishing $\eta$ profile we recover the dilaton domain wall solution of \cite{Kehagias:1999tr,Gubser:1999pk}. The only scale in the system is given by $\tilde{\phi}_4$ which preserves $R$ symmetry but breaks SUSY.  
The warp factor and the $\phi$ profile can be written in $z$ coordinates as
\begin{equation}
F(z)=\left(1-\frac{\tilde{\phi}_4^2z^{8}}{6}\right)^{1/2}\ ,\qquad \phi(z)=\sqrt{6}\ \mathrm{arctanh} \left(\frac{\tilde{\phi}_4z^4}{\sqrt{6}} \right)\label{ddw}\ .
\end{equation}
From the analytic solution one can see that $\tilde{\phi}_4$ determines the position of the singularity which is inversely proportional to $\tilde{\phi}_4^{1/4}$. The dashed line in Figure \ref{Bulletto} corresponds to a particular solution of this class where $z_{\text{sing}}=1$ and $\tilde{\phi}_4=\sqrt{6}$, accordingly. 

\subsubsection*{\it Dilaton like background}

The other solutions in the two dimensional subspace can be thought as generalizations to non vanishing $\eta$ of
the dilaton domain wall solution \eqref{ddw}. Indeed we showed that the behavior near the
singularity is the same as in the pure dilatonic solution. A way of seeing this fact is to consider the $\eta$ profile as a
perturbation over the background \eqref{ddw} as it was considered in \cite{Argurio:2012cd}. In this approximation 
$\tilde\phi_4$ fixes the overall scale of the background (i.e.~the location of the singularity). It is then obvious that the
second order equation for $\eta$ has generically two independent solutions, but only one linear combination will be vanishing at the
singularity. This selects a particular ratio between $\eta_0$ and
$\tilde \eta_2$, the boundary parameters characterizing the $\eta$ profile.
Since this is certainly true for small values of $\eta_0$ and
$\tilde \eta_2$, it remains true also for arbitrary values. 

\begin{figure}[h]
\centering
\includegraphics[width=50mm]{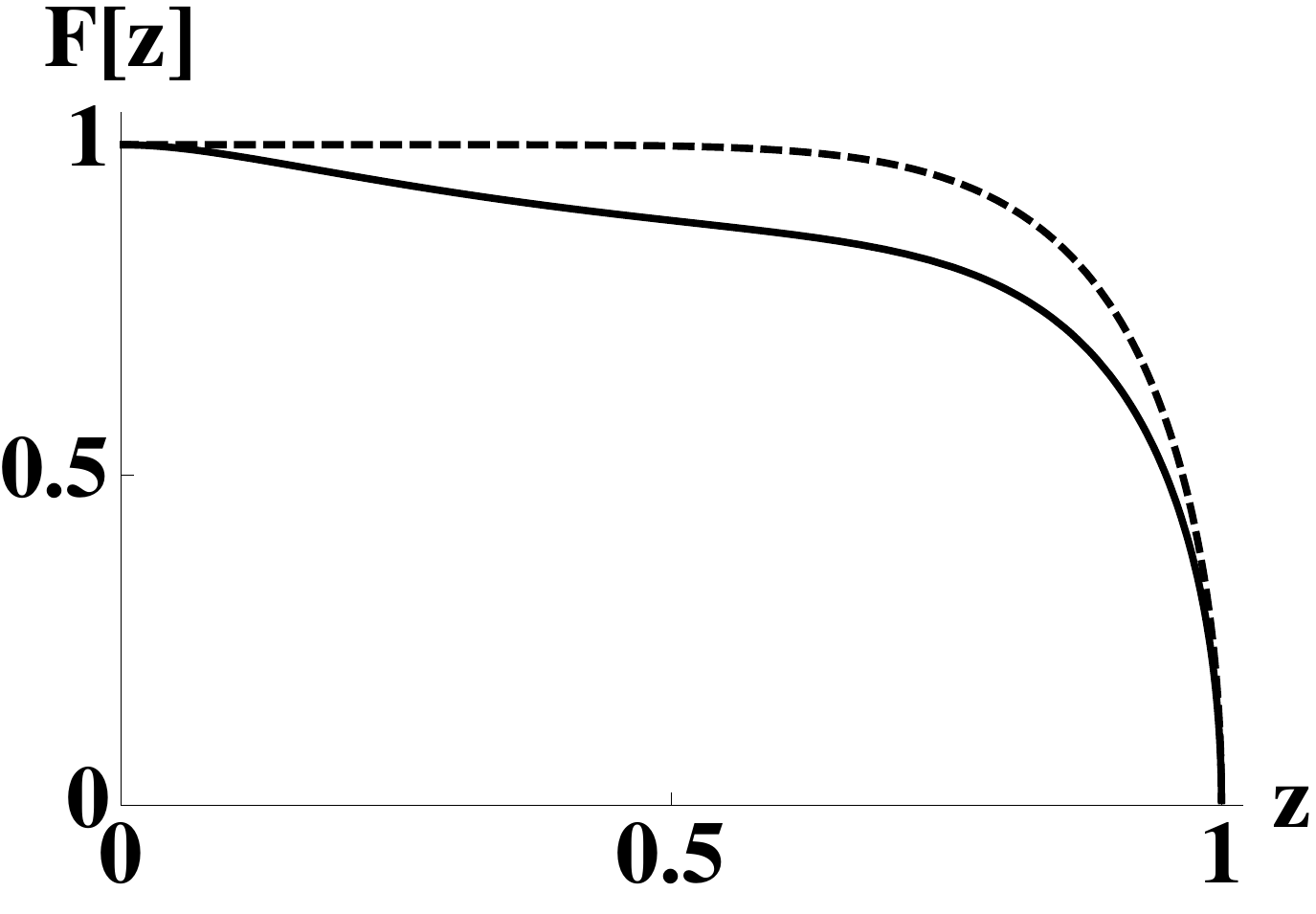} \hspace{0.2cm}
\includegraphics[width=50mm]{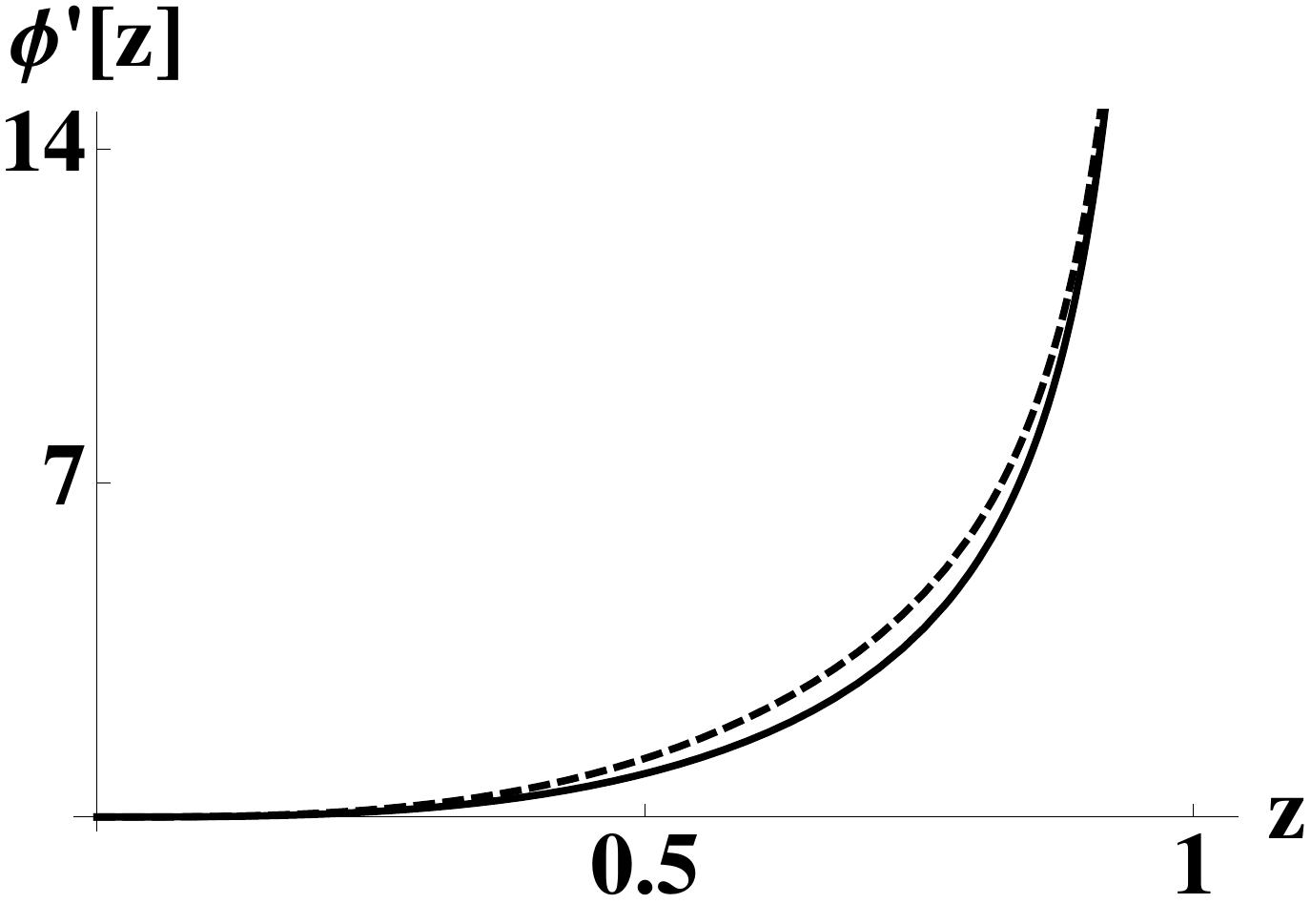}\hspace{0.2cm}
\includegraphics[width=50mm]{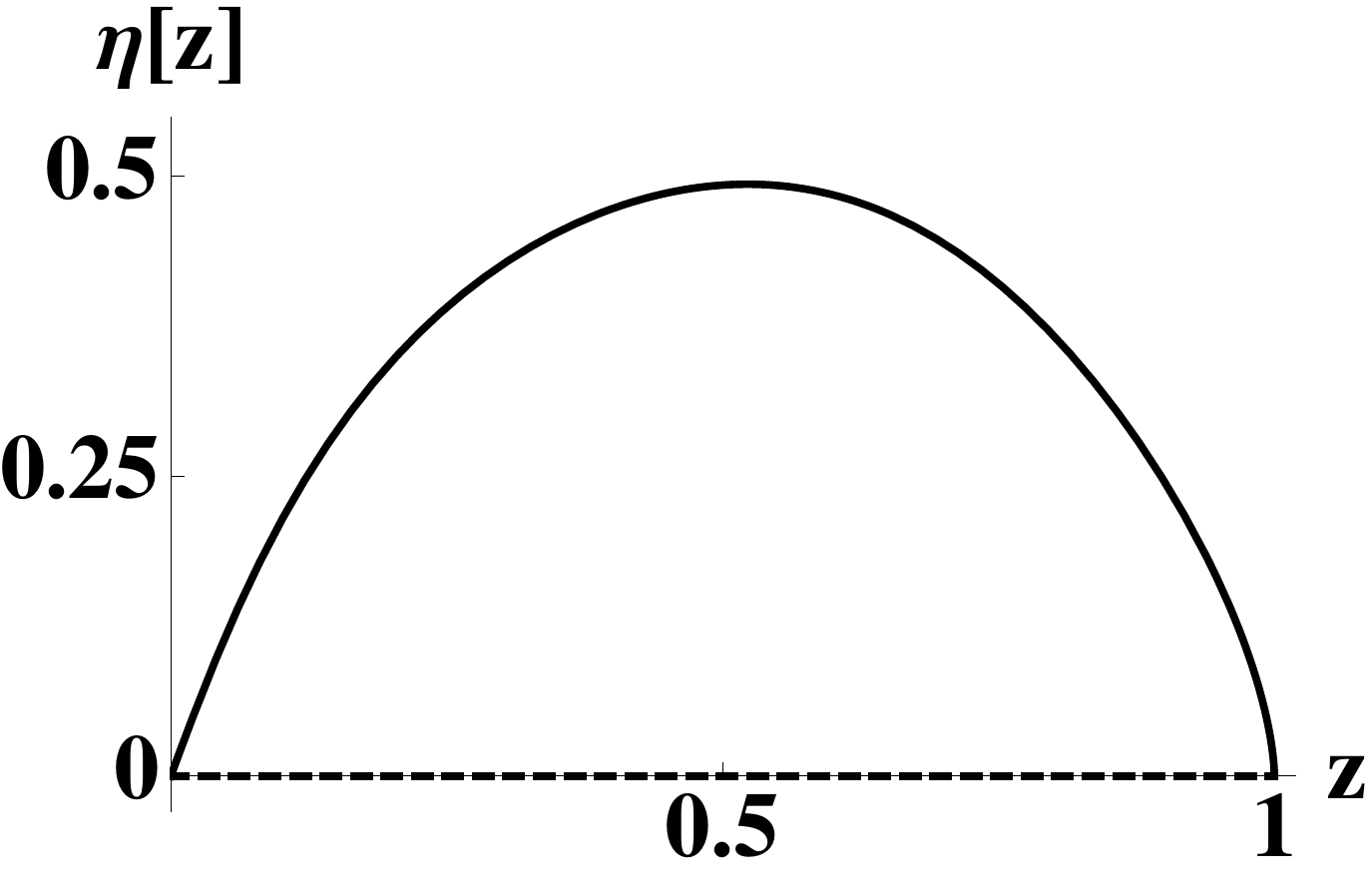}
\caption{$F(z)$, $\phi'(z)$ and $\eta(z)$ profiles for a dilaton like background. Extracting the UV parameters numerically we find $\eta_0=67$, $\tilde{\eta}_2=2.5$ and $\tilde{\phi}_4=1.8$. The dashed lines represent the dilaton domain wall solution \eqref{ddw} with $\tilde{\phi}_4=\sqrt{6}$.}
\label{Bulletto}
\end{figure}

Numerically, this means one has to build these backgrounds by imposing
the boundary conditions near the singularity, since it will be
impossible to find conditions on the $z=0$ boundary that fall on the
zero measure subspace of parameters that corresponds to this class. A natural choice to describe the $2d$ subspace of dilaton like solutions is to take as free parameters the position of the singularity $z_{\text{sing}}$ and the coefficient in front of the leading term in the expansion of $\eta$ at the singularity 
\begin{equation}
\eta\sim \eta_w x^{\sqrt{3/8}}\ .\label{etawdef}
\end{equation}
Once the solution is found numerically for a particular value of $\{z_{\text{sing}},\eta_{w}\}$, one can extract from its UV expansion the values of $\tilde{\phi}_4$, $\eta_0$ and $\tilde{\eta}_2$. The solid line in Figure \ref{Bulletto} shows a particular solution with $z_{\text{sing}}=1$ and $\eta_{w}=1$ which is compared with the dilaton domain wall solution with $\tilde{\phi}_4=\sqrt{6}$ discussed before.   

\subsection{Non singular backgrounds}\label{Nonsingular}
Non singular backgrounds are solutions for which the coordinate $z$
extends from the boundary at $z=0$ all the way to $z\to
\infty$. Alternatively, one can use a more symmetric set of coordinates where the variable spans from $+\infty$ to
$-\infty$. This second set of coordinates has the advantage of making easier to visualize the regular flows.

In order to define the new set of coordinates we introduce an alternative ansatz for the metric
\be
ds^2=dy^2 +
e^{2A(y)}\eta_{\mu\nu}dx^\mu dx^\nu\ . \label{parametrization2} 
\ee
This parametrization is simply related to the previous one by
$z=e^{-y}$ and $e^{2A}=\frac{F}{z^2}$. Note that the boundary is at
$y\to \infty$ and that there $A\sim y$.  In these coordinates, Eqs.~\eqref{Fdprime}--\eqref{eqphi} read:
\begin{subequations}
\begin{align}
&-4\ddot A-4\dot A^2=2{\dot\eta}^2+\frac{1}{2}\cosh^2\eta {\dot\phi}^2+\frac{1}{2}\left(\cosh^22\eta
-4 \cosh 2\eta-5\right)\\
&12\dot A^2+\frac{3}{2}\left(\cosh^22\eta
-4 \cosh 2\eta-5\right)=2{\dot\eta}^2+\frac{1}{2}\cosh^2\eta
{\dot\phi}^2\\
& \ddot\eta+4\dot A\dot\eta=\frac{1}{8}\sinh 2\eta {\dot\phi}^2
+\frac{3}{2}\sinh 2\eta (\cosh 2\eta-2)\\
&\partial_y 
\left(e^{4A}\cosh^2\eta \dot\phi\right)=0\ ,
\end{align}
\end{subequations}
where $\dot{}\equiv \partial_y$. 
Again, the first integration of
the last equation is trivial and gives $\dot \phi$ as a function of $A$
and $\eta$,
\begin{equation}
 \dot\phi(y) = -\frac{4\, \tilde{\phi}_4}{e^{4A(y)} \cosh[\eta(y)]^2}\ ,
 \label{exactphi2}
\end{equation}
where $\tilde{\phi}_4$ is the same integration constant as before. 

In order to have a regular solution in our system, the deep bulk metric must asymptote
to $AdS$, but for the flow to be non trivial, it must describe the geometry
associated to the other extremum of the potential, at $\bar\eta$ such
that $\cosh 2 \bar \eta= 2$. We are thus looking for a class of
solutions comprising, and possibly generalizing, the solution of
\cite{Distler:1998gb}. 

If we normalize the $AdS$ radius in the $\eta=0$ vacuum to have $L=1$,
then at the $\eta=\bar\eta$ vacuum the radius is given by $\bar L=
2\sqrt{2}/3$. Consistently with the fact that $\bar L< L$, the only possible
flow is  from the $\eta=0$ vacuum near the boundary to the
$\eta=\bar\eta$ vacuum in the deep bulk. The boundary conditions on the metric function and for $\eta$ will be
\be
A(y) \to y, \quad \eta \to 0 \qquad \mbox{for} \qquad y \to +\infty \label{UVfixedpoint}
\ee
and
\be
A(y) \to \frac{3}{2\sqrt{2}}y , \quad  \eta \to \bar \eta \qquad \mbox{for} \qquad y \to -\infty\ . \label{IRfixedpoint}
\ee
From \eqref{exactphi2} we immediately see that, if $\tilde\phi_4\neq
0$, then for $y \to -\infty$ we have that $\phi \propto
e^{-3\sqrt{2}y}$ and it blows up. This would be incompatible with
the other equations of motion where $\dot\phi^2$ appears. These terms must
be negligible in the deep bulk for the $AdS$ solution to be
asymptotically obtained. Hence we conclude that we must have
$\tilde\phi_4=0$ in the class of non singular solutions. 

\begin{figure}[h]
\centering
\includegraphics[width=50mm]{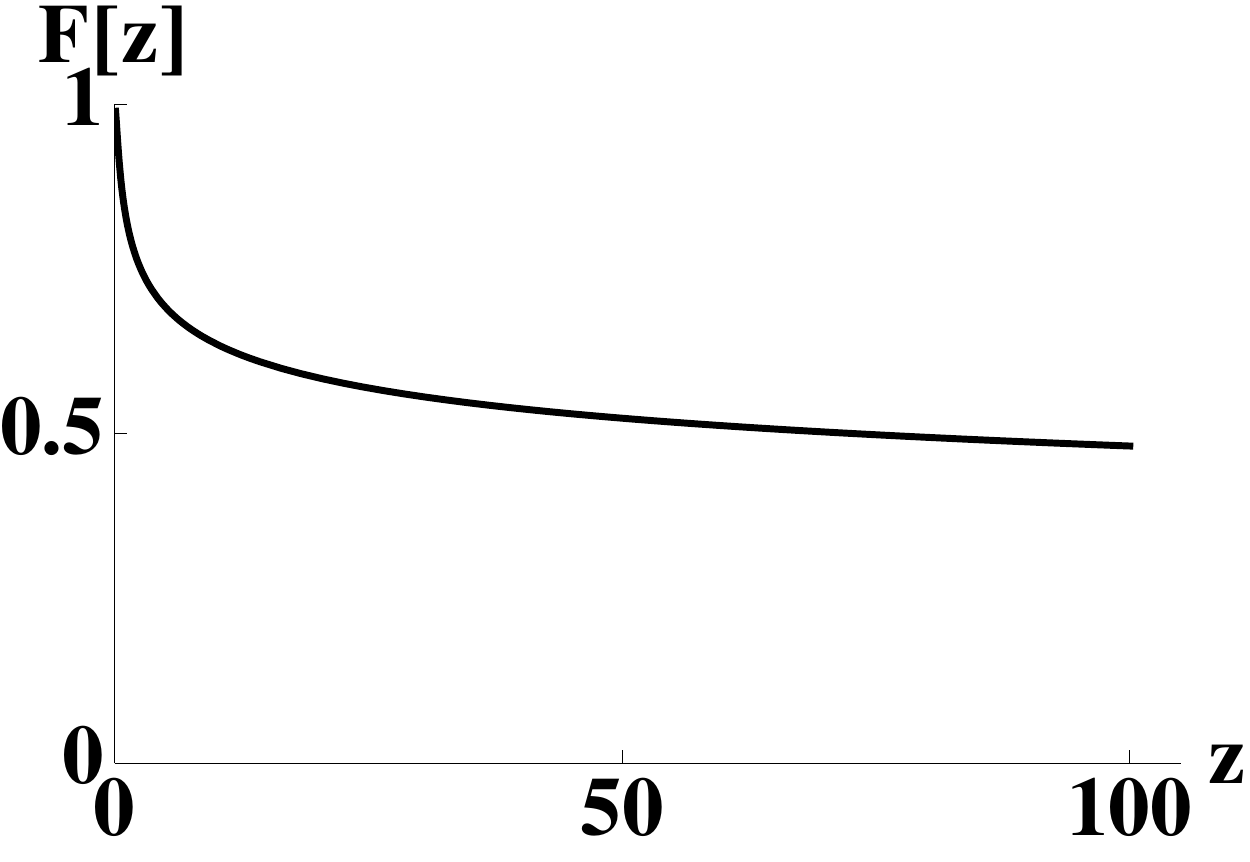} \hspace{0.2cm}
\includegraphics[width=50mm]{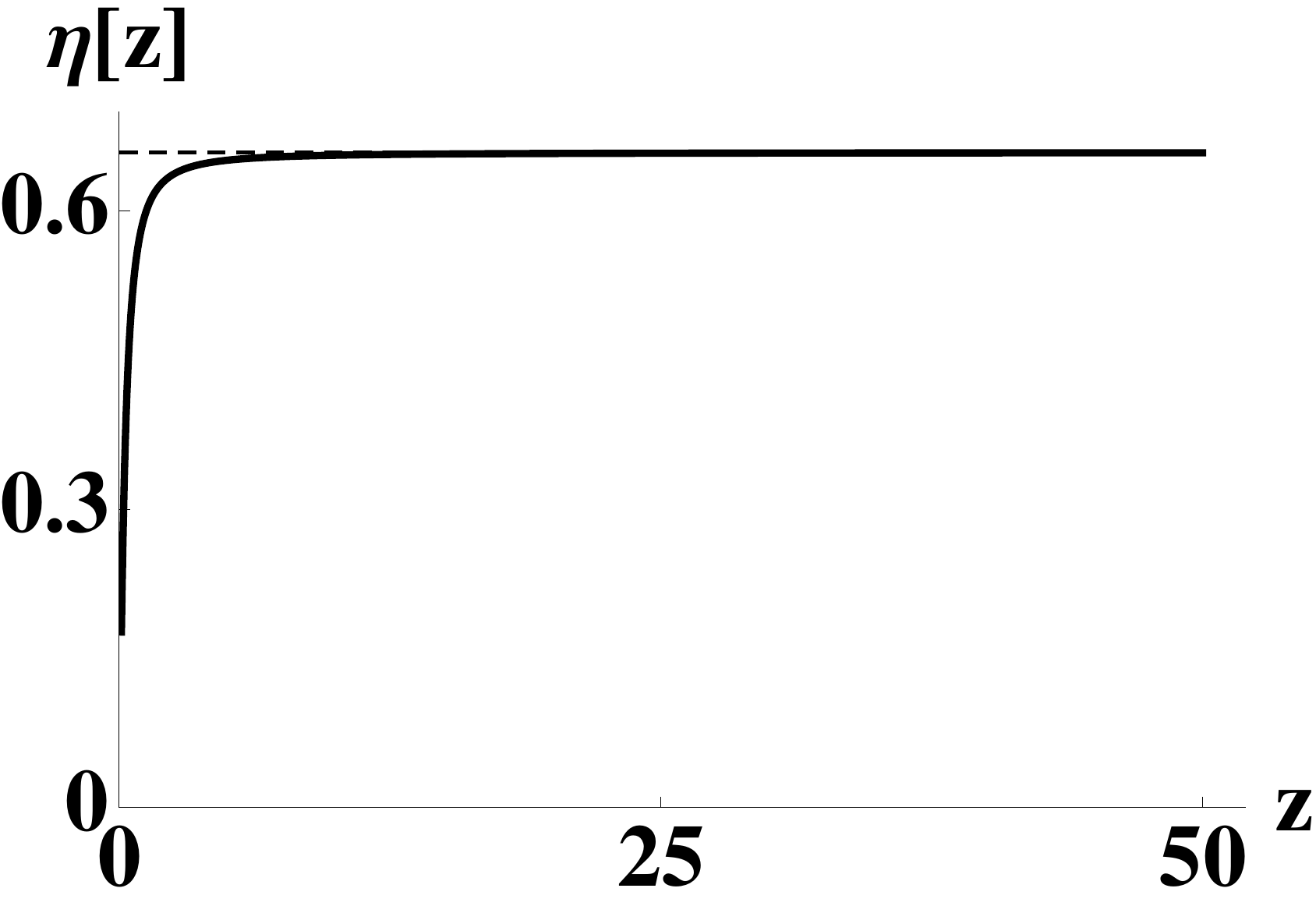}\hspace{0.2cm}
\includegraphics[width=50mm]{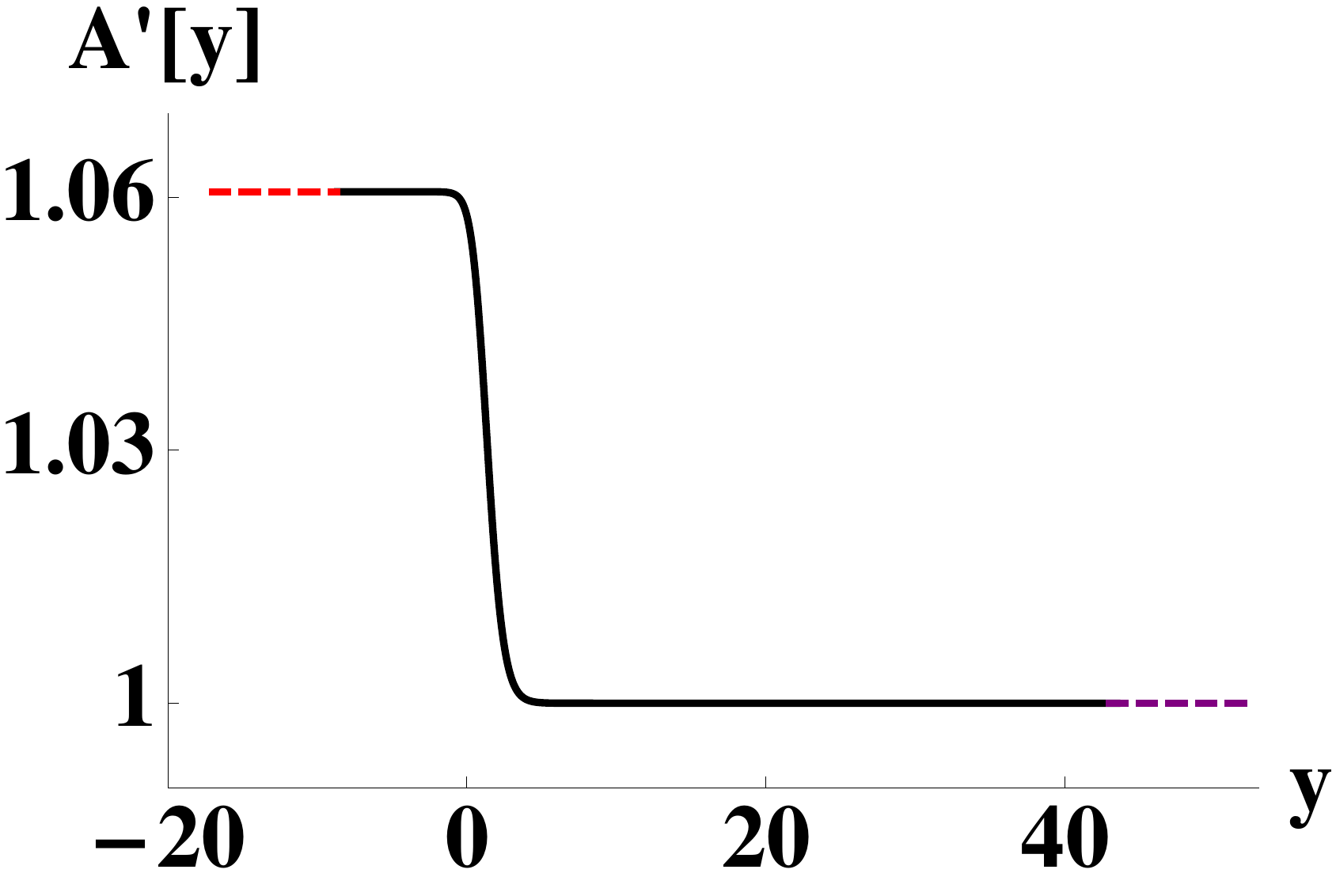}
\caption{$F(z)$ and $\eta(z)$ profiles for a non singular background. $\phi'(z)$ is set to zero.
On the right we display a plot for $A'(y)$ defined in \eqref{parametrization2}. The violet dashed line corresponds to the $AdS$ radius at the UV fixed point \eqref{UVfixedpoint} which we fixed to be one. The red dashed line corresponds to the $AdS$ radius of the IR fixed point \eqref{IRfixedpoint}.}
\label{DZ}
\end{figure}

We can now study how $\eta$ can approach its extremum value of $\bar
\eta$ for $y\to-\infty$. If we write $\eta =\bar\eta +\delta\eta$, we obtain
two independent asymptotic solutions $\delta\eta=e^{n_\pm y}$
with
\be
n_\pm = \frac{3}{\sqrt{2}}(-1\pm \sqrt{3})\ .
\ee
In order to have  $\delta \eta\to 0  $ for $y\to -\infty$, we need to
select the solution with $n_+$. This means that only one combination of the
two independent solutions for $\eta$ corresponds to  a non singular
background. In other words, the values of $\eta_0$ and $\tilde\eta_2$
giving rise to a non singular flow must be related. We are thus
reproducing the solution of \cite{Distler:1998gb} and nothing
more. Out of the 3 dimensional parameter space, we select a line in
the $(\eta_0,\tilde\eta_2)$ plane at $\tilde\phi_4=0$. The only
parameter left is essentially related to the value of $y$ (or, in RG
flow language, to the scale) at which the transition between the two
vacua takes place (i.e.~the maximum value of $|\ddot A|$).
The ratio between $\eta_0$ and $\tilde\eta_2$ that selects this
interpolating flow can be determined numerically. For the solution displayed in Figure \ref{DZ} we find ${\eta_{0}=0.215,\tilde{\eta}_2=0.036}$.
We reproduce one such flow in Figure \ref{DZ} while we plot the line of non singular solutions in our $3d$ parameter space in Figure \ref{THECUBE}.

\subsection{Classification}\label{classification}
\begin{figure}[h]
\centering
\includegraphics[width= 0.8\textwidth]{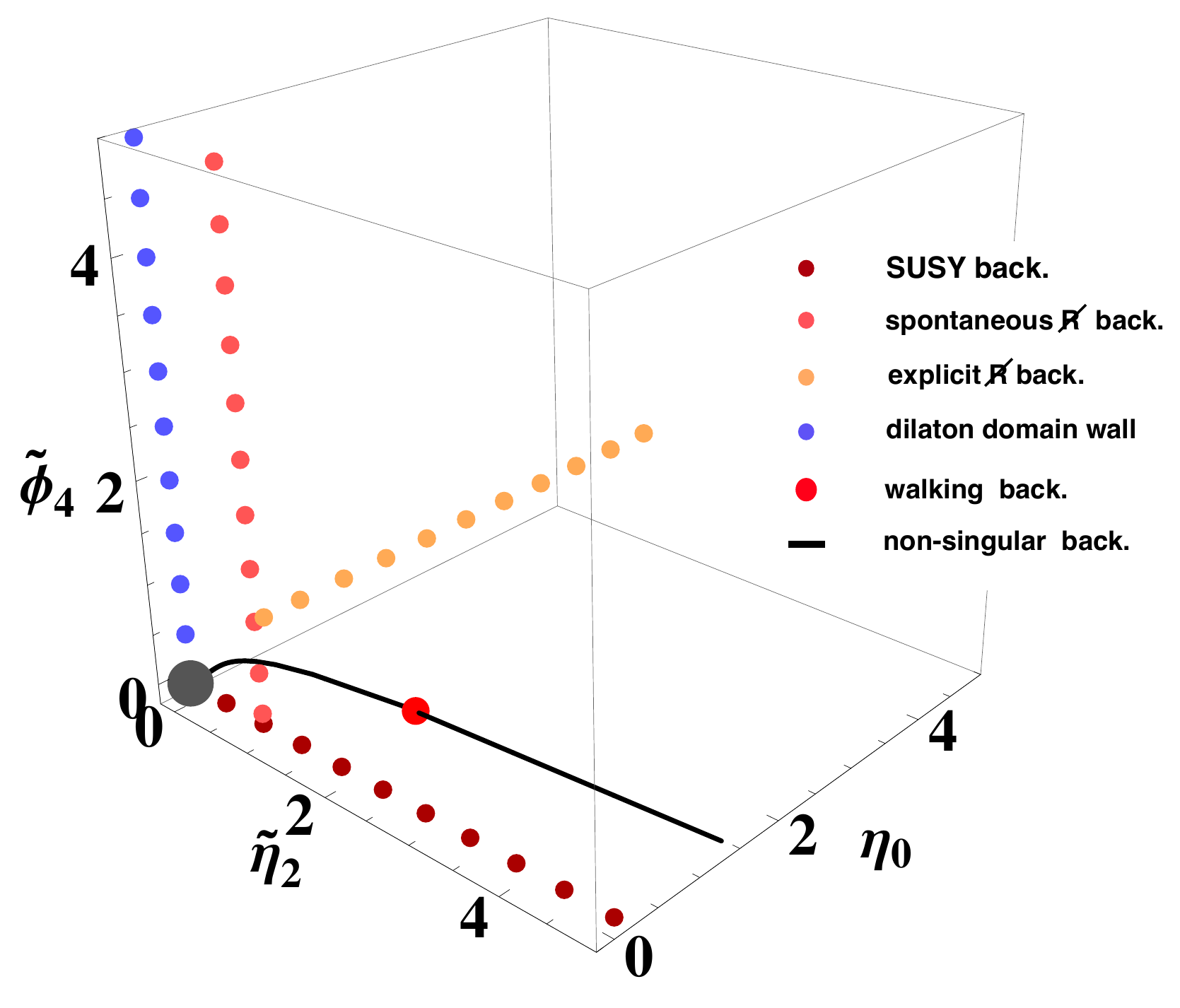}
\caption{A pictorial description of the $3d$ parameter space \eqref{parspace}. The different solutions discussed in the previous sections correspond to different color in the $3d$ cube. For each type of solution we pick up a line in the parameter space whose physical features will be described in Section \ref{Results}.}
\label{THECUBE}
\end{figure}

To summarize, we display in Figure \ref{THECUBE} a portion of the $3d$ parameter space which is covered by our solutions.

Our classification of the backgrounds is the following:
\begin{itemize}
\item[{\it i)}] For generic values of $(\eta_0, \tilde \eta_2,
  \tilde\phi_4)$, we have a singular background where for some value
  of $z$ the warp factor goes to zero, $\eta$ blows up and the dilaton
  kinks to a finite value. This class of solutions can be thought as obtained from the line $\eta_0=\tilde{\phi}_4=0$  corresponding to the supersymmetric GPPZ solution. By switching on either the SUSY breaking VEV $\tilde{\phi}_4$ or the SUSY breaking soft term $\eta_0$ or both, one can span the entire volume. A particularly interesting region is the plane where $\eta_0=0$ and only VEVs are present. We select a line in this plane with $\tilde{\eta}_2=1$ in order to study how the physics changes at increasing values of the SUSY breaking parameters $\tilde{\phi}_4$. In order to study the effect of $\eta_0$, which is breaking both SUSY and $U(1)_{R}$ explicitly, we select a line with $\tilde{\eta}_2=\tilde{\phi}_4=1$.\footnote{Taking $\tilde{\phi}_4=0$ in order to have a SUSY preserving vacuum seems artificial in a theory where a soft term has been switched on.} This concludes our analysis of the generic singular backgrounds.  
   
\item[{\it ii)}] For non vanishing values of $\tilde\phi_4$
  and $\eta_0$ a fixed function of  $\tilde \eta_2$, we have singular
  backgrounds where at some value of $z$ the warp factor goes to zero,
  the dilaton blows up and $\eta$ goes to zero. This class of backgrounds can be thought as a deformation of the dilaton domain wall line of solutions where $\eta_0=\tilde{\eta}_2=0$. The ratio between $\tilde \eta_2$ and $\eta_0$ should be tuned in order to have a vanishing $\eta$ profile at the singularity. For that reason the parametrization  \eqref{parspace} is not well suited to describe this class of solutions.
  
\item[{\it iii)}] For $\tilde\phi_4=0$ and $\eta_0$ a specific function  of
   $\tilde \eta_2$, we have the non singular backgrounds
  interpolating between the $\eta=0$ $AdS$ solution near $z=0$ to the
  $\cosh2\eta=2$ $AdS$ solution for $z\to \infty$ found by Distler and Zamora. As shown in Figure \ref{THECUBE} this line of solutions crosses the  $\eta_0=0$ axis just in one point where also $\tilde{\eta}_2=0$. This is consistent with the fact that we have just one superconformal fixed point in our truncation which corresponds to the $AdS_{5}$ solution with $L=1$. Generic solutions close to the line of non singular backgrounds display a walking behavior, where the singularity is pushed far away in the $z$ coordinate as in Figure \ref{walk}. 
\end{itemize}
We will now consider the salient physical properties of all the
above backgrounds, including their stability, through the two point correlators of the gauge invariant operators discussed in Section \ref{sugra}.

\section{Physical properties of the RG flows from the correlators}\label{Results}
The physical interpretation of SUGRA backgrounds describing 
flows from a SCFT in the UV to an IR theory with a gapped phase
becomes tricky because of the generic presence of naked singularities in the deep interior of the bulk. Several proposals have been made in the literature to distinguish the background singularities which are physically acceptable or not \cite{Gubser:2000nd, Maldacena:2000mw, Apreda:2003sy}.

From our point of view, a background singularity in a supergravity solution describing a flow in a field theory is allowed if it satisfies two physically motivated requirements. 

The first requirement is that the supergravity background is
describing a RG flow which respects the holographic $c$ theorem. This
poses a condition on the domain wall warp factor
\cite{Girardello:1998pd,Freedman:1999gp} which can be summarized most
simply in the parametrization \eqref{parametrization2} by $\ddot A<0$. This is 
a way to ensure that the number of degrees of freedom is decreasing
along the RG flow.  The condition holds for all the possible flows in
our model, since it descends directly from the equations of
motion for the background.  We will thus not need to worry about it in
the following.
 
The second requirement is that the supergravity background is
\textit{stable} within our truncation. This
can be studied by considering the linearized fluctuations of all the fields in our SUGRA.  Linearized fluctuations of a given SUGRA field in an $AAdS$ background can be mapped via the holographic dictionary to a two point correlator for the corresponding operator in the boundary theory. The two point correlators can have poles and branch cuts corresponding to one particle exchanges of bound states or multi particle exchanges respectively. Since our theory is strongly coupled we expect the one particle exchanges to be dominant and computing the two point correlators will give us information about the masses and the spectral density of the resonances. In order for a background to be \textit{stable} within the truncation we should not find any tachyonic resonance, namely no pole for $k^2>0$ (i.e. in Euclidean momentum space).  

Two remarks are in order about this second requirement. The first is that
overall stability depends on the truncation. For instance, a
background could be stable in our model, but there could be
unstable modes if one were to consider additional fields that belong
to a more general SUGRA. So it is possible that some of the
backgrounds that are found to be stable  in our truncation, are not
stable if considered as backgrounds of the full gauged ${\cal N}=8$
supergravity. This is for example the case for the background of
\cite{Distler:1998gb} whose IR fixed point is known to be unstable in $\mathcal{N}=8$ \cite{Distler:1999tr}.

The second comment concerns the fact that the
singularity introduces some freedom in the bulk (IR) boundary conditions that
one can impose on the fluctuations. One might be worried that the
spectrum will depend on them. Most of the arbitrariness can be fixed by requiring normalizability (near the singularity) of the fluctuations. The remaining freedom can be then fixed using arguments based on symmetry, both SUSY and the $U(1)$s.


In the following, we are going to inspect the physical properties of the backgrounds essentially by directly probing their spectrum of resonances. In particular, we will be looking for massless or light resonances, associated to exact or approximate symmetries. We will extract the spectrum from two point correlators of the gauge invariant operators discussed in Section \ref{sugra}. These correlators will also allow us to verify Ward identities and their violation, in particular those derived from SUSY. 

For backgrounds such as the ones that we consider, which realize RG flows to vacua of strongly coupled gauge theories, the tool we have at hand to compute correlators is holographic renormalization  \cite{Mueck:2001cy,Bianchi:2001de,Bianchi:2001kw,Skenderis:2002wp}.
In this section we will present all the physical results, while we defer to the next section the discussion of the framework and some details of the computations that are needed.

The correlators that we are going to consider are the ones that
involve the operators discussed in Section~\ref{sugra}. It is
convenient to give a complete parametrization of the two point
correlators for  the FZ multiplet and the linear multiplet. Using just
Poincar\'e invariance we can rewrite the correlators in terms of
dimensionless scalar form factors. For the FZ multiplet we  follow the treatment of \cite{Argurio:2013uba}, to which we refer for the complete list. 
The correlators we will be concerned with are the bosonic real ones:
\begin{subequations}
\begin{align}
&\langle T_{\mu\nu}(k)\,T_{\rho\sigma}(-k)\rangle = -\frac 1 8 X_{\mu\nu\rho\sigma}\, C_2(k^2) -\frac{1} {12} \frac{m^2}{k^2}P_{\mu\nu}P_{\rho\sigma}\,F_2 (k^2) \ , \label{<TT>} \\
&\langle j_\mu^R(k)\,j_\nu^R(-k)\rangle = -P_{\mu\nu}\, C_{1R}(k^2) - \frac{1}{3}\,m^2\frac{k_\mu k_\nu}{k^2}\,F_1(k^2)\label{<jj>}\ , \\
&\langle x(k)\,x^*(-k)\rangle = \frac{2}{3}\,m^2 \,F_0(k^2) \label{<xxdagger>}\ ,
\end{align}
\end{subequations}
where $P_{\mu\nu}=k^2\eta_{\mu\nu}-k_\mu k_\nu$ is the transverse projector,
$X_{\mu\nu\rho\sigma}=P_{\mu\nu}P_{\rho\sigma}-3P_{\rho(\mu}P_{\nu)\sigma}$
is the transverse and traceless projector, and we recall that we are using indices $\mu=1,\dots 4$
for the $4d$ spacetime, assuming a Wick rotation to Euclidean signature.
The form factors $C_2$ and $C_{1R}$ are associated to the traceless and divergenceless component of the operators $T_{\mu\nu}$ and $j_\mu^R$, respectively, 
and they determine the central charge $c$ at a conformal fixed point.\footnote{Note that with respect to  \cite{Argurio:2013uba}, we have redefined here the ``transverse" form factors $C_s$ as $C_s^\mathrm{here}=C_s^{ \mbox{\tiny\cite{Argurio:2013uba}}}+\frac{m^2}{3k^2}F_s$. In the present definition they are strictly only affected by transverse degrees of freedom. In the previous definition they prevented the correlator from displaying spurious massless poles.}
The form factors $F_2$, $F_1$ and $F_0$ determine instead the correlators of the trace operators 
which are non vanishing if and only if there is a scale $m$ where conformality is broken explicitly.
If SUSY is preserved both the $C_s$ and the $F_s$ must be equal:
\begin{equation}
C_{1R}=C_{2}=C_{R SUSY}\ ,\qquad F_0=F_{1R}=F_{2}=F_{SUSY}\ .\label{WardSUSY1}
\end{equation} 

Supersymmetry breaking and $R$ symmetry breaking are  well characterized by the study of 
the above form factors.  In particular, we can check whether the breaking 
of $R$ symmetry or conformality along a given RG flow is spontaneous or explicit 
by finding the corresponding Goldstone mode in the two point correlators. SUSY breaking will introduce a violation in the Ward identities \eqref{WardSUSY1} equating the different form factors. Since the breaking has to be soft (i.e triggered by relevant operators), at high external momenta compared to the SUSY breaking parameters the Ward identities \eqref{WardSUSY1} are effectively recovered. 
Moreover, these correlators are 
essential to determine the stability of any vacuum configuration.

We will not consider here  correlators of the supercurrent $S_\mu$. Their parametrization is discussed in \cite{Argurio:2013uba}, and their interest resides in that they potentially contain information about the fermionic 
Goldstone mode related to spontaneous breaking of supersymmetry. These
correlators will be studied in a slightly different setting in a forthcoming work \cite{inpreparation}. 

For the linear multiplet,
the correlators in terms of dimensionless form factors were first written in \cite{Meade:2008wd}: 
\begin{subequations}
\begin{align}
&\langle J(k)J(-k)\rangle  =  C_0(k^2)\ , \label{czero}\\
&\langle j_\alpha(k)\bar j_{\dot\alpha}(-k)\rangle  = 
-\sigma^{\mu}_{\alpha\dot\alpha} k_\mu C_{1/2}(k^2) \ ,\label{chalf}\\
&\langle j_{\mu}(k)\bar j_{\nu}(-k)\rangle  =
-P_{\mu\nu}C_1(k^2)\ ,\label{cone}\\
&\langle j_\alpha(k)j_\beta(-k)\rangle  = 
\epsilon_{\alpha\beta}M B_{1/2}(k^2)\ ,\label{bhalf}
\end{align}
\end{subequations}
where $C_s$ are dimensionless real functions, $B_{1/2}$ is a complex
function and $M$ is a mass parameter related to the scale at which
supersymmetry is broken (usually not the same $m$ as in the previous expressions). 
These form factors characterize the spectrum of states in the strongly coupled gauge theory which are charged under 
the global $U(1)_F$.
If SUSY is preserved all $C_s$ are required to be equal and $B_{1/2}$ has to vanish: 
\begin{equation}
C_{0}=C_{1/2}=C_{1}=C_{SUSY}\ ,\qquad B_{1/2}=0\ .\label{WardSUSY2}
\end{equation}
Again, any departure from these requirements reflects SUSY breaking. If, as
desirable, SUSY breaking happens below a certain scale $M$ (both if explicit or spontaneous), then the functions $C_s$ and $B_{1/2}$ will
depart from their SUSY values only at momenta $k\lesssim M$.
In order to see this feature we will plot the combination of real form factors 
\begin{equation}
A\equiv -(C_0-4C_{1/2}+3C_1)\ ,\label{GGMA}
\end{equation}
which is  the combination that enters in the expression for
sfermion soft masses in General Gauge Mediation (GGM)\cite{Meade:2008wd} as we will discuss in Section \ref{GGMSpectrum}.\footnote{In GGM the strongly
  coupled theory we describe holographically is identified to the SUSY breaking hidden sector. The knowledge of $A$ and $B_{1/2}$ is enough to determine the soft terms in the visible sector once the $U(1)_F$ is weakly gauged by visible sector gauge degrees of freedom.}  
Let us finally notice that the fermionic form factors $C_{1/2}$ and $B_{1/2}$ can also probe some properties related to  R symmetry. In particular $B_{1/2}$ can be non zero in a SUSY breaking flow provided that also the $U(1)_R$ symmetry is broken, while it is zero if $R$ symmetry is preserved. 

To summarize, for each broad class of backgrounds, we will study the
form factors $C_2$, $C_{1R}$, $C_1$, $C_0$, $C_{1/2}$ and $B_{1/2}$,
with a particular regard to their pole structure. Moreover, we will
also consider the pole structure of all the other bosonic form factors,
to make sure that there are no tachyons in the spectrum. For the sake
of brevity, we will only present plots of the form factors that are
most important to the physics of each background. We have numerically checked that
no tachyonic poles appear in any of the form factors, for several values of the parameters characterizing the backgrounds. By continuity of the physical spectrum, we take this as evidence that there is a vast region of the parameter space where the backgrounds are stable and physically viable according to our criterion. A further indication of the absence of tachyons is provided by the fact that in a specific region of the parameter space we find SUSY solutions.

We now study representative backgrounds from each class following the classification summarized in Figure \ref{THECUBE}.

\subsection{Backgrouds with $\eta$ blowing up}\label{etablowingup2}

We start by considering the class of RG flows which are dual to backgrounds where $\eta$ is blowing up. As discussed in \ref{etablowingup} these are backgrounds which can be obtained for a generic point in the parameter space \eqref{parspace}. A nice way of understanding the physics of these backgrounds is to interpret them as a deformation of the SUSY solution \eqref{GPPZback} with which determines the behavior of the background at the singularity. 

\subsubsection*{\it{SUSY backgrounds}}

\begin{figure}[h]
\centering
\includegraphics[width= 1\textwidth]{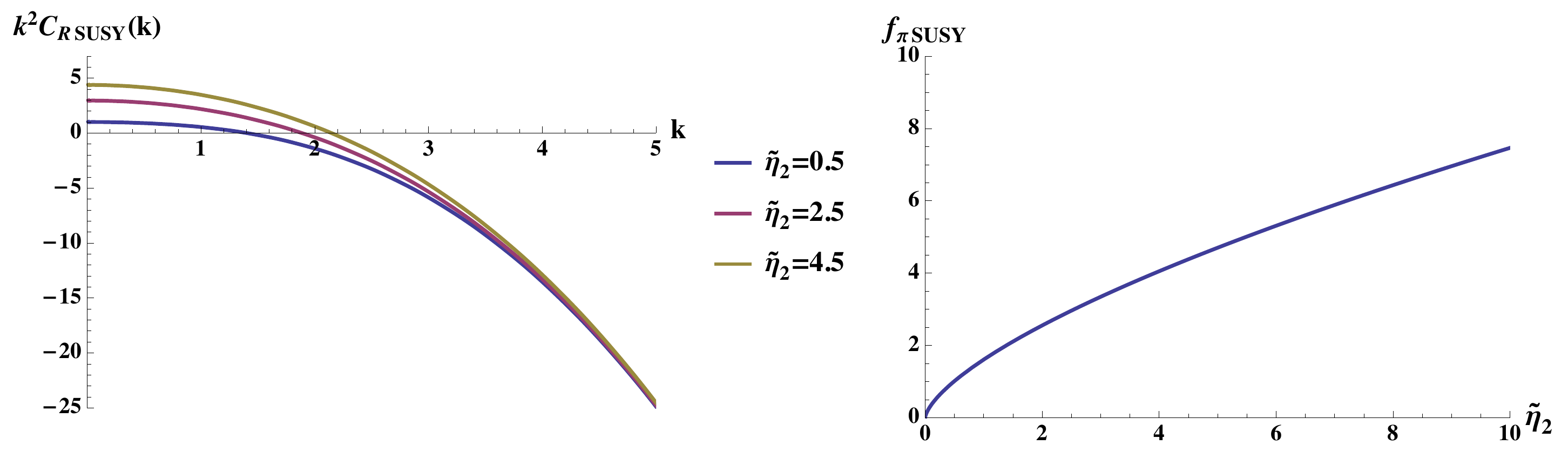}
\caption{On the left: $k^2C_{RSUSY }$ for different values of $\tilde{\eta}_2$. On the right: dependence of $f_{\pi SUSY}$ (i.e. the residue of the $1/k^2$ pole in $C_{RSUSY }$) on $\tilde{\eta}_2$.} 
\label{GPPZ_C_1C_2}
\end{figure}

Having set all the SUSY breaking scales to zero we are left with a one parameter class of solutions distinguished by the value of $\tilde\eta_2$. The latter breaks
conformality and $R$ symmetry spontaneously, but preserves SUSY.  The traceless form factors of the stress energy tensor multiplet are all equal if SUSY is preserved. $C_{RSUSY}$ defined in \eqref{WardSUSY1} displays a $\tfrac{1}{k^2}$ pole which corresponds to the massless SUSY multiplet of Goldstone modes associated to the spontaneous breaking of conformal symmetry. In Figure \ref{GPPZ_C_1C_2} we show how the value of the residue $f_{\pi SUSY}$ of the Goldstone multiplet depends on the breaking scale $\tilde{\eta}_2$. For dimensional reasons we get $f_{\pi SUSY}\sim \tilde{\eta}_2^{2/3}$, where the order one coefficient in front is found numerically to be $1.6$ by fitting the values of the residue for different choices of $\tilde{\eta}_2$. From the left panel of Figure \ref{GPPZ_C_1C_2} we see that the conformal behavior of $C_{RSUSY}$ is recovered at high momenta $k^2>\tilde{\eta}_2^{2/3}$. 

The behavior of linear multiplet correlators is also fixed by the single supersymmetric form factor $C_{SUSY}$ defined in \eqref{WardSUSY2}. $B_{1/2}$ vanishes because of superysmmetry. Note that the vanishing of $B_{1/2}$ is non trivial in a background which breaks $R$ symmetry and it can be taken as a further evidence of the reliability of our numerical computation.

\begin{figure}[h]
\centering
\includegraphics[width= 1\textwidth]{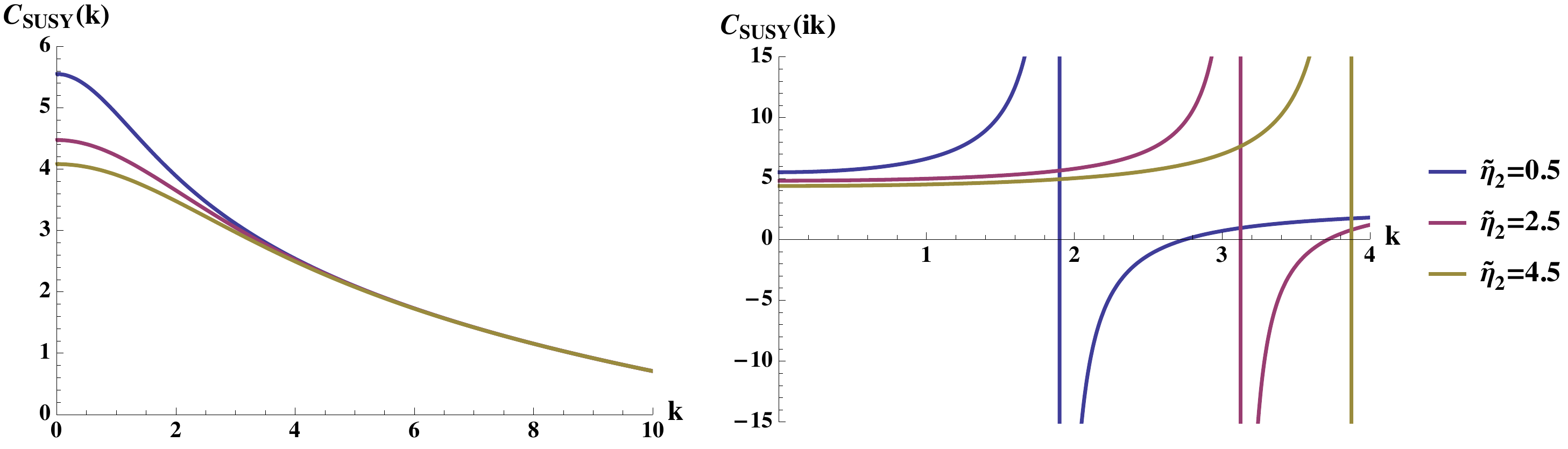}
\caption{On the left: $C_{SUSY }$ for $k^2>0$ for different values of $\tilde{\eta}_2$. On the right: First pole in $C_{SUSY }$ for $k^2<0$ for different values of $\tilde{\eta}_2$. }
\label{GPPZ_CGGM}
\end{figure}

In Figure \ref{GPPZ_CGGM} we show that the value of $\tilde{\eta}_2$ determines the size of the supersymmetric mass gap for $k^2>0$. Looking at the same form factor for $k^2<0$ we see explicitly how $\tilde{\eta}_2$ controls the mass of the first pole in the current-current correlator. The larger is the value of $\tilde{\eta}_2$ the heavier is the first pole in $C_{SUSY}$. Accordingly, the value of $C_{SUSY}$ at $k^2=0$ gets smaller increasing the value of $\tilde{\eta}_2$. 

\subsubsection*{\it{Spontaneous R symmetry breaking backgrounds}}
\begin{figure}[h]
\centering
\includegraphics[width= 0.5\textwidth]{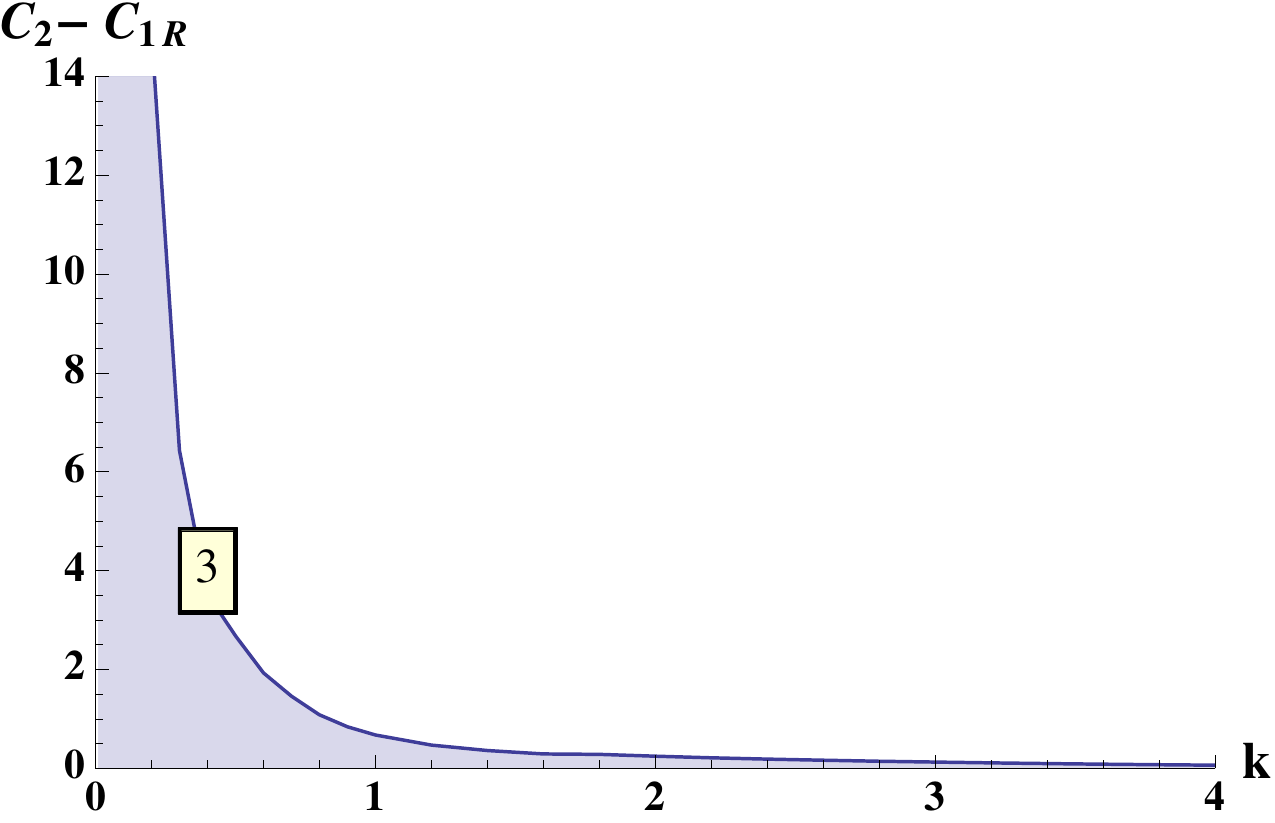}\hfill 
\includegraphics[width= 0.5\textwidth]{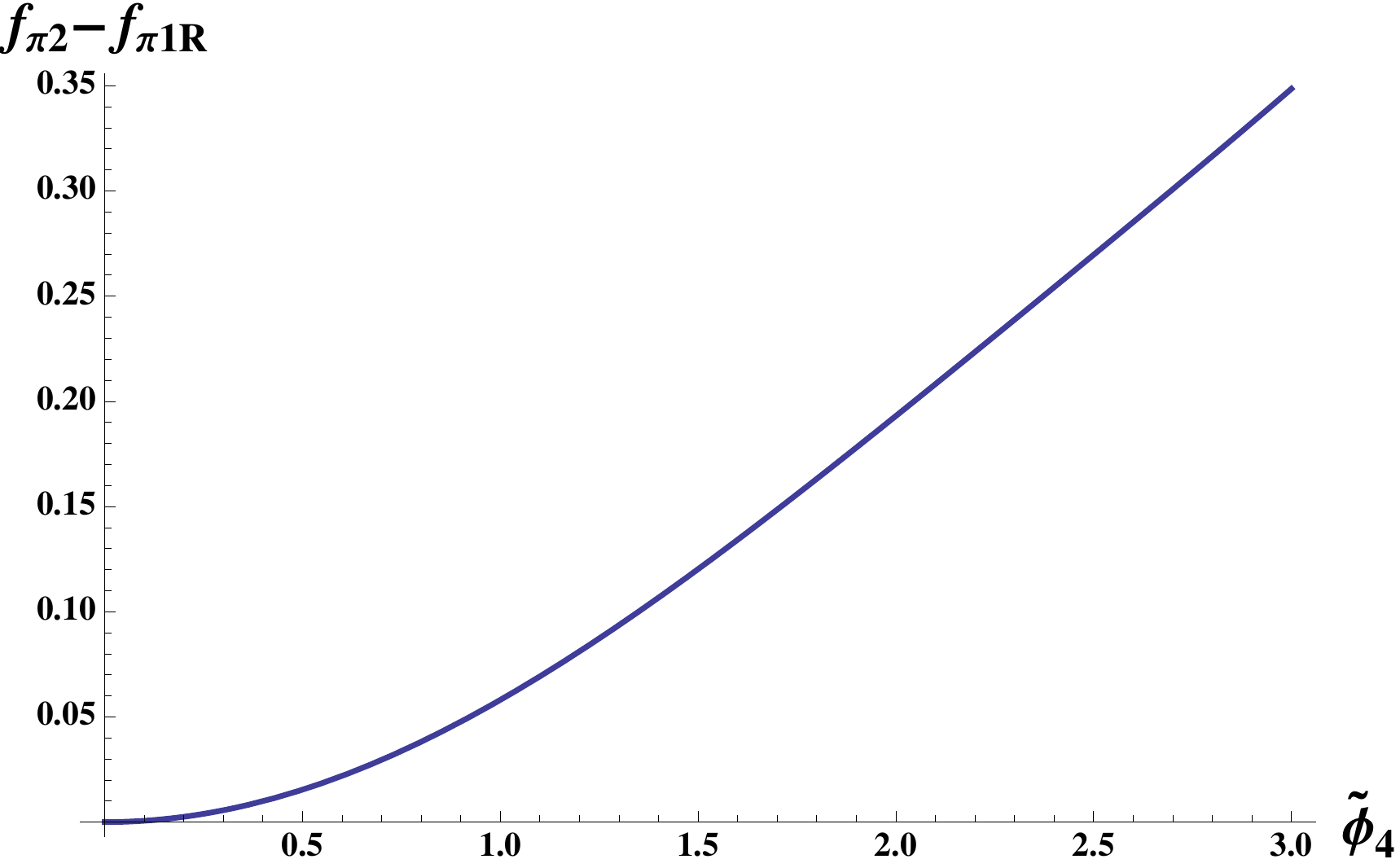}
\caption{On the left: $C_{2}-C_{1R}$ for a sample point with $\tilde{\phi}_4=3$. The function goes to zero for $k^2>\tilde{\phi}_4^{1/2}$ and it has a massless pole $1/k^2$ at $k^2=0$. On the right: The difference between the dilaton residue $f_{\pi2}$ and the R axion one $f_{\pi1R}$ as a function of $\tilde{\phi}_4$.}
\label{spontaneous_C1R_C_2}
\end{figure}

Switching on a non zero $\tilde\phi_4$ over the SUSY background, we obtain a two parameter class of solutions where all the symmetries, including SUSY, are broken spontaneously. Like in the supersymmetric case, broken conformality should lead
to the presence of a massless dilaton, that shows up as a pole at
$k^2=0$ in $C_2$. Broken $R$ symmetry gives rise to a massless Goldstone
boson, the R axion, that produces a massless pole in $C_{1R}$. Since
SUSY is also broken, $C_2$ and $C_{1R}$ should differ as shown on the left panel of Figure \ref{spontaneous_C1R_C_2} for a sample point with $\tilde{\eta}_2=1$ and $\tilde{\phi}_4=3$. In particular, the values of the dilaton residue $f_{\pi 2}$ and the one of the R axion $f_{\pi 1R}$ should be different, their difference being proportional to the SUSY breaking parameter $\tilde{\phi}_4$. Indeed, fixing $\tilde{\eta}_2=1$, we show in Figure \ref{spontaneous_C1R_C_2} (right) that $f_{\pi 2}-f_{\pi 1R}$ goes to zero for $\tilde{\phi}_4=0$ where SUSY is restored. Interestingly, the difference between $C_{2}$ and $C_{1R}$ goes to zero at high momenta (i.e $k^2>\tilde{\phi}_4^{1/2}$). This is the expected UV behavior of correlators in a SUSY breaking theory obtained deforming the SUSY theory by means of a SUSY breaking VEV.  

\begin{figure}[h]
\centering
\includegraphics[width= 0.5\textwidth]{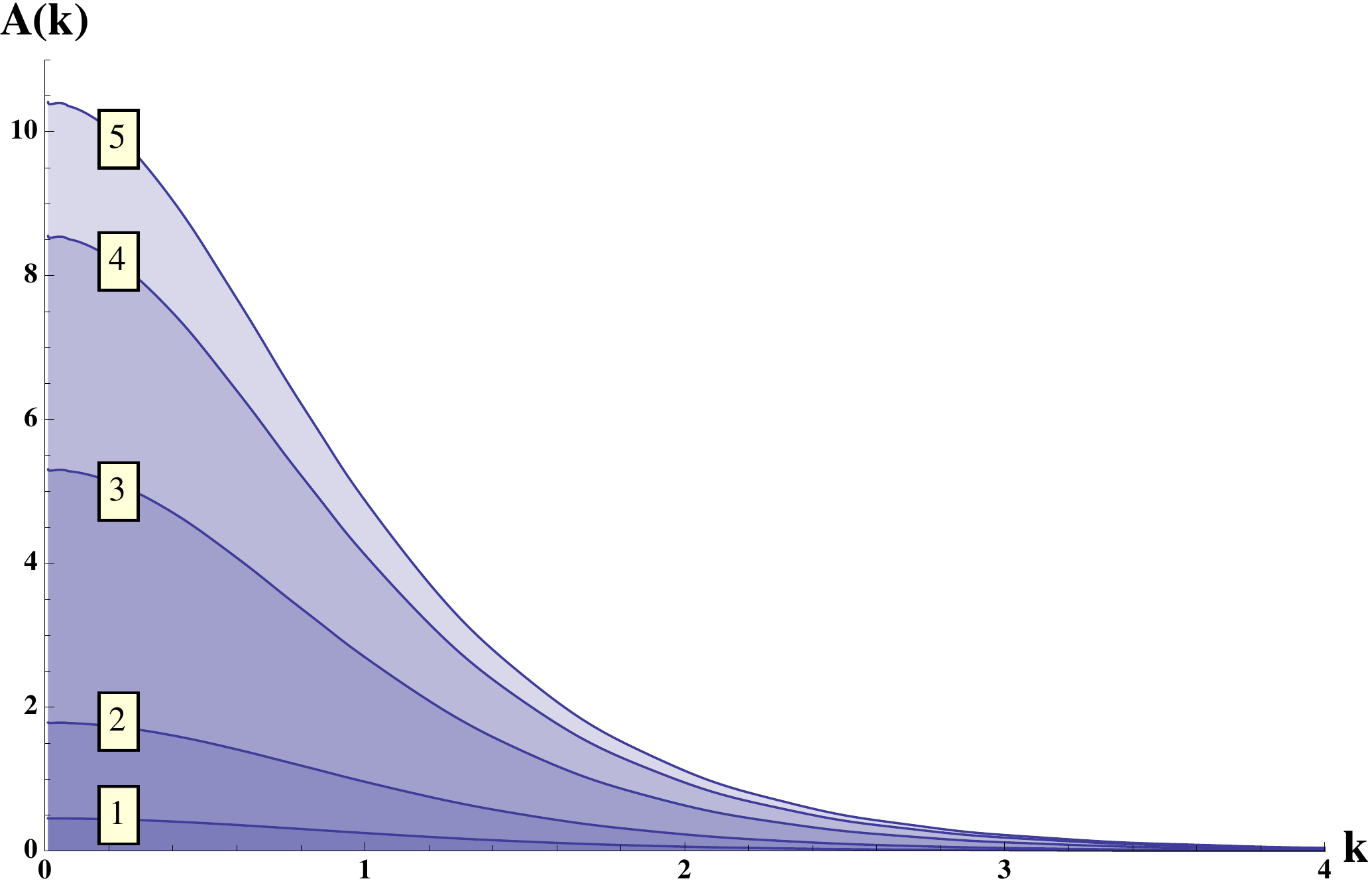}\hfill 
\includegraphics[width= 0.5\textwidth]{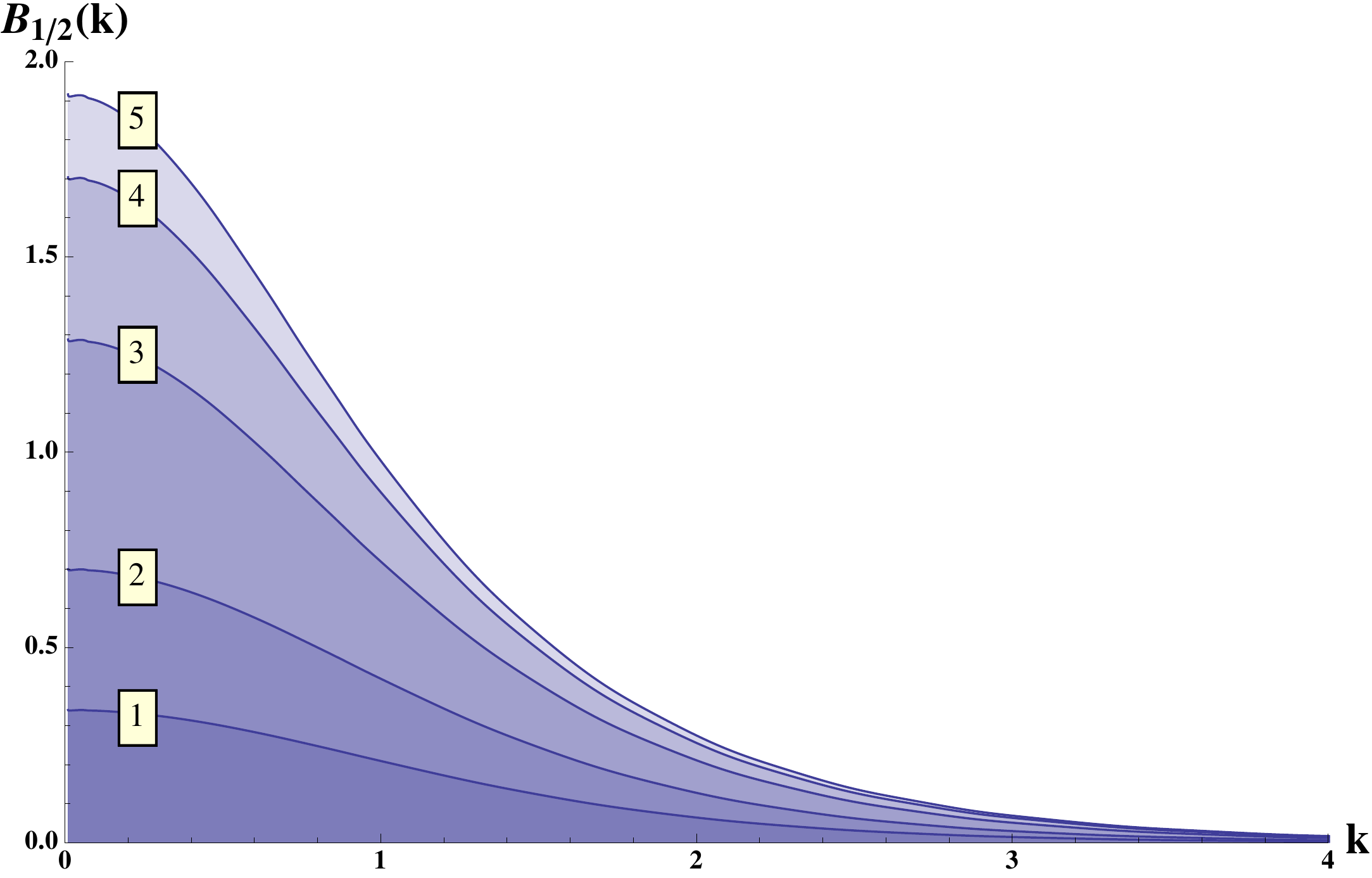}
\caption{On the left: $A$ defined in \eqref{GGMA} for $\tilde{\eta}_2=1$ at increasing values of $\tilde{\phi}_4$. On the right: $B_{1/2}$ for $\tilde{\eta}_2=1$ at increasing values of $\tilde{\phi}_4$. }
\label{spontaneous_C1_C_2}
\end{figure}

In Figure \ref{spontaneous_C1_C_2} we show the behavior of the linear multiplet correlators, plotting $A$ and $B_{1/2}$ for different values of $\tilde{\phi}_4$ keeping $\tilde{\eta}_2=1$. First of all, both $A$ and $B_{1/2}$ are gapped functions in the IR since both the dilaton and the
R axion are not charged under the unbroken $U(1)_F$. The value of both $A$ and $B_{1/2}$ at $k^2=0$ increases when the SUSY breaking parameters $\tilde{\phi}_4$ increases (recall that both $A$ and $B_{1/2}$ vanish in the SUSY limit). The behavior of $B_{1/2}$ follows that of $A$ since $R$ symmetry is spontaneously broken by the VEV of a SUSY operator and hence no further symmetries are protecting $B_{1/2}$ from the SUSY breaking dynamics. Both $A$ and $B_{1/2}$ fall very rapidly to zero in the UV, in agreement with the fact that $\tilde{\phi}_4$ is the VEV of a $\Delta=4$ complex operator which 
enters as its modulus squared in the current-current OPE. 

Finally, let us notice that since also SUSY seems to be broken by VEVs, we should also
expect a pole in the correlator of the supercurrent. However, here we
meet a difficulty of this specific subclass of backgrounds, that we
can trace back to the large $N$ limit and the underlying assumptions of
the present paper. Indeed, since only VEVs are present, conformality
is only spontaneously broken and the trace of the stress energy tensor
has to be trivially zero. Hence the stress energy tensor itself cannot
acquire a VEV. Now the massless Goldstino pole in the supercurrent
correlator is proportional to this VEV, and so cannot arise in our
backgrounds. The reason we can explain this is to believe that the
SUSY breaking dynamics can be ascribed at least partly to operators 
that acquire large dimensions in the large $N$ limit, and that we have
neglected in our approach (see also \cite{Gubser:1999pk} for further discussion about this point). We refer to \cite{inpreparation} for the
study of a setting where the Goldstino pole can be treated.

\subsubsection*{\it{Explicit R symmetry breaking backgrounds}}
\begin{figure}[h]
\centering
\includegraphics[width= 1\textwidth]{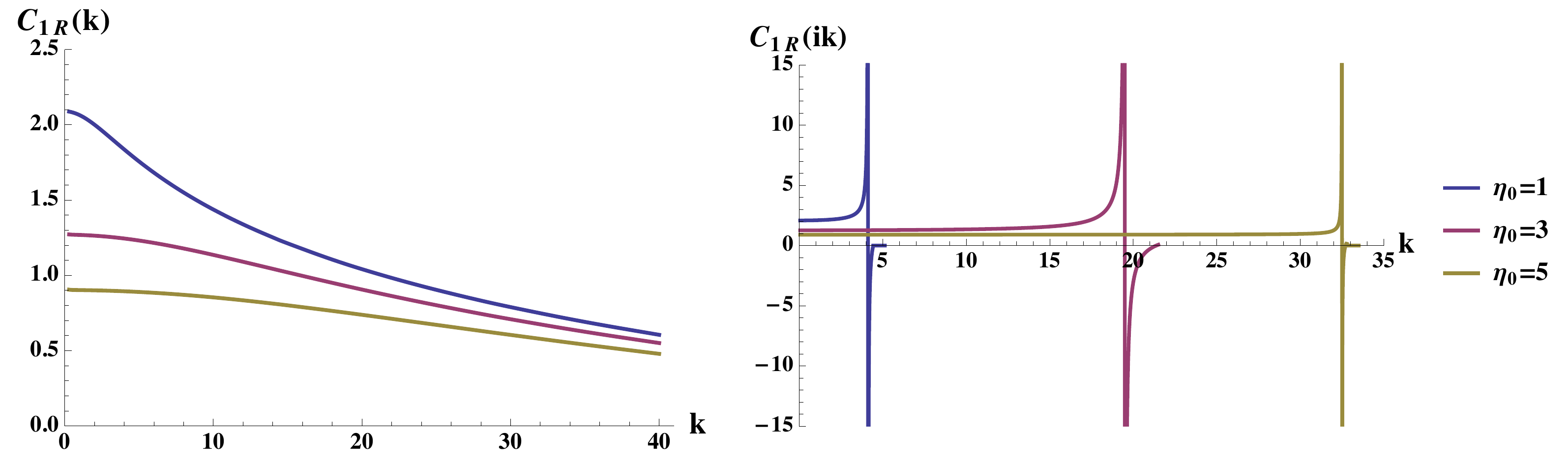}
\caption{On the left: $C_{1R}$ for $k^2>0$ for different values of $\eta_0$. On the right: 1st pole in $C_{1R}$ for $k^2<0$ for different values of $\eta_0$. The subsequent poles for the $\eta_0=1$ and the $\eta_0=3$ cases are not displayed in order to make the plot more readable.}
\label{explicit_C1R}
\end{figure}

Starting from a background where $\tilde{\eta}_2=\tilde{\phi}_4=1$ we want to switch on a source term proportional to $\eta_0$ which breaks both conformality and $R$ symmetry explicitly. According to that we expect the massless Goldstone modes of the spontaneous case to get a mass proportional to $\eta_0$. In Figure \ref{explicit_C1R} we show that this is indeed the case plotting $C_{1R}$ for $k^2>0$ (left) and $k^2<0$ (right). From the left panel of  Figure \ref{explicit_C1R} we see that $C_{1R}$ is a gapped function and its value at $k^2=0$ is inversely proportional to $\eta_0$. Consistently, from the right panel of  Figure \ref{explicit_C1R} we see that the mass of the first pole in $C_{1R}$ grows at increasing $\eta_0$. We already encountered the relation between the value of a gapped function at $k^2=0$ and the mass of the first pole at $k^2<0$ discussing Figure \ref{GPPZ_CGGM} in the SUSY case. What is new about Figure \ref{explicit_C1R} is that we clearly see the uplifting of the massless Goldstone mode while an explicit breaking parameter is switched on. 

Obtaining this result from the holographic point of view has a number of subtleties related to the holographic renormalization procedure and to the presence of spurious poles due to the parametrization of the form factors whose discussion is deferred to the next section. 

\begin{figure}[h]
\centering
\includegraphics[width= 0.49\textwidth]{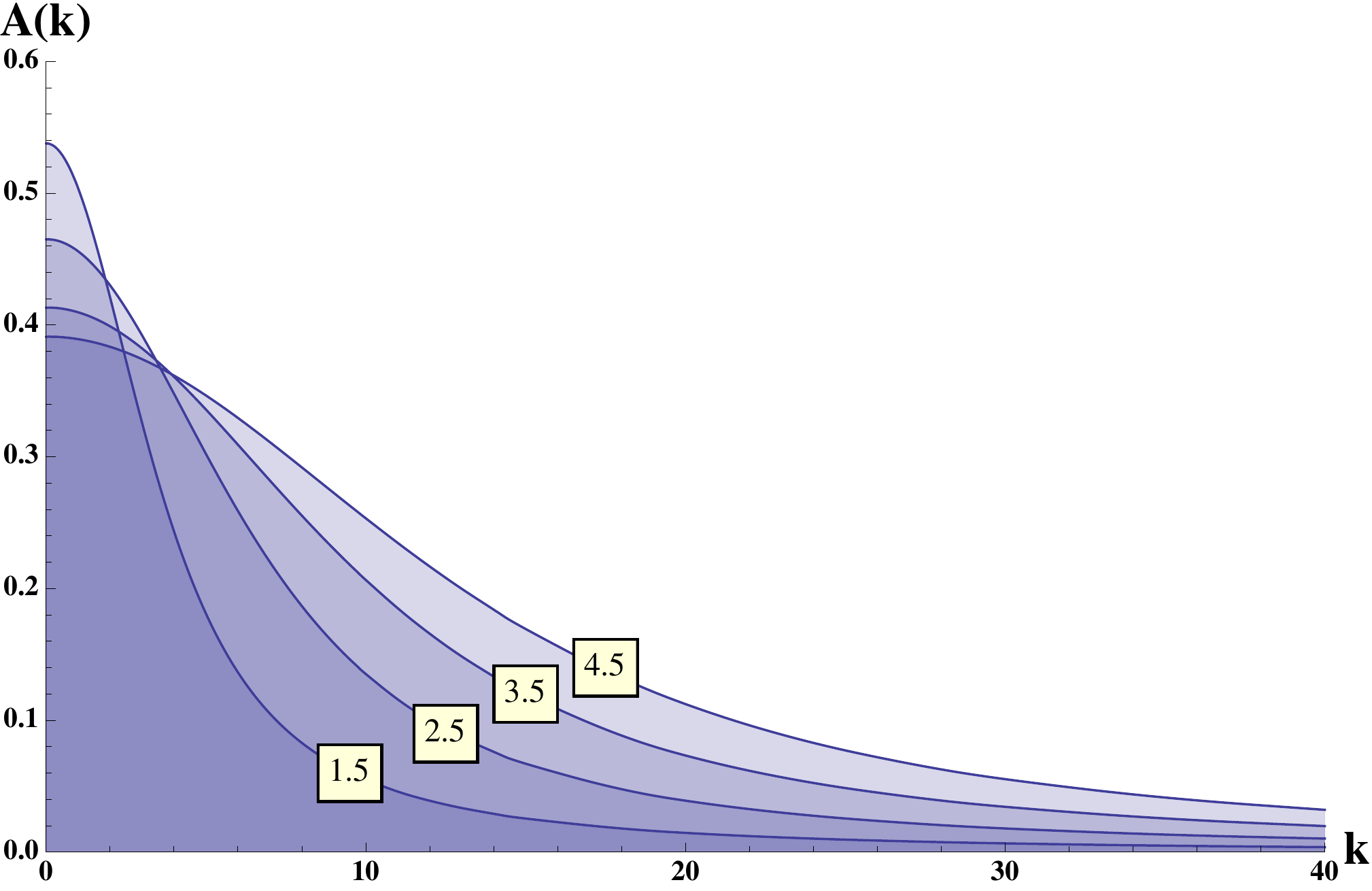}\hfill 
\includegraphics[width= 0.49\textwidth]{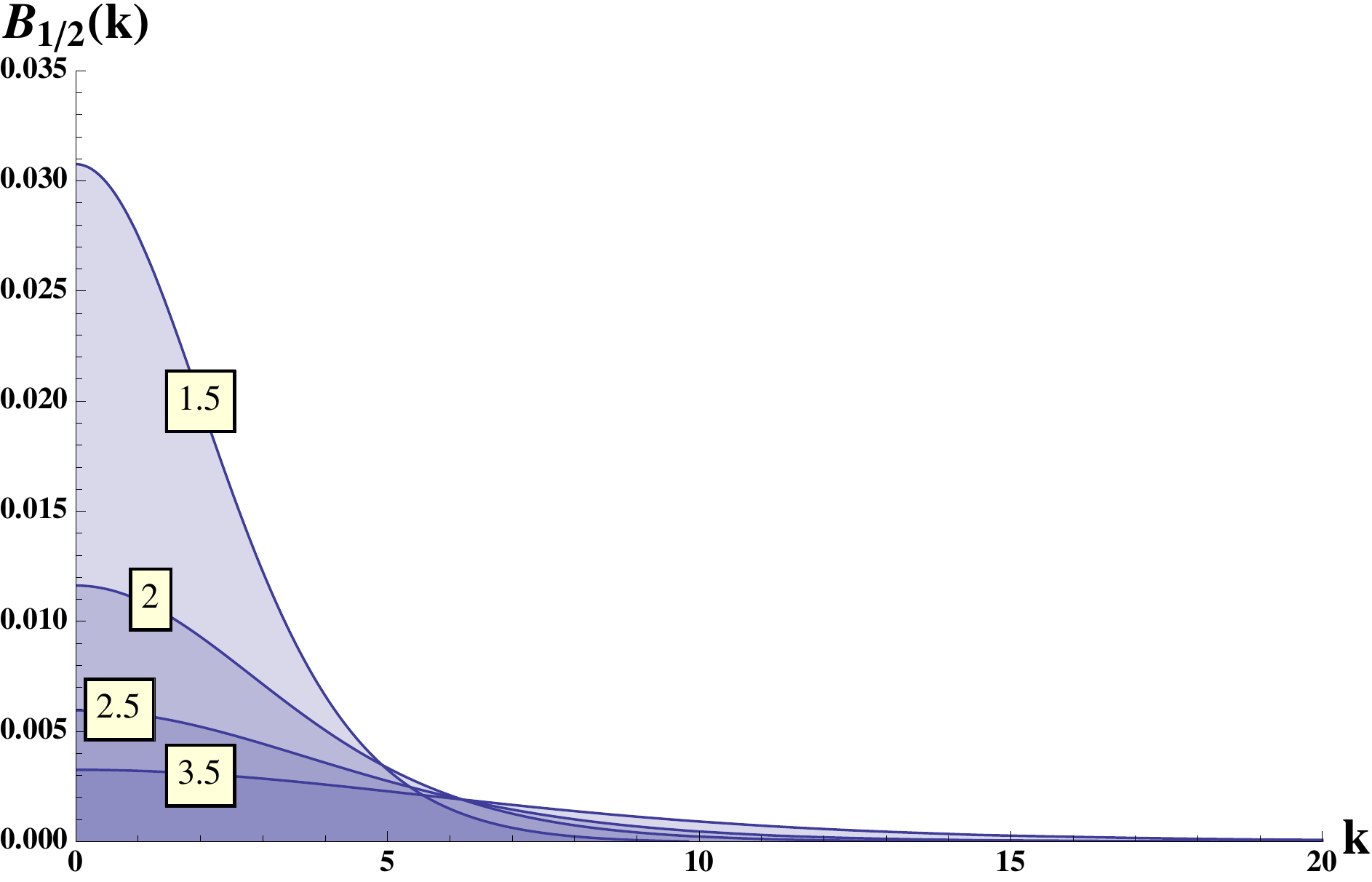}
\caption{On the left: $A$ for $k^2>0$ for different values of $\eta_0$. On the right: $B_{1/2}$ for $k^2>0$ for different values of $\eta_0$. The behavior of $B_{1/2}$ shows that the  gaugino mass is \emph{screened} at increasing values of $\eta_0$ (see also Figure \ref{softmasses} in Section 6).}
\label{explicit_A_B}
\end{figure}

In Figure \ref{explicit_A_B} we plot $A$ (left) and $B_{1/2}$ (right) for increasing values of $\eta_0$. Note that we are interested in a regime where the explicit breaking parameter is parametrically larger than the other two. Increasing the value of $\eta_0$ we are effectively getting closer and closer to a situation where only the explicit breaking term is triggering the dynamics of the flow. In agreement with this picture we can see that $A$ for $k^2>0$ is a gapped function whose IR value at $k^2=0$ decreases while $\eta_0$ increases (again $\eta_0$ is controlling the position of the mass gap). More interestingly, the UV behavior of $A$ feels strongly the presence of $\eta_0$ which is a dimension one parameter that enters in the current-current OPE as a relevant operator. 
$B_{1/2}$ in Figure \ref{explicit_A_B} is instead decreasing while $\eta_0$ is increasing. In particular one can show that taking $\tilde{\phi}_4=0$ one gets $B_{1/2}=0$. This result has no direct explanation in terms of classical symmetries of the theory. However, we seem to recover a strongly coupled example of the \emph{gaugino mass screening} phenomenon in gauge mediation \cite{ArkaniHamed:1998kj} (we will see in Section 6 that $B_{1/2}$ is indeed the parameter controlling the gaugino mass in the visible sector). Our setup looks similar to the semi direct gauge mediation one discussed in \cite{Argurio:2009ge} where the hidden sector gaugino gets a mass but the gaugino mass in the visible sector is ``screened''. However, there is no way of defining a messenger sector here and we have to rely on our numerical analysis. 

\subsection{Backgrounds with dilaton blowing up}\label{dilatonblowingup2}
We now consider the backgrounds where the dilaton blows up at the
singularity and $\eta$ tends to zero.

As discussed in \ref{dilatonback}, the backgrounds where the dilaton blows up define a $2d$ subspace of the $3d$ one. The whole class of backgrounds can be understood as deformations of the dilaton domain wall background, where the values of $\eta_0$ and $\tilde{\eta}_2$ are extremely fine tuned in order to fulfil the requirement of vanishing $\eta$ at the singularity.

\subsubsection*{\it{Dilaton domain wall background}}
When $\eta=0$ we realize a one parameter class of solutions parametrized by $\tilde{\phi}_4$. These solutions preserve $R$ symmetry while breaking both conformality and SUSY spontaneously.   

\begin{figure}[h]
\centering
\includegraphics[width= 0.49\textwidth]{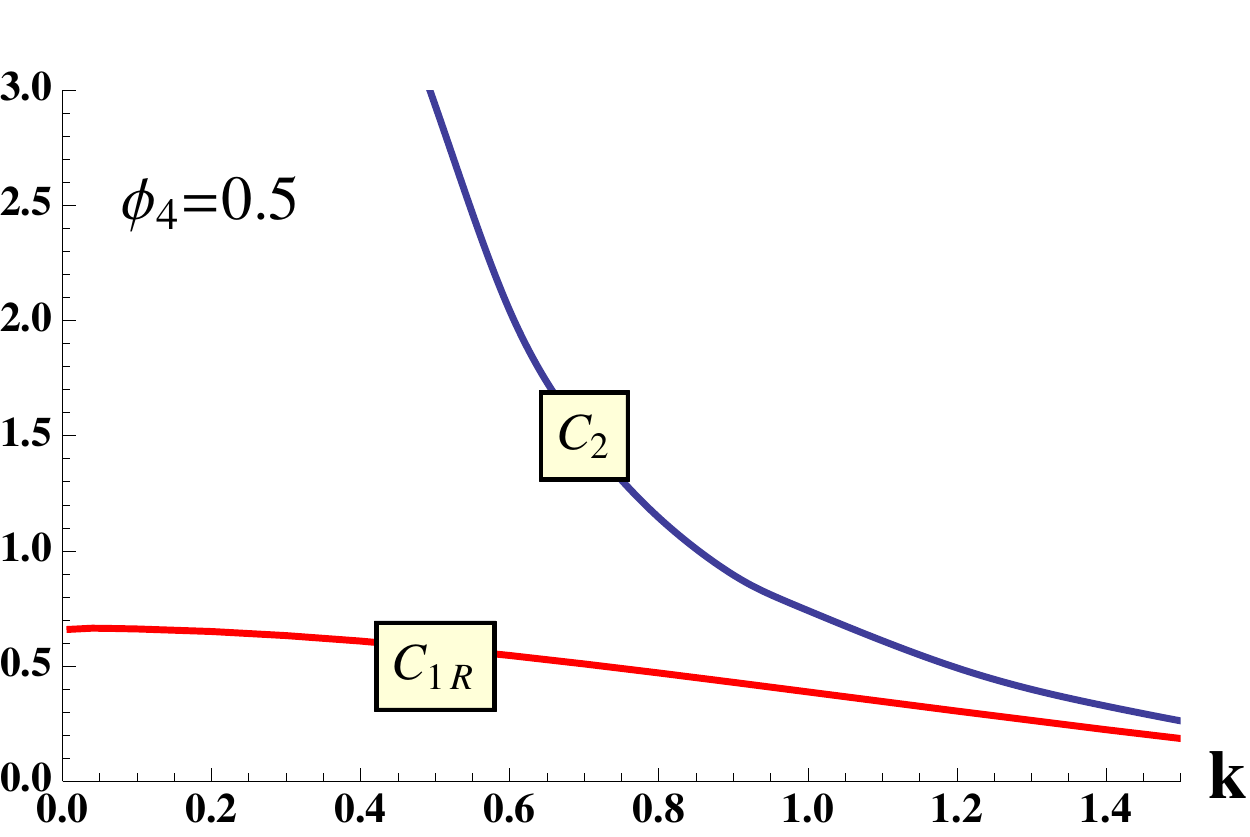}\hfill 
\includegraphics[width= 0.49\textwidth]{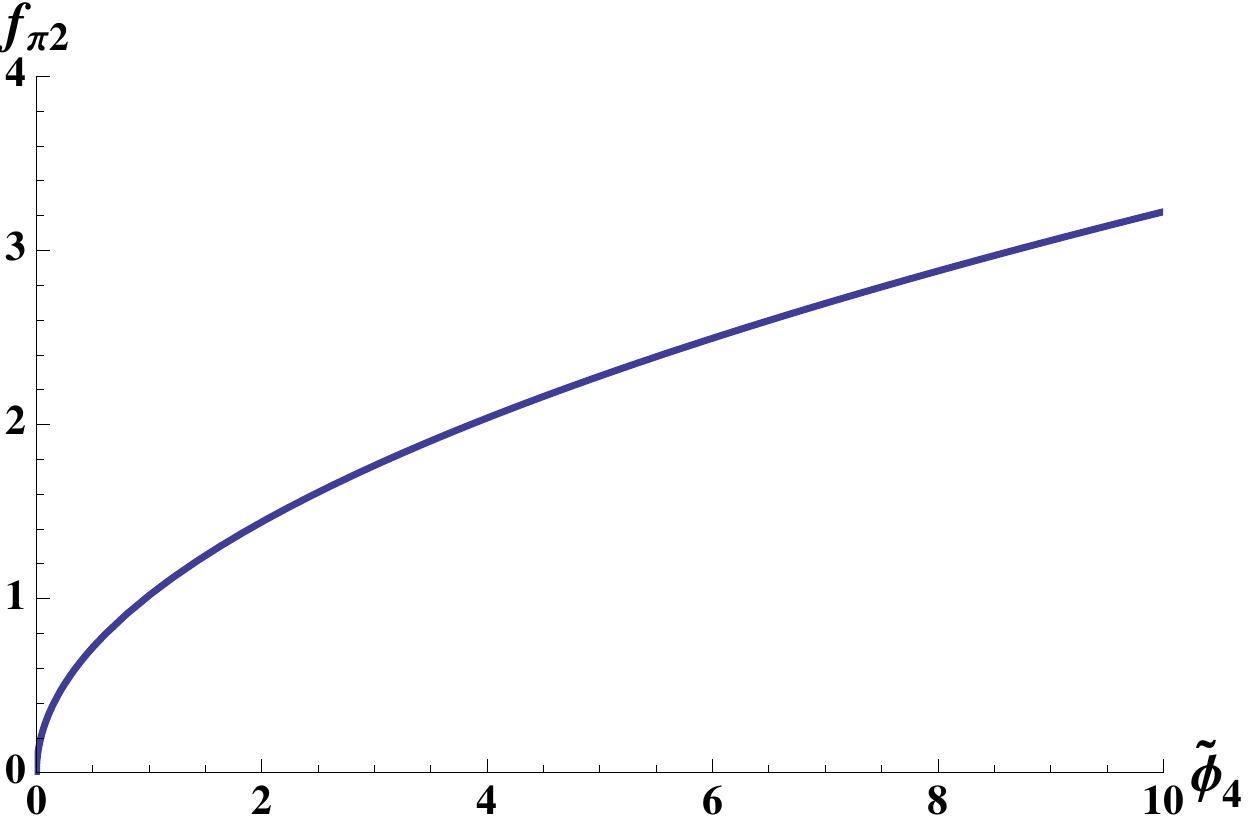}
\caption{On the left: $C_{1R}$ and $C_{2}$ for $k^2>0$ on the dilaton domain wall background for $\tilde{\phi}_4=0.5$. On the right: Dependence of the dilaton residue $f_{\pi 2}$ on $\tilde{\phi}_4$.}
\label{C1R_C2_Gubser}
\end{figure}

In the left panel of Figure \ref{C1R_C2_Gubser} we show that $C_2$ has the
correct massless pole for spontaneously broken conformal symmetry,
while $C_{1R}$ is a gapped function since $R$ symmetry is preserved and SUSY
is broken. In the right panel of Figure  \ref{C1R_C2_Gubser} we show the dependence of the dilaton residue $f_{\pi 2}$ on $\tilde{\phi}_4$. We derive this by extracting $f_{\pi 2}$ from the IR behavior of $C_{2}$ for different values of $\tilde{\phi}_4$. As we could have guessed by dimensional analysis $f_{\pi 2}\sim \tilde{\phi}_4^{1/2}$ while the coefficient in front comes out to be $1$ from the numerical fit. 

\begin{figure}[h]
\centering
\includegraphics[width= 0.5\textwidth]{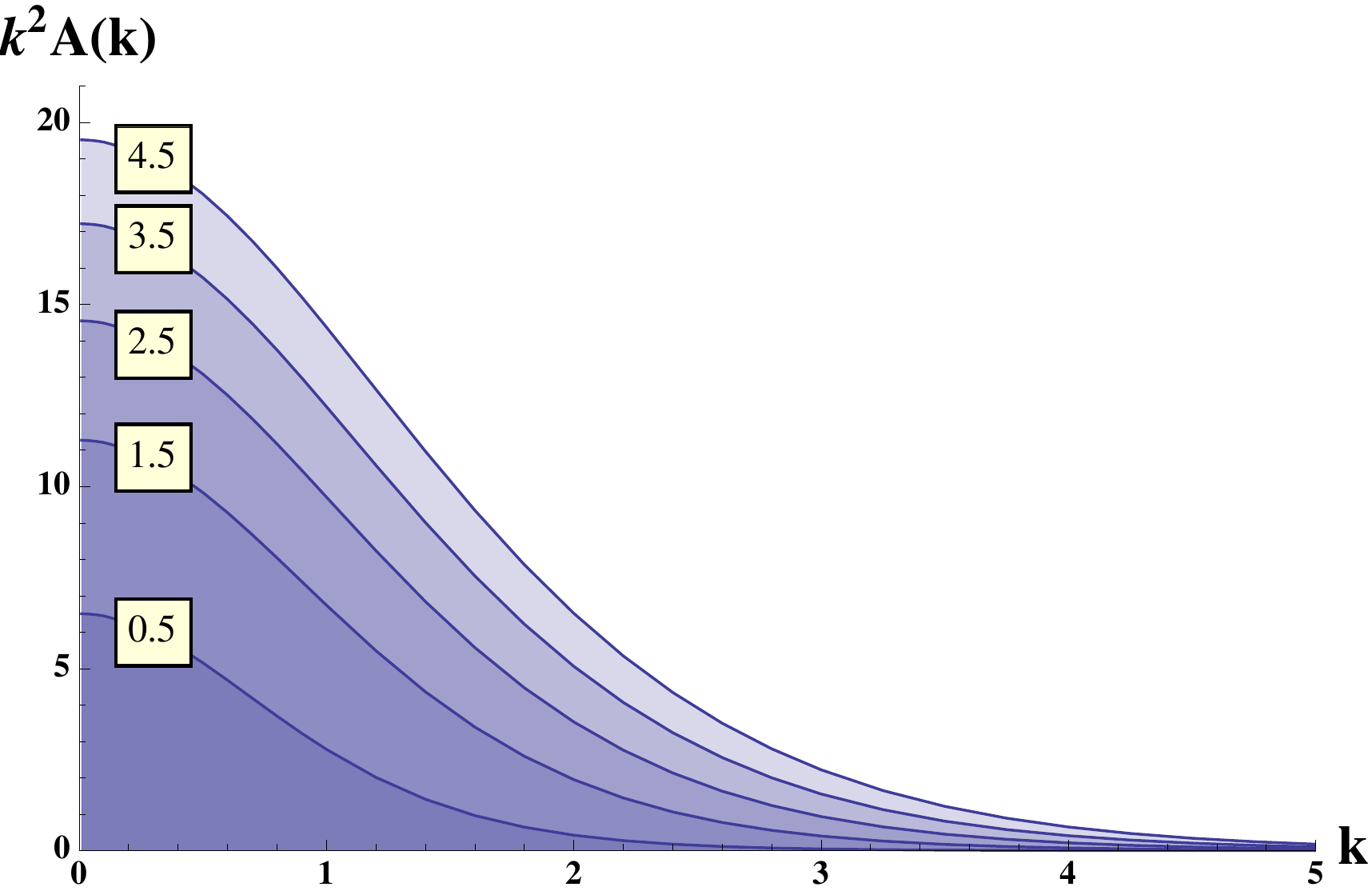}\hfill 
\includegraphics[width= 0.5\textwidth]{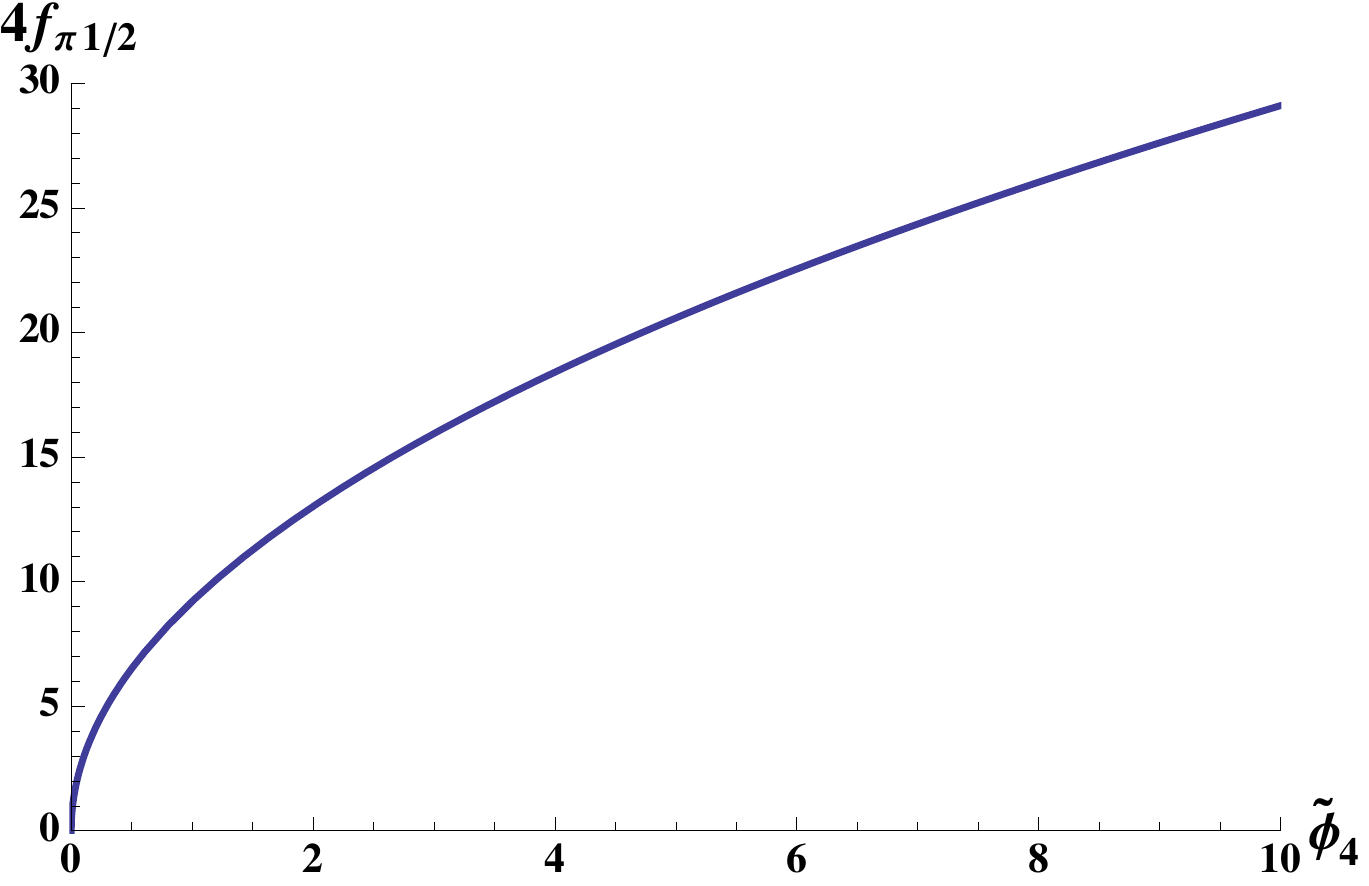}
\caption{On the left: $k^2A$ for $k^2>0$ on the dilaton domain wall background for different values of $\tilde{\phi}$. $A$ displays a $1/k^2$ behavior which comes from a simple pole in $C_{1/2}$. On the right: Dependence on $\tilde{\phi}_4$ of the 't~Hooft fermion residue $f_{\pi 1/2}$ in $C_{1/2}$.}
\label{C1_C2_Gubser}
\end{figure}

In Figure \ref{C1_C2_Gubser} we show the dependence of $A$ on $\tilde{\phi}_4$ while 
$B_{1/2}$ vanishes trivially because of $R$ symmetry.  As already
noticed in \cite{Argurio:2012cd} we see that $A$ has a $1/k^2$ pole at $k^2=0$. The latter arises because there is a massless
pole in $C_{1/2}$ that is associated to massless 't~Hooft fermions
required to compensate in the IR the $U(1)_R$ global anomaly. This massless mode turns out to be charged under $U(1)_F$ because $U(1)_{R}$ and $U(1)_{F}$ mix along the flow.
In the right panel of Figure \ref{C1_C2_Gubser} we see that the dependence of the 't~Hooft fermion residue $f_{\pi 1/2}$ is again proportional to $\tilde{\phi}_4^{1/2}$ while the numerical coefficient is found to be $2.3$. 

From the QFT point of view, the fact that a massless pole could arise in $C_{1/2}$ was first realized in \cite{Buican:2009vv}. We give here a holographic realization of this mechanism which gives rise to soft masses  of the Dirac type for gauginos. In Section 6 we will comment further on the phenomenological relevance of this mechanism.\footnote{It would be interesting to study a possible realization of this mechanism from the pure field theoretical perspective in some model of Dynamical Supersymmetry Breaking (DSB). This would give a UV complete realization of the Dirac gaugino scenario in gauge mediation where the fermionic partners of the visible sector gauginos arise naturally from the hidden sector dynamics. See \cite{Abel:2011dc} for similar ideas along those lines.}  

\subsubsection*{\it{Dilaton like backgrounds}}

In \ref{dilatonback} we discussed how a non trivial profile for $\eta$ can be switched on over the dilaton domain wall background.  Backgrounds in this class behave like the dilaton domain wall ones close to the singularity and define a $2d$ subspace of the near boundary parameters which is better described in terms of the position of the singularity $z_{\text{sing}}$ and $\eta_{w}$ defined in \eqref{etawdef}. Both of these parameters should be thought as functions of $\eta_0$, $\tilde{\eta}_2$ and $\tilde{\phi}_4$. In Table \ref{Bulletto_points} we list some benchmarks of the dilaton like solutions with $z_{\text{sing}}=1.2$ and different $\eta_{w}$. The position of the singularity is mostly controlled by the value of $\tilde{\phi}_4$ in agreement with the fact that its appearance is triggered by the dilaton profile blowing up. The values of $\eta_{0}$ and $\tilde{\eta}_2$ grow at increasing $\eta_{w}$ and are fine tuned such that the $\eta$ profile vanishes at the singularity as depicted in Figure \ref{Bulletto}. Let us notice that the solutions we find are the backreacted version of those presented in \cite{Argurio:2012cd} where $\eta$ was treated as a small perturbation over the dilaton domain wall background. 

  \begin{figure}[h]
\centering
\includegraphics[width= 0.5\textwidth]{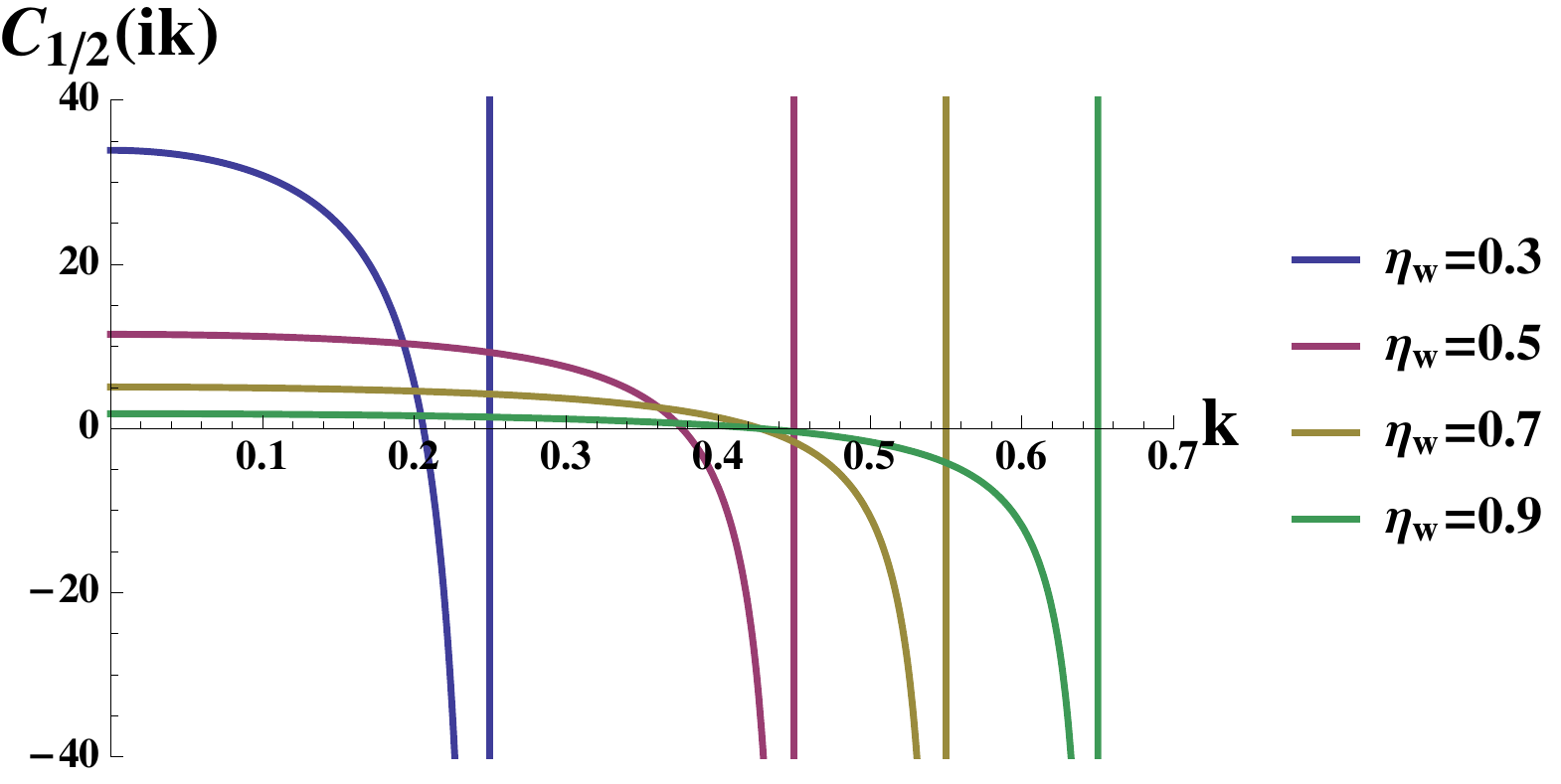}\hfill 
\includegraphics[width= 0.5\textwidth]{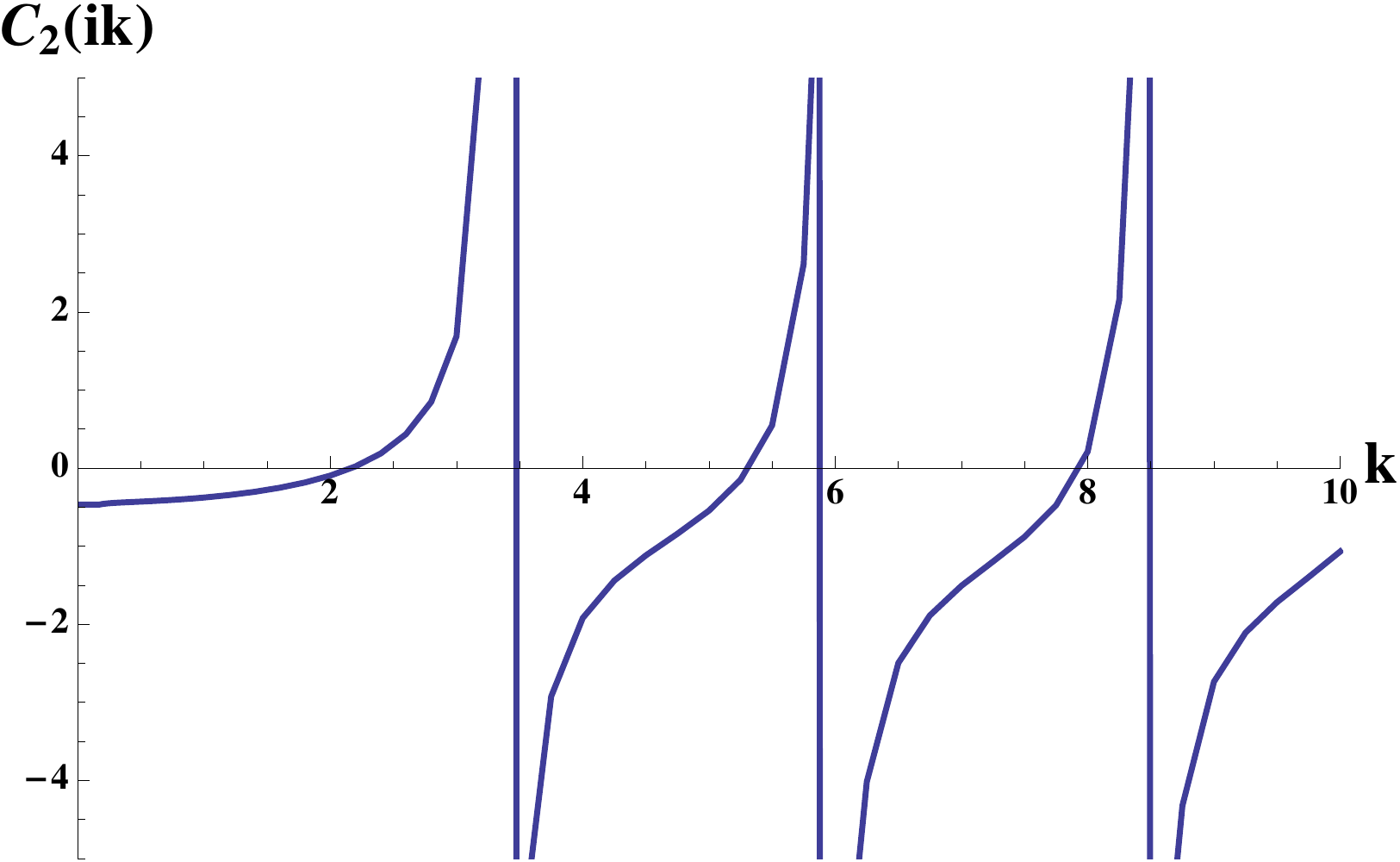}
\caption{On the left: position of the first pole in $C_{1/2}$ for $k^2<0$ for different values of $\eta_{w}$. The associated values of the UV parameters are reported in Table \ref{Bulletto_points}. The same pole appears in $B_{1/2}$ for $k^2<0$ and it is then associated to a one particle state which is charged under $R$ symmetry.  On the right: pole structure of $C_{2}$ for $k^2<0$ and $\eta_{w}=0.3$. The massless dilaton of the dilaton domain wall solution is uplifted in the dilaton like backgrounds.}
\label{Bulletto_1}
\end{figure}

\begin{table}[hc]
\begin{center}
 \caption{Values of the UV parameters \eqref{parspace} for dilaton like backgrounds with $z_{\text{sing}}=1.2$ and varying $\eta_{w}$. }\label{Bulletto_points}
\begin{tabular}{@{}ccccc@{}}
\toprule
  $\eta_{w}$ & $\tilde{\phi}_4$& $\tilde{\eta}_2$ & $\eta_0$ \\ \midrule 
          0.3 &      1.14     &   0.44 & -0.2\\\midrule
          0.5 &       1.06      &   0.82 & 0.6\\\midrule
          0.7 &         0.95      &    1.44&  8.37    \\\midrule
          0.9 &           0.79      &    3.14 &    151.86 \\
          
            \bottomrule
    \end{tabular}
  \end{center}
  \end{table}

In Figure \ref{Bulletto_1} (left) we show that the $1/k^2$ pole in $C_{1/2}$ presented in Figure \ref{C1_C2_Gubser} for the dilaton domain wall solution is now taking a mass. The latter  increases for increasing values of $\eta_{w}$. This is coherent with the field theory interpretation of the $1/k^2$ pole as a 't~Hooft fermion coming from an unbroken $U(1)_{R}$ which is anomalous in the UV and mixes with the $U(1)_{F}$ flavor symmetry along the flow. The non trivial $\eta$ profile controlled by $\eta_{w}$ is breaking $R$ symmetry giving a non zero Majorana mass to the massless 't~Hooft fermion. The latter remains parametrically small since the $\eta$ profile is forced to vanish at  the singularity. The first pole in $C_{1/2}$ at $k^2<0$ is then naturally interpreted as the uplifted mass of the 't~Hooft fermion which would become massless again in the limit $\eta_{w}\to0$. The same pole appears also in $B_{1/2}$, consistently with the fact that it is associated to a resonance which carries a non zero $R$ charge.   

\begin{figure}[h!]
\centering
\includegraphics[width= 0.5\textwidth]{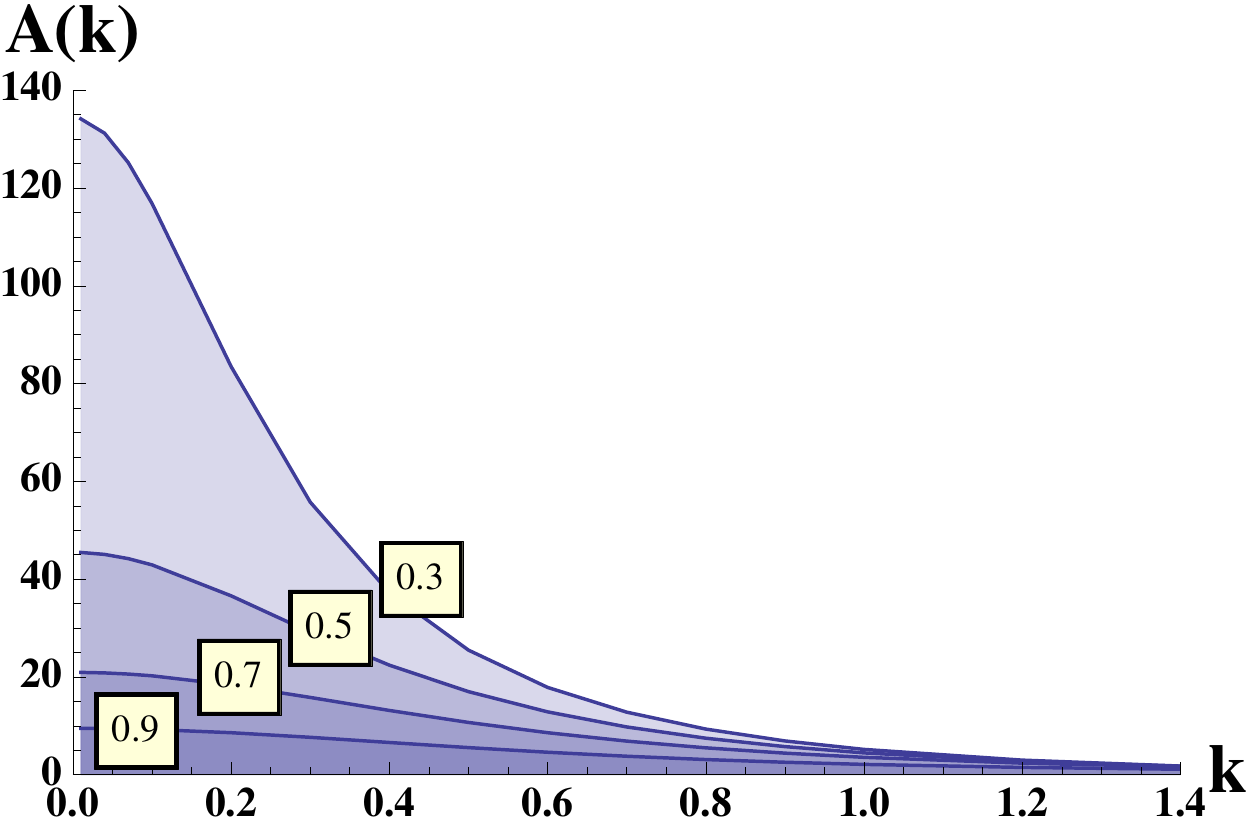}\hfill 
\includegraphics[width= 0.5\textwidth]{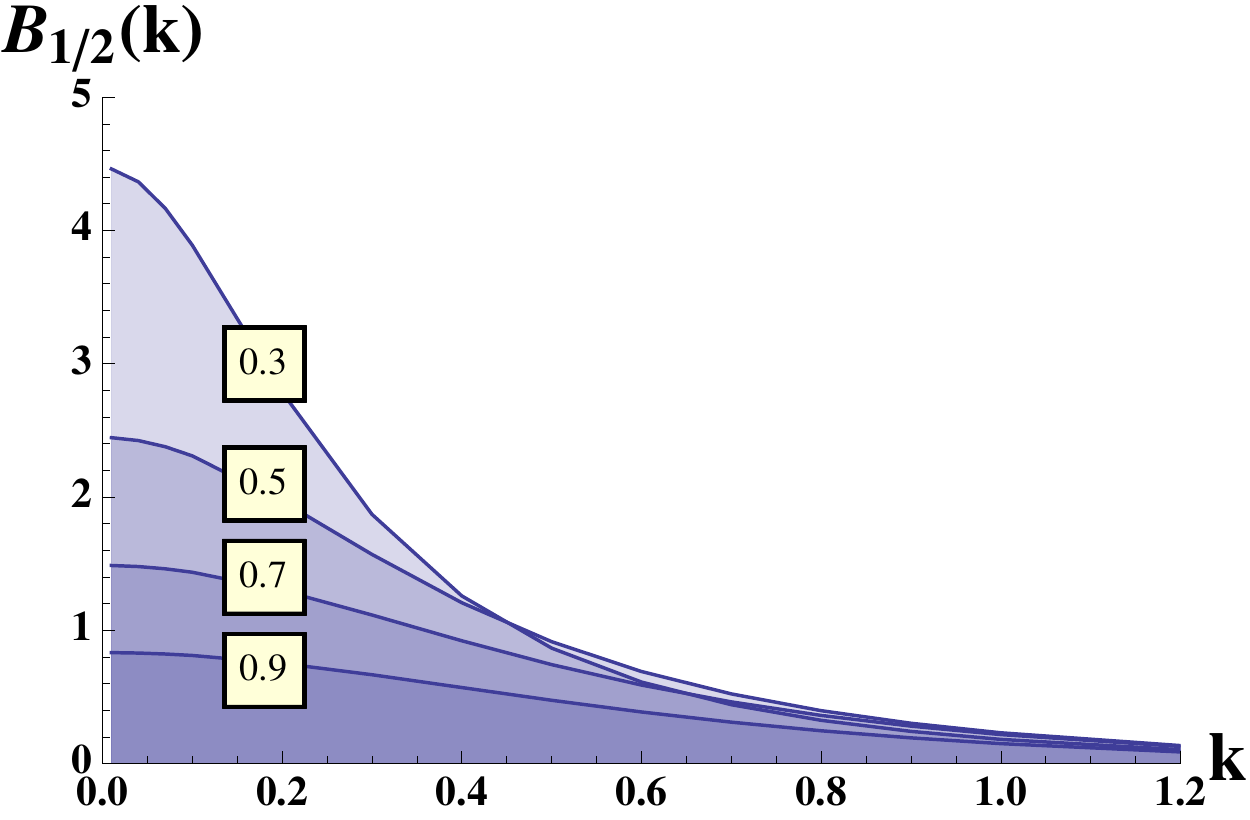}
\caption{On the left: $A$ for $k^2>0$ on the dilaton like background for different values of $\eta_w$ at fixed $z_{\text{sing}}=1.2$. On the right: $B_{1/2}$ for $k^2>0$ on the dilaton like background for different values of $\eta_w$ at fixed  $z_{\text{sing}}=1.2$. Each line corresponds to a different value of $\eta_w$ which has been defined in \eqref{etawdef}. The corresponding values of the UV parameters \eqref{parspace} can be extracted from the UV behavior of the solution and are reported in Table \ref{Bulletto_points}.}
\label{Bulletto_2}
\end{figure}

In Figure \ref{Bulletto_1} (right) we show that the presence of a non trivial $\eta$ profile is also giving a mass to the  dilaton of the dilaton domain wall solution presented in Figure \ref{C1R_C2_Gubser}. The uplifting of the dilaton pole is unavoidable in the dilaton like solutions since both $\eta_{0}$ and $\tilde{\eta}_2$ are forced to be non zero as discussed in \ref{dilatonback}. 

Finally, we display in Figure \ref{Bulletto_2} the behavior of both $A$ and $B_{1/2}$ for $k^2>0$. Both functions are gapped as expected. However, the presence of a light resonance both in $C_{1/2}$ and in $B_{1/2}$ makes $A$ and $B$ very peaked at low momenta. In particular, we see that the value of $A$ and $B$ at $k^2=0$ is decreasing with $\eta_{w}$ consistently with the fact that the 't~Hooft fermion Majorana mass becomes bigger.

\subsection{Non singular \& Walking backgrounds}

\subsubsection*{\it{Flow to an IR fixed point}} 
As discussed in Section \ref{Nonsingular}, tuning exactly $\eta_0$ and $\tilde\eta_2$ and setting $\tilde{\phi}_4=0$ one obtains a line of 
non singular solutions which is depicted in Figure \ref{THECUBE}. 
From the QFT point of view, the non singular backgrounds are realizing interpolating flows between an $\mathcal{N}=1$ SCFT in the UV and a non SUSY fixed point in the IR. We should recall again that this class of backgrounds was already obtained from the top down perspective in \cite{Distler:1998gb} and it is known to be unstable in the full $\mathcal{N}=8$ SUGRA. Within our truncation these flows are perfectly healthy because the $\mathcal{N}=8$ scalar which would acquire a $m^2<-4$ in the IR (leading to a non unitary theory) is not included in the definition of our ``effective'' CFT in the UV. 

\begin{figure}[h!]
\centering
\includegraphics[width= 0.5\textwidth]{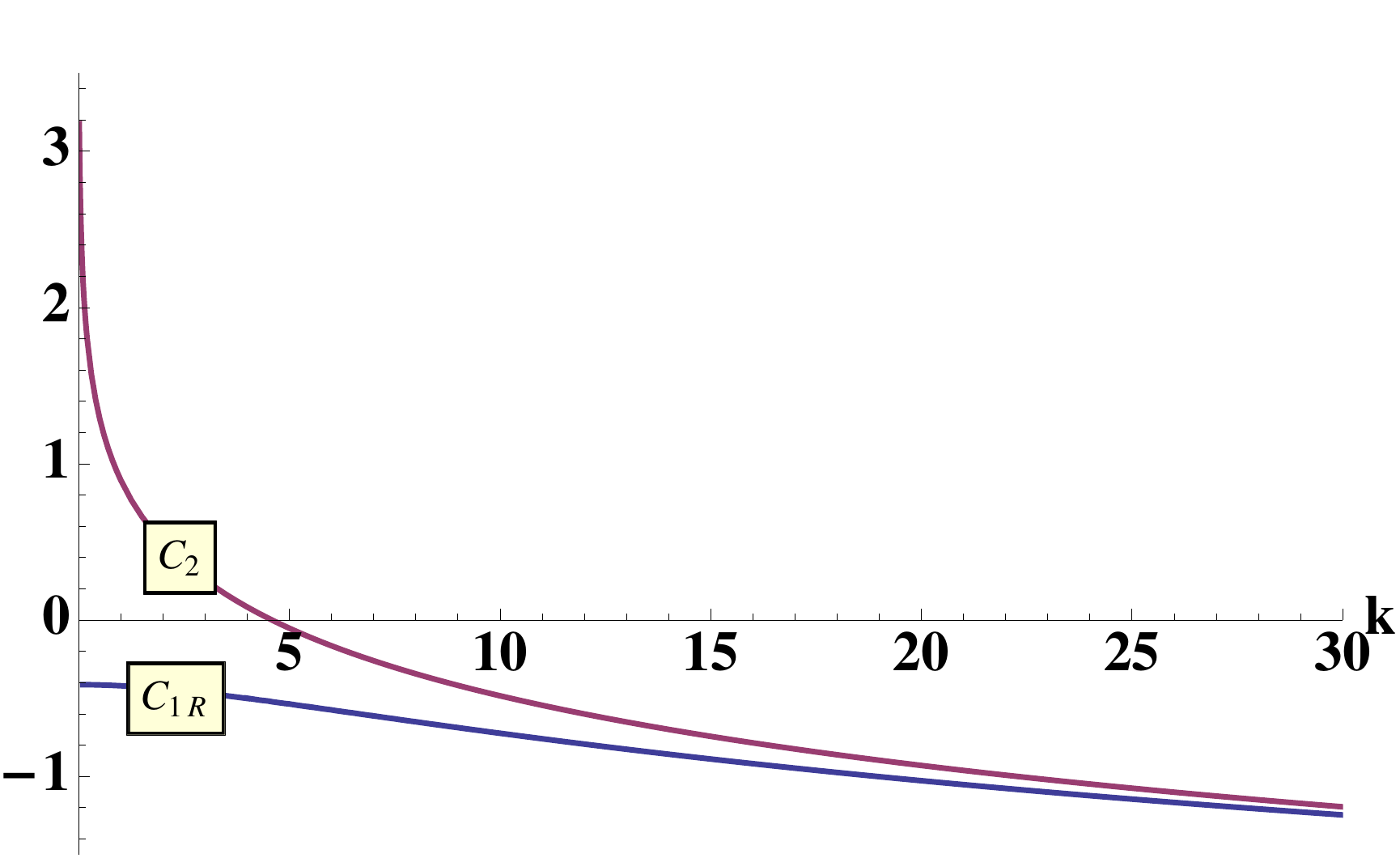}\hfill
\includegraphics[width= 0.5\textwidth]{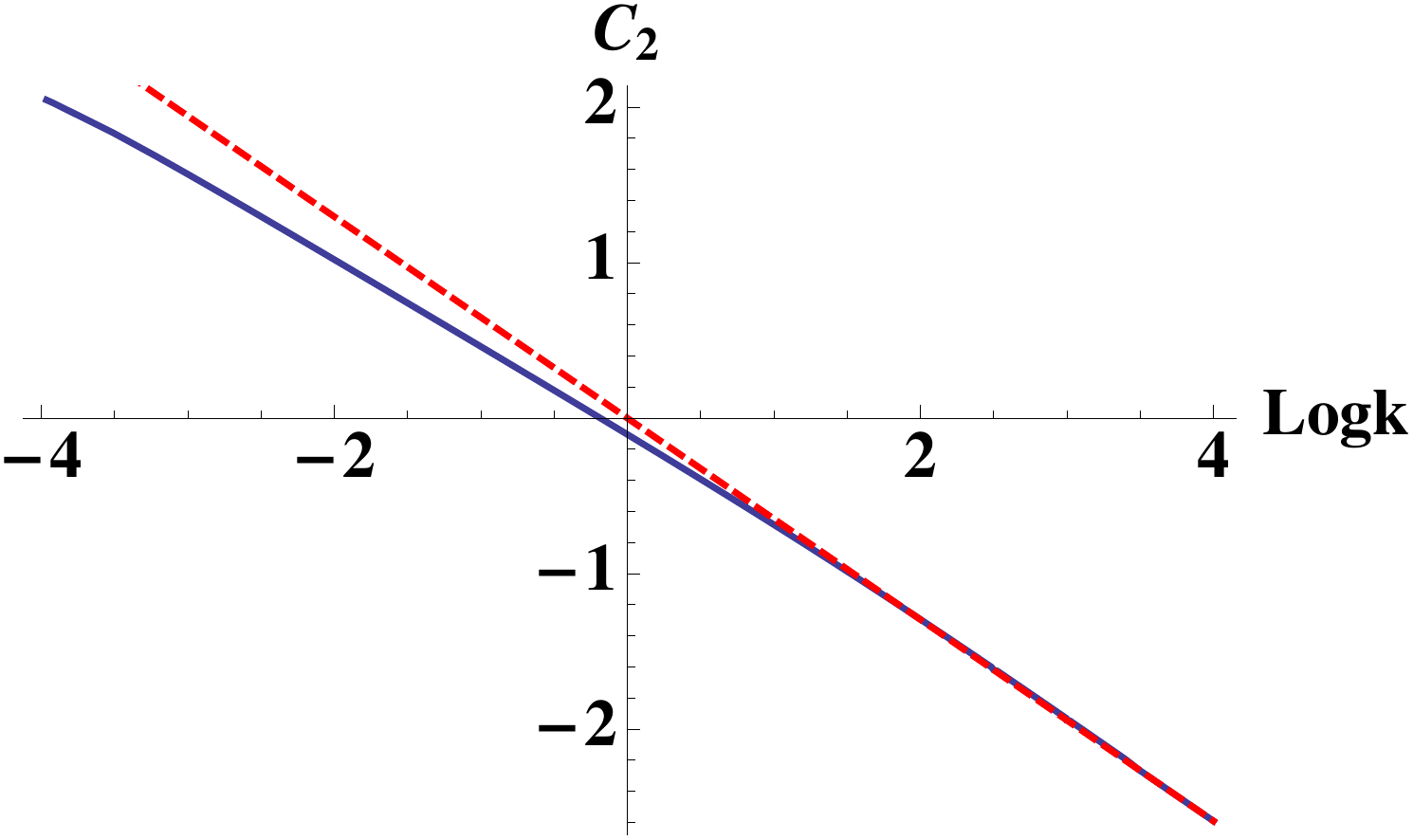}
\caption{On the left: $C_{2}$ and $C_{1R}$ for $k^2>0$ over the non singular background solution described in Section \ref{Nonsingular}. On the right: $C_{2}$ as a function of $\log k$. The red dashed line corresponds to the $AdS_{5}$ solution with $L=1$ while the blue solid line correspond to the interpolating flow between the UV $AdS_{5}$ fixed point with $L=1$ and the IR one with $L=\tfrac{2\sqrt{2}}{3}$.}
\label{DZplots}
\end{figure}

Correlators of the stress energy tensor are particularly important in the study of QFTs interpolating between pairs of CFTs. The main reason comes from the fact that external gravity can be used as a probe of the changing of degrees of freedom along the flow \cite{Anselmi:1997rd,Anselmi:1997am}. 
In Figure \ref{DZplots} (left) we show the behavior of $C_{2}$ and $C_{1R}$ along the flow. $C_{2}$ displays a logarithmic behavior at both high and small external momenta. This is indeed the expected behavior of the stress energy tensor two point function at the UV and IR fixed points, where only the identity operator can have a non trivial expectation value.  $C_{1R}$ is instead a gapped function in the IR where $R$ symmetry and also SUSY are explicitly broken. In the UV $C_{1R}$ approaches the logarithmic behavior of $C_{2}$ consistently with the restoration of SUSY Ward identities at high momenta.    

In Figure \ref{DZplots} (right) we show more clearly the behavior of $C_{2}$, comparing it with the pure $AdS_{5}$ case. The two point function of the stress energy tensor at the fixed point is fully determined by the value of the $c$ central charge which appears in front of the logarithm.\footnote{In a SUSY fixed point this behavior is obviously shared by the two point functions of the R current and the supercurrent in agreement with the SUSY Ward identity \eqref{WardSUSY1}.} Plotting $C_{2}$ as a function of $\log(k)$ we can probe the value of the central charge as the slope of the $C_{2}$ line.    

From Figure \ref{DZplots} (right) we see clearly that $C_{2}$ is a good candidate for an interpolating function between the UV and IR central charges. Moreover we see that the value of the central charge is decreasing going towards the IR since the solid blue line stays always below the dashed red one which represents the pure $AdS_{5}$ case. This is in agreement with the holographic $c$ theorem \cite{Girardello:1998pd,Freedman:1999gp}. The ratio between the slopes of $C_{2}$ for high and small momenta should be equal to the ratio between the UV and IR central charges. This ratio can be determined analytically from the value of the potential \eqref{poteta} at the stationary points
\begin{equation}
\frac{c_{UV}}{c_{IR}}=\left(\frac{V_{UV}}{V_{IR}}\right)^{ 3/2}=\frac{27}{16\sqrt{2}}\ .
\end{equation}   
Extracting numerically the slopes of $C_{2}$ we get agreement with this result at the percent level. An analytical example of the interpolating behavior of $C_{2}$ in holography has been constructed in \cite{Anselmi:2000fu} (see also \cite{Mueck:2008gv,Muck:2010uy} for further studies about the properties of holographic flows interpolating between two CFTs).   

\subsubsection*{\it{Walking backgrounds}}
We now consider backgrounds where the choice of parameters is mildly
tuned. We set $\tilde\phi_4=0$ for simplicity, and tune the choice of
$\eta_0$ and $\tilde\eta_2$ so that we find a background where the
$\eta$ profile lingers for a long range around the minimum of the
potential, before exploding into the singularity. 

\begin{figure}[h!]
\centering
\includegraphics[width= 0.5\textwidth]{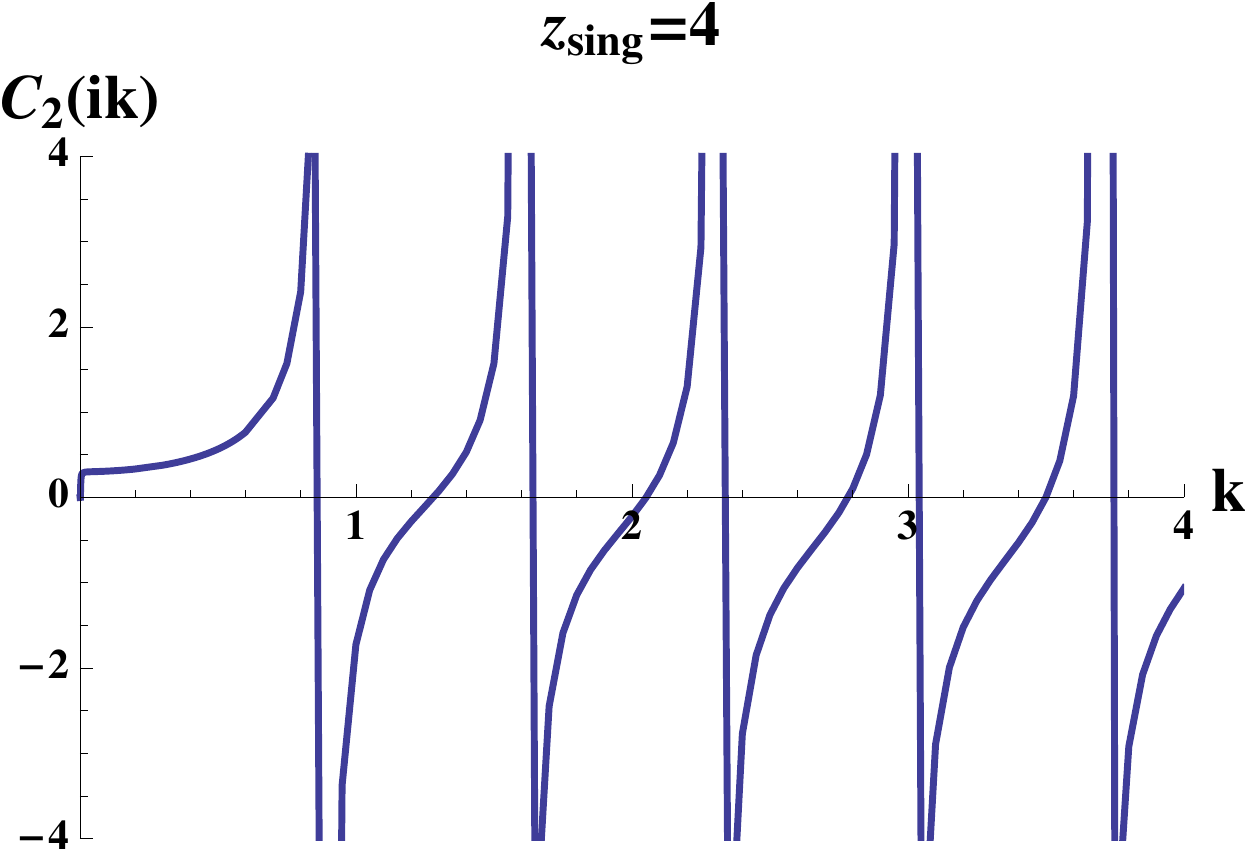}\hfill
\includegraphics[width= 0.5\textwidth]{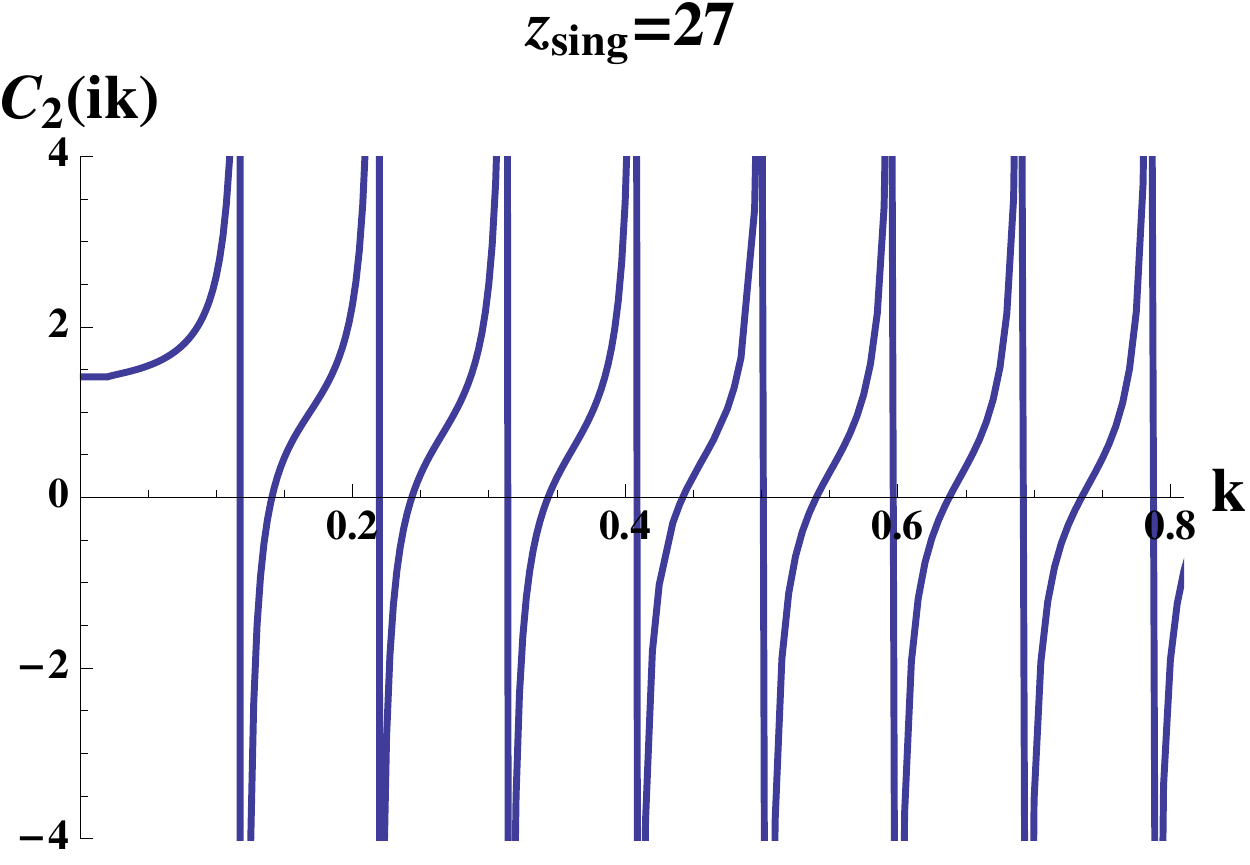}
\caption{On the left: $C_{2}$ for $k^2<0$ over a singular background solution with $z_{\text{sing}}=4$. On the right: $C_{2}$ for $k^2<0$ over a singular background solution with $z_{\text{sing}}=27$. This corresponds to the solution displayed in Figure \ref{walk} where  $F$ and $\eta$ display a walking behavior.}
\label{Walkingplots}
\end{figure}

In Figure \ref{Walkingplots} we show the pole structure of $C_{2}$ for $k^2<0$ over two different backgrounds very close to the non singular background presented in Figure \ref{DZ}. Fine tuning the UV parameters one can push $z_{\text{sing}}$ arbitrarily far. Accordingly the lightest mode in the spectrum gets lighter. However, comparing the two plots in Figure \ref{Walkingplots} we see that no hierarchy is generated between the lightest mode and the other modes in the spectrum. Pushing $z_{\text{sing}}$ to larger values corresponds simply to an overall rescaling of the spectrum in this case. This is not surprising since, from the perspective of the IR fixed point, conformality is broken by an irrelevant operator which is not related to a (approximately) flat direction in the scalar potential. As a consequence, the breaking of conformal symmetry is not parametrically small in any sense. 
 
Our example here shows how having a SUGRA solution with a walking behavior does not imply straightforwardly the presence of a naturally light dilaton, parametrically lighter than the rest of the spectrum. A thorough analysis of the necessary conditions to obtain a light dilaton from a deformation of a UV fixed point has been recently performed in \cite{Coradeschi:2013gda}, and holographic examples based on nearly marginal deformation have been constructed in \cite{Coradeschi:2013gda,Bellazzini:2013fga,Megias:2014iwa,Cox:2014zea}.  

\section{Holographic renormalization and (super)symmetry breaking RG flows}
In this section we aim at providing the theoretical and technical
framework in which the correlators that we have discussed in the
previous section have been obtained. 

Since we are focusing on two point functions of gauge invariant
operators in field theories with an $AAdS$ gravity dual,
the framework for the computations is that of holographic
renormalization
\cite{Mueck:2001cy,Bianchi:2001de,Bianchi:2001kw,Skenderis:2002wp}. In
a nutshell, this is a systematic procedure that allows one to establish
in all generality (i.e.~in a solution independent way) the generating
function for the correlators. This is the on shell supergravity action
expressed as a boundary term, after one has properly subtracted
divergencies associated to the infinite volume of $AdS$. In order to
extract the two point correlators one then has to fluctuate the bulk
fields dual to the operators one is interested in, with boundary
conditions which amount to fixing the leading mode at the boundary and
asking for regularity in the bulk.

This is a well known and much used procedure that we will apply to our generic backgrounds which have two scalars with a non trivial profile. 
Also, as already reviewed in the previous section, the most important features in the correlators that we
compute are associated to symmetry, and supersymmetry, breaking. We
will thus pay a particular attention to how holographic renormalization deals with symmetry breaking.
Since we will be interested in two point correlators, we can actually simplify the holographic renormalization machinery by separating non trivial background profiles and fluctuations from the start. This approach allows to use the simpler linearized equations of motion for the fluctuations instead of the full non linear equations of motion. 

The holographic renormalization of the model presented in Section
\ref{sugra} is complicated due to the number of fields involved,
and because we want to consider a rather generic background. The
details tend to obscure the simple physical insight concerning the
consequences of symmetry breaking. We thus begin this section with a
toy model of a vector and an axion-like scalar and study its
holographic renormalization when the symmetry is broken either
 explicitly or spontaneously.\footnote{Essentially similar models are
  discussed in \cite{Bianchi:2001kw}. However there the spontaneous
  and explicit cases are discussed in two different models, while here
  we aim at giving a unified treatment.}
 
\subsection{A toy model for symmetry breaking in holography}
In this section we present a simplified model of a vector coupled to a
scalar in the bulk of  $AdS$. From the point of view of the bulk
supergravity, as soon as the scalar has a non trivial profile, the
gauge symmetry associated to the vector is spontaneously broken, in an
$AdS_5$ version of the Brout-Englert-Higgs mechanism. From the boundary theory point
of view on the other hand, the physics depends on the specific profile
of the scalar. If the profile is non normalizable (i.e.~there is a
source term), then the global symmetry dual to the bulk gauged
symmetry is explicitly broken by the presence of a non invariant
operator with a non vanishing coupling. In this case we expect
non trivial Ward identities involving the non conservation of the current to be implemented on the correlators. If on the other hand the profile is normalizable (i.e.~it corresponds to a pure VEV), then the global symmetry is spontaneously broken by the VEV of the same operator. 
We expect a massless Goldstone boson to show up in the correlators of the conserved current.

Our starting point is the following action:
\begin{equation}
{\cal S}=\int d^5x \sqrt{G}\left[\frac{1}{4}F^{MN}F_{MN}+(\partial_M \Phi-iA_M \Phi)(\partial^M \Phi^*+iA^M \Phi^*)+\mu^2\Phi\Phi^*\right]. \label{scomplM}
\end{equation}
We can parametrize the complex scalar $\Phi$ in terms of two real fields, the modulus and the phase:
\begin{equation}
\Phi=\frac{1}{\sqrt{2}}me^{i\alpha},
\end{equation}
so that the action becomes
\begin{equation}
{\cal S}=\int d^5x \sqrt{G}\left[\frac{1}{4}F^{MN}F_{MN}+\frac{1}{2}\partial_M m \partial^M m
+\frac{1}{2}m^2(\partial_M \alpha-A_M)(\partial^M \alpha -A^M)+\frac{1}{2}\mu^2 m^2\right].\label{stuckel}
\end{equation}
We see that the scalar $m$ is not charged under the gauge symmetry, which operates as $A_M\to A_M +\partial_M \lambda$ together with $\alpha \to\alpha +\lambda$.  We will be referring to $\alpha$  as the ``axion''. $m$ actually couples to the gravitational sector but for the sake of simplicity here we will not consider neither backreaction nor fluctuations of the modulus $m$. We will just assume that it has a profile which depends on its mass $\mu$. 
Note that the present parametrization is not valid for a trivial (vanishing)
profile for $m$ (the degree of freedom
associated to $\alpha$ disappears
altogether from the action). In that case one should just expand to
quadratic order  the action  \eqref{scomplM}, which splits into two
decoupled free actions for $A_M$  and $\Phi$.

We will assume the background metric to be just the $AdS$ one
\begin{equation}
ds^2=\frac{1}{z^2}\left(dz^2+dx_\mu^2\right).
\end{equation}
For $\mu^2=-3$, to which we will stick henceforth, the most general Poincar\'e invariant profile is 
\begin{equation}
m=m_0 z +\tilde m_2 z^3.
\end{equation}
Note that this set up is similar to many phenomenologically motivated bottom up constructions, such as hard and soft wall models. Here we use it merely as a toy model, 
in principle full backreaction can be contemplated but the main
features detailed below remain unchanged, as will be clear when we will
apply the same procedure to the full SUGRA model.

We  restrict ourselves to the quadratic action for the system of vector and axion fluctuations, given by
\begin{equation}
{\cal S}=\int d^5x
\sqrt{G}\left[\frac{1}{4}F^{MN}F_{MN}
+\frac{1}{2}m(z)^2(\partial_M \alpha-A_M)(\partial^M \alpha
-A^M)\right].
\label{Salpha}
\end{equation}

We now split the indices into those which are respectively tangential
and perpendicular to the boundary. We further separate the 4d vector into transverse and longitudinal parts
\begin{equation}
A_\mu=A_\mu^t +\partial_\mu A_l, \qquad \partial^\mu A_\mu^t=0.
\end{equation}

The equations of motion derived from \eqref{Salpha}  split into transverse and longitudinal subsystems. The equations for the transverse fluctuations are
\begin{equation}
-\partial_z\left(\frac{1}{z}{A_\mu^t}'\right)-\frac{1}{z}\Box A_\mu^t+\frac{1}{z^3}m^2 A_\mu^t=0. \label{eqtrans}
\end{equation}
The system for the longitudinal fluctuations is
\begin{subequations}
\begin{align}
&-\partial_z\left(\frac{1}{z}A_l'\right)+\partial_z\left(\frac{1}{z}A_z\right)-\frac{1}{z^3}m^2  \alpha+\frac{1}{z^3}m^2 A_l=0 ,\label{longeq1}\\
&\frac{1}{z}\Box A_l'-\frac{1}{z}\Box A_z-\frac{1}{z^3}m^2 \alpha'+\frac{1}{z^3}m^2 A_z=0, \label{boxeq}\\
&-\partial_z\left(\frac{1}{z^3}m^2\alpha'\right) + \partial_z\left(\frac{1}{z^3}m^2A_z\right)-\frac{1}{z^3}m^2\Box\alpha+\frac{1}{z^3}m^2 \Box A_l=0.\label{longeq3}
\end{align}
\end{subequations}
Note that we are not fixing any gauge whatsoever.

The above equations allow us to rewrite the action \eqref{Salpha} on
shell as a boundary action, that we call  the regularized action
because we assume that it is evaluated at a very small but finite
value of $z$:
\begin{equation}
{\cal S}_\mathrm{reg}=-\int_{z=\epsilon} d^4x
\left\{\frac{1}{2z}{A_\mu^t}{A_\mu^t}'+\frac{1}{2z^3}m^2(\alpha-A_l)(\alpha'
-A_z)\right\}.
\end{equation}
It clearly splits into a transverse and a longitudinal piece. We now implement holographic renormalization, specializing in turn to the explicit and spontaneously broken cases.

\subsubsection{Explicit breaking: $m_0\neq 0$}
We first write the fluctuations as a near boundary expansion. The transverse fluctuations become
\begin{equation}
A_\mu^t=a_{0\mu}^t+a_{2\mu}^tz^2 \log z  + \tilde a_{2\mu}^tz^2  +\dots
\end{equation}
The leading term $a_{0\mu}^t$ is the source for the transverse
(i.e.~divergence free) part of a boundary global symmetry current $J_\mu^t$.

In the longitudinal sector the expansions of the fluctuating fields are as follows:
\begin{subequations}
\begin{align}
&\alpha= \alpha_0 +  \alpha_2 z^2 \log z+ \tilde \alpha_2 z^2+\dots \label{expalpha}\\
&A_l  =  a_{l0}+  a_{l2} z^2 \log z+ \tilde a_{l2}z^2 +\dots \label{expal} \\
&A_z  = 2a_{z2}z\log z + (2\tilde a_{z2}+a_{z2})z+\dots \label{expaz}
\end{align}
\end{subequations}
The expansion of $A_z$ is motivated by how it enters the equations.

Note that at this stage we have not
fixed the gauge in any way. The unitary gauge would amount to choose
$\alpha=0$, while the axial gauge sets $A_z=0$. We will keep the gauge
unfixed unless it is needed, in which case
we will choose the axial gauge which is the most suited one for
holography. Indeed it is difficult to associate a holographic dual
operator to $A_z$, while an operator related to $\alpha$ can be found
in the following way:
\begin{equation}
\int d^4x \Phi_0 O_m +c.c. = \sqrt{2}\int d^4x\, ( m_0\, \mathrm{Re}\, O_m - m_0\alpha_0\,  \mathrm{Im}\, O_m)+\dots.\label{explcoupl}
\end{equation}
Strictly speaking, the above relation is telling that the source for
$\mathrm{Im}\, O_m$, when $m_0\neq 0$, is
$\beta_0=m_0\alpha_0$. 

Let us anticipate that in the spontaneous case, when $m_0=0$ and thus $m=\tilde m_2
z^3$, we will have to write the coupling as follows:
\begin{equation}
- \sqrt{2}\int d^4x\, \tilde m_2\alpha_0\,  \mathrm{Im}\, O_m,
\label{spontcoupl}
\end{equation}
so that now the source for $\mathrm{Im}\, O_m$ is $\tilde
\beta_0=\tilde m_2\alpha_0$. 
Note that while the coupling in \eqref{explcoupl} implies that
$\alpha_0$ is dimensionless, in \eqref{spontcoupl} it is implied on
the other hand that $\alpha_0$ has dimension minus two, in order to
compensate the dimension of $\tilde m_2$. We will thus have to
pay attention to the fact that the expansion for $\alpha$ is different
depending on the profile of $m$, corresponding to explicit and
spontaneous breaking of the symmetry.

We finally remark that the operator sourced by $a_{l0}$ is $\partial_\mu J^\mu$:
\begin{equation}
\int d^4x a_{0\mu}J^\mu \supset \int d^4x \partial_\mu a_{l0}J^\mu =
-\int d^4x a_{l0}\partial_\mu J^\mu.
\end{equation}

The equations for the fluctuations are
\eqref{eqtrans}--\eqref{longeq3}. When  $m=m_0z$, they imply the following relations among the parameters in the expansion. In the transverse sector
\begin{equation}
a_{2\mu}^t = \frac{1}{2}(-\Box +m_0^2) a_{0\mu}^t. \label{a2trans}
\end{equation}

In the longitudinal sector there are actually three coupled equations (one of which is a consequence of the other two) for the linear combinations $A_l'-A_z$, $\alpha'-A_z$ and $\alpha-A_l$. Solving for the equations with the expansions \eqref{expalpha}--\eqref{expaz}, one finds the following relation (among other redundant ones)
\begin{equation}
\alpha_2-a_{z2}=-\frac{1}{2}\Box(\alpha_0-a_{l0}).
\end{equation}

The regularized action reads
\begin{align}
{\cal S}_\mathrm{reg}=\int_{z=\epsilon} d^4x &\left\{\log z \frac{1}{2}a_{0\mu}^t(\Box -m_0^2) a_{0\mu}^t +
m_0^2 \log z \frac{1}{2}(\alpha_0-a_{l0})\Box(\alpha_0-a_{l0})\right. \nonumber\\
&\quad- a_{0\mu}^t \tilde a_{2\mu}^t -m_0^2(\alpha_0-a_{l0})(\tilde \alpha_2-\tilde a_{z2})\nonumber\\
&\quad\left.- \frac{1}{4}a_{0\mu}^t(-\Box +m_0^2) a_{0\mu}^t + m_0^2 \frac{1}{4}(\alpha_0-a_{l0})\Box(\alpha_0-a_{l0})
\right\}.
\end{align}
The first line contains the logarthmically divergent terms, the second line has potentially non-local terms depending on the bulk profile of the fluctuations (through $\tilde a_{2i}^t$, $\tilde \alpha_2$ and $\tilde a_{z2}$) while the third line displays finite local terms.

The divergent terms must be cancelled by  local counter terms. Note that finite counter terms can also affect the local finite piece of the action. The counter terms must be written in a way that respects the symmetries of the boundary action. We thus have
\begin{eqnarray}
{\cal S}_\mathrm{ct}&=&\int_{z=\epsilon} d^4x\sqrt{\gamma} \log z \left\{\frac{1}{4}F_{\mu\nu}F^{\mu\nu} +\frac{1}{2}m^2(\partial_\mu \alpha -A_\mu)(\partial^\mu\alpha-A^\mu) \right\} \nonumber \\
&=& \int_{z=\epsilon} d^4x \log z \frac{1}{2}\left\{ -a^t_{0\mu} \Box a^t_{0\mu}+m_0^2 a_{0\mu}^t a_{0\mu}^t -m_0^2 (\alpha_0-a_{l0})\Box(\alpha_0-a_{l0})\right\} \label{scttoy}
\end{eqnarray}
so that we see that both the transverse and longitudinal counter terms
are part of the same covariant and gauge invariant counter term. 

Given that the counter term action can also contain  finite terms, we end up with a renormalized action which still depends on free parameters in front of the finite local terms:
\begin{align}
{\cal S}_\mathrm{ren}=\int d^4k &\left\{
-a_{0\mu}^t \tilde a_{2\mu}^t -m_0^2(\alpha_0-a_{l0})\tilde \alpha_2 \right.\nonumber\\
&\quad\left.- \frac{1}{2} a_{0\mu}^t(\chi k^2 +\xi m_0^2) a_{0\mu}^t -\frac{\xi}{2}m_0^2(\alpha_0-a_{l0})k^2(\alpha_0-a_{l0})
\right\},\label{anyscheme}
\end{align}
where we have now made the transition to momentum space, and
we have set $A_z=0$ (and hence $\tilde a_{z2}=0$) to be in the
axial gauge. 
The coefficient $\chi$ is related to the possibility to add finite counter terms proportional to $F_{\mu\nu}F^{\mu\nu}$, while the coefficient $\xi$ corresponds to finite counter terms proportional to the terms with $m^2$ in \eqref{scttoy}. Let us recall that a particular choice of the finite counterterms defines a scheme in the dual QFT.

We are now finally ready to compute the correlators.
In a notation similar to the one of \eqref{<jj>}, 
the correlator of the current $J_\mu$ dual to the bulk field $A_\mu$ has an expression given by
\begin{equation}
\langle J_\mu (k)
J_\nu(-k)\rangle = - (k^2 \delta_{\mu\nu}-k_\mu k_\nu)C(k^2)-m_0^2 \frac{k_\mu k_\nu}{k^2}F(k^2).
\label{toyparam}
\end{equation}
We now remark that  both the transverse and the longitudinal parts of the correlator above can have a spurious massless pole while the full correlator has none. If $F$ is finite at $k^2=0$ then it seems that the correlator of the longitudinal
part of the current has a massless pole. However if $C$ also has a massless pole with residue $m_0^2 F(0)$, then the massless pole cancels from the full correlator. Indeed, precisely
when $m_0\neq 0$, we do not expect massless poles in the current correlator,
since the symmetry is explicitly broken while a massless pole would
be associated to a Goldstone mode.
We now show that a massless pole in $C$, and at the same time a constant part of $F$, can actually be eliminated
altogether by a scheme choice, thus establishing their spurious nature. 

The holographic computation of the transverse part of the correlator \eqref{toyparam} is given by
\begin{equation}
\langle J_\mu^t (k) J_\nu^t(-k)\rangle =-\frac{\delta^2 {\cal S}_\mathrm{ren}}{\delta a_{0\mu}^t \delta a_{0\nu}^t}=
\frac{\delta \tilde a_{2\mu}^t}{\delta a_{0\nu}^t}+\frac{\delta \tilde
  a_{2\nu}^t}{\delta a_{0\mu}^t}+
(\chi k^2+\xi m_0^2)(\delta_{\mu\nu}-\frac{k_\mu k_\nu}{k^2}).
\end{equation}
The projector appears since we are taking variations with respect to
transverse vectors. We see that the parameter $\xi$ can 
be fixed in order to cancel an unphysical massless pole. The value of $\xi$
will depend on the solution of the bulk fluctuations, i.e.~on the
limit of $\tilde a_{2\mu}^t$ for $k \to 0$. Note that independently on
the value one chooses for $\xi$, all other information extracted from
the above correlator is 
physical (for instance, the poles for imaginary values of
$k$).\footnote{In \cite{Bianchi:2001kw}, no particular choice of
  scheme (i.e.~of our $\xi$) was made, it was only checked that the massless
  poles cancel in the complete correlator.}

In the simple model above, we can actually obtain an analytical expression for $\tilde a_{2\mu}^t$. The equation for the transverse fluctuation being the same as for a massless vector upon the substitution $k^2 \to k^2 +m_0^2$, we then have
\begin{equation}
\tilde{a}_{2 \mu}^t=\left[\frac{k^2+m_0^2}{4}\log\left(k^2+m_0^2\right)+\frac{k^2+m_0^2}{2}\left(\log 2+\gamma+\frac{1}{2}\right)\right]a_{0 \mu}^t.
\end{equation}
We thus get
\begin{equation}
C(k^2)_{\chi,\xi}=-\frac{k^2+m_0^2}{k^2}\left[\frac{1}{2}\log\left(k^2+m_0^2\right)+\left(\log
2+\gamma+\frac{1}{2}\right)\right]-\chi - \xi\frac{ m_0^2}{k^2}
\end{equation}
and so in order to cancel the pole at $k^2=0$ we need to fix 
\begin{equation}
\xi=-\frac{1}{2}\log m_0^2-\left(\log
2+\gamma+\frac{1}{2}\right)\ . \label{valuexi}
\end{equation}
We eventually have the following form factor, finite at the origin
\begin{equation}
C(k^2)=\frac{k^2+m_0^2}{2k^2}\log\left(\frac{m_0^2}{k^2+m_0^2}\right),
\end{equation}
where we further cancelled the constant term by tuning $\chi$. (Note that we recover a $\log$-dependence in the superconformal limit.)
In a more general background, the value of $\xi$ can be fixed in a similar way in
order to cancel the massless pole that arises in the transverse correlator.
The choice of $\xi$ will also be
relevant for the longitudinal part, that we now consider.

An important feature of the renormalized action \eqref{anyscheme} in the longitudinal sector is that it depends
only on the combination $\alpha_0-a_{l0}$. Indeed, as observed
previously, the equations (and the bulk boundary conditions) 
also depend only on $\alpha-A_l$ so that $\tilde \alpha_2$ will depend only on $\alpha_0-a_{l0}$ as well. Taking
this into account, and substituting $\alpha_0$ for the source
$\beta_0=m_0\alpha_0$, the renormalized action for the longitudinal sector becomes 
\begin{equation}
{\cal S}_\mathrm{ren}=-\int d^4k\,  (\beta_0-m_0a_{l0})f(k^2)(\beta_0-m_0a_{l0}),
\label{srentoyexp}
\end{equation}
for some typically non local function $f(k^2)$, which also depends on the choice of scheme \eqref{valuexi}.
 
As a consequence, we have that
\begin{equation}
\frac{\delta {\cal S}_\mathrm{ren}}{\delta a_{l0}}=-m_0\frac{\delta {\cal S}_\mathrm{ren}}{\beta_0}.
\end{equation}
This is nothing but the holographic implementation of the Ward identity, 
relating the non conservation of the current to the operator that breaks the symmetry. 
Indeed, relating the zero modes to the dual operators that they source, we derive the identity
\begin{equation}
\langle \partial_\mu J^\mu\rangle = -\sqrt{2} m_0 \langle \mathrm{Im}\, O_m\rangle,
\end{equation}
which is actually true for any insertion in a correlator and therefore the closest thing to an operator identity to be derived holographically.

We can now complete the analysis of this toy model by computing the two point function
\begin{equation}
-\frac{\delta^2 {\cal S}^l_\mathrm{ren}}{\delta \beta_0\delta \beta_0}=2 \langle \mathrm{Im} O_m\mathrm{Im} O_m\rangle= \frac{1}{m_0^2}\langle k_\mu J^\mu k_\nu J^\nu\rangle=2 f(k^2)=-k^2F(k^2)\ ,
\end{equation}
where the value of $f(k^2)$  is obtained
by solving the equations in the bulk, and the last equality refers to the parametrization in \eqref{toyparam}. The value we get in the toy
model is not particularly illuminating, we refer to our main model for more physically interesting examples. Note however that the finite counter terms in \eqref{anyscheme} can precisely cancel a finite value $F(0)$.

\subsubsection{Spontaneous breaking: $m_0=0$}
We now consider the case $m_0=0$, and so we take $m=\tilde m_2 z^3$.

As we already remarked, the expansions in the longitudinal sector have to be modified.
In this case it turns out that unnecessary ambiguities are eliminated
right away by fixing the axial gauge $A_z=0$. Then, instead of \eqref{expalpha}--\eqref{expaz}, we  expand the
remaining fields as follows:
\begin{subequations}
\begin{align}
&\alpha =\frac{\alpha_0}{z^2} +  \alpha_2  \log z+ \tilde \alpha_2 +\dots \\
&A_l = a_{l0}+  a_{l2} z^2 \log z+ \tilde a_{l2}z^2 +\dots 
\end{align}
\end{subequations}
In particular we underline that the leading term in $\alpha$ goes like $z^{-2}$.
Note that because of the gauge fixing, $\alpha_0$ and $\alpha_2$ are
actually invariant under residual gauge transformations, while $\tilde
\alpha_2$ transforms with the same shift as $a_{l0}$, since
$\alpha-A_l$ is gauge invariant.

By inserting the expansions in the equations, one again learns  the
 relations among the various modes. For the transverse fluctuations, the same relation 
\eqref{a2trans} is valid but with $m_0=0$.
In the longitudinal sector, the only relevant relation in order
to compute the regularized action is the following:
\begin{equation}
\alpha_2=-\frac{1}{2}\Box\alpha_0.
\end{equation}

As far as the transverse sector is concerned, the procedure of holographic renormalization is exactly the same as in the explicit case, we just have to set $m_0=0$ in all the expressions. We thus concentrate on the longitudinal sector, where we get 
\begin{equation}
{\cal S}_\mathrm{reg}^l=\int_{z=\epsilon} d^4x \ \tilde m_2^2\left\{ \frac{1}{z^2}\alpha_0^2-\frac{1}{2}\log z \alpha_0\Box\alpha_0
+\alpha_0\tilde \alpha_2+  \frac{1}{4}\alpha_0\Box\alpha_0-\alpha_0 a_{l0} \right\}.
\end{equation}
The counter terms are easily determined to be
\begin{align}
{\cal S}^l_\mathrm{ct}&=\int_{z=\epsilon} d^4x\, \sqrt{\gamma}\ m^2\left\{-\alpha^2-
\frac{1}{2}\log z\alpha \Box \alpha\right\}\notag\\
&=
\int_{z=\epsilon} d^4x\ \tilde m_2^2\left\{ -\frac{1}{z^2}\alpha_0^2 +\frac{1}{2}\log z \alpha_0\Box\alpha_0-2
\alpha_0\tilde \alpha_2\right\} .
\end{align}
We thus get a renormalized action given by
\begin{equation}
{\cal S}^l_\mathrm{ren}=-\int d^4k\ \tilde m_2^2 \left\{\alpha_0\tilde
  \alpha_2+\frac{1}{4}\alpha_0 k^2 \alpha_0 +\alpha_0 a_{l0} \right\}.
\label{Srenalpha}
\end{equation}
Note that the coefficient of the second term can be again modified by
a finite local counter term. On the other hand, there is no finite
{\em local} counter term that can be used to change the coefficient of
the last term, even if it depends only on the sources $\alpha_0$ and
$a_{l0}$. 

When solving for the fluctuations in the bulk, 
it is the combination $\tilde \alpha_2 -a_{l0}$
that will be solved in terms of $\alpha_0$. This can be seen from the
equations, and from the fact that gauge invariance requires us to take
boundary conditions in the bulk for the combination $\alpha-A_l$ only, so that we have 
$\tilde \alpha_2= a_{l0}+\tilde f(k^2) \alpha_0$, where $\tilde f(k^2)$ is a possibly non local function of $k^2$.
In terms of the source $\tilde \beta_0=\tilde m_2 \alpha_0$,  the
renormalized action can be rewritten as
\begin{equation}
{\cal S}^l_\mathrm{ren}=-\int d^4k\left\{\tilde \beta_0\tilde f(k^2)\tilde
  \beta_0+ 2\tilde m_2\tilde\beta_0 a_{l0} \right\}, \label{srentoyspont}
\end{equation}
where we have 
included in $\tilde f(k^2)$ the local term present in \eqref{Srenalpha} and possible finite counter terms. Notice the differences with respect to \eqref{srentoyexp}.

We now turn to the correlators, where the massless Goldstone mode should appear. 
Concerning the transverse sector, the choice of counter terms
is exactly as in the unbroken (i.e. purely conformal) case, where the choice of $\chi$ in
\eqref{anyscheme} only
affects constant terms in the form factor $C$. The non trivial part of
the computation will stem from solving the equations for the
fluctuations in a background with $m=\tilde m_2 z^3$. We expect  that in such a background
$\tilde a_{2\mu}^t$ will be such that $C$ has  a massless pole
associated to a Goldstone mode.  This is indeed what we find in the
complete models discussed later. Note that since $m_0=0$, there are no finite counter terms that can cancel this massless pole.

Turning to the longitudinal sector, we note that the last term in \eqref{srentoyspont} is crucial. It leads to a non trivial correlator
between $\partial_\mu J^\mu$ and $\mathrm{Im}\, O_m$, even if all other
correlators involving $\partial_\mu J^\mu$, i.e.~the longitudinal part of
$J_\mu$, are vanishing. Eventually, this leads to the
expression
\begin{equation}
\langle J_\mu(k) \mathrm{Im}\, O_m (-k)\rangle =
\sqrt{2}\tilde m_2 \frac{k_\mu}{k^2},\label{schwt}
\end{equation}
featuring the massless pole related to the Goldstone particle. Note
that here the massless pole arises purely from considerations of the
UV expansion, and the only information we need is the presence of
the mode $\tilde m_2$ dual to the VEV of $\mathrm{Re}\, O_m$. On the
other hand, the massless pole in the form factor appearing in the
transverse current correlator appears after computing the fluctuations
of the (transverse) vector in the backreacted bulk.

This is not a surprise. The above correlator is purely given in terms
of a Schwinger term, hence also from the field theory side it can be
completely determined once the symmetry breaking VEV is known.

The other correlators are 
\begin{equation}
\langle J_\mu^l(k) J_\nu^l (-k)\rangle =0 ,
\end{equation}
and
\begin{equation}
\langle \mathrm{Im}\, O_m(k) \mathrm{Im}\, O_m (-k)\rangle =
2\tilde f(k^2).
\end{equation}
The  above correlator will contain the massless
Goldstone pole and possibly other non local terms.

We now want to make a comment on gauge invariance.
We first remark that ${\cal S}_\mathrm{reg}^l$ is gauge invariant.
On the contrary, because of its last term, ${\cal S}^l_\mathrm{ct}$ is not gauge
invariant. Then the same is true for ${\cal S}^l_\mathrm{ren}$. This
is an expected feature, since the non gauge invariance of ${\cal
  S}^l_\mathrm{ren}$ is needed in order to reproduce the fact that the
VEV of $\mathrm{Im}\, O_m$ is not invariant:
\begin{equation}
\delta \langle \mathrm{Im}\, O_m\rangle =\lambda  \langle
\mathrm{Re}\, O_m\rangle =- \sqrt{2}\lambda \tilde m_2,
\end{equation}
where $\lambda$ is the parameter of the transformation. This in turn
is the reason why the Schwinger term appears in the correlator \eqref{schwt}.

\subsubsection{Lessons from the toy model}
The practical lesson that we want to highlight from this toy model is the following. When a symmetry is broken spontaneously, we expect a massless Goldstone boson pole to show up in the correlators of the current (which is transverse) with itself, and of the operator $\mathrm{Im}\, O_m$ with itself. Such poles arise because the non-trivial backreaction in the bulk affects the fluctuations of the dual fields, and do not depend on the details of the holographic renormalization one performs at the boundary. In addition, the massless pole also shows up in the correlator of the longitudinal part of the current with $\mathrm{Im}\, O_m$, purely because the VEV of $\mathrm{Re}\, O_m$ produces a non vanishing Schwinger term. The holographic manifestation of the latter is completely dependent on rightfully carrying out the renormalization procedure.

When the symmetry is broken explicitly we expect no massless poles in
the current correlators. On the other hand we expect Ward identities
between the longitudinal part of the current and the operator
$\mathrm{Im}\, O_m$. The absence of spurious massless poles in the
transverse current two point function can be enforced by fixing the choice of  finite counter terms. This depends both on the holographic renormalization and on the solution to the fluctuation equations in the bulk. The Ward identities on the other hand are completely implemented in the renormalized action through the boundary analysis.

We are now ready to attack the full problem, comprising all the fields
in the gravity multiplet, the hypermultiplet and the vector multiplet
of the supergravity discussed in Section \ref{sugra}.
As far as the vector multiplet is concerned, the procedure of
holographic renormalization was carried out in \cite{Argurio:2012cd} for generic backgrounds
such as the ones considered here, including also the fermionic
fields. Note that the symmetry associated to this vector field is
always unbroken and so the analysis does not involve the subtleties
discussed above. In  \cite{Argurio:2013uba} the gravity multiplet of
the same ${\cal N}=2$ supergravity was considered, but the holographic
renormalization was carried out only for the transverse parts of the
currents, in purely $AdS$ backgrounds. Below we will generalize to
backgrounds which are only asymptotically $AdS$, and where we have to
take into account the mixing of the components of the gravity
multiplet with the
components of the hypermultiplet. In the present paper we will only
consider the bosonic fields in the gravity and hypermultiplet sectors. 

We will divide our discussion in two parts, considering in turn the graviphoton $R_M$ and the scalars $C_0$
and $\alpha$ with which it
mixes, and the graviton $h_{MN}$ and the ``active'' scalars $\phi$ and
$\eta$ that mix with it. These two subsets of fields do not mix so that we can consider them separately.

\subsection{The graviphoton sector}
We consider first the sector consisting of the graviphoton $R_M$, together
with the two scalars that mix with it, that is $\alpha$, the phase
associated to the scalar $\eta$, and $C_0$, the imaginary partner of the dilaton
$\phi$. This system is very similar to the one of the toy model just
considered, and it differs just by the additional complication of the
presence of $C_0$, and by the fact that the couplings to $\eta$ and
$\phi$ have a less trivial functional dependence.

The action before any gauge choice is
\begin{align}
\mathcal{S}_{\text{graviphoton sect.}}=\int d^5x
  \sqrt{G}&\left[
\frac{1}{4}\mathcal{R}^{MN}\mathcal{R}_{MN}+\frac{3}{2}\sinh^22\eta R^{M}R_{M}+\frac{1}{4}\sinh^22\eta\partial^{M}\alpha
\partial_{M}\alpha\right. \notag \\ 
&\ \ +
\frac{1}{4}e^{2\phi}\cosh^4\eta\partial_{M}C_0\partial^{M}C_0
 +\frac{1}{4} e^{\phi}\sinh^22\eta\partial_{M}C_0\partial^{M}\alpha
 \notag \\
&\ \ \left.-\frac{\sqrt{6}}{4}e^{\phi}\sinh^2 2\eta\partial^{M}C_0 R_{M}-\frac{\sqrt{6}}{2}\sinh^22\eta\partial^{M}\alpha R_{M}
\right]\ . \label{Srmusect}
\end{align}
The metric and the fields $\eta$ and $\phi$ take their background
values in the action above, i.e.~they do not fluctuate.

Let us first write the equations of motion derived from
\eqref{Srmusect}, where for simplicity we already impose the axial
gauge $R_z=0$, and we have decomposed the tangential part of the
graviphoton into a transverse and a longitudinal piece,
$R_\mu=R_\mu^t+\partial_\mu r$:
\begin{subequations}
\begin{align}
&\frac{z^3}{F}\partial_z\left(\frac{F}{z}{R_{\mu}^t}'\right)+\frac{z^2}{F}\Box{R_{\mu}^t}-3\sinh^2 2\eta {R_{\mu}^t}=0\ , \\
&\frac{z^3}{F}\partial_z\left(\frac{F}{z} r'\right)+\sqrt{\frac{3}{2}}\sinh^22\eta\left(\alpha+\frac{1}{2}e^{\phi}C_0-\sqrt{6}r\right)=0\ ,\\
& \frac{z^2}{F}\Box r'-\sqrt{\frac{3}{2}}\sinh^22\eta\left(\alpha'+\frac{1}{2}e^{\phi} C_{0}'\right)=0 \label{constraint2}\ ,\\
&\frac{z^3}{F}\partial_{z}\left(\frac{F^2}{z^3}e^{2\phi}\cosh^2\eta
  C_0'\right)+e^{2\phi}\cosh^2\eta \Box C_0
+\frac{1}{2}Fe^{\phi}\phi'\sinh^22\eta\left(\alpha'+\frac{1}{2}e^{\phi} C_{0}'\right)=0\ .
\end{align}
\end{subequations}
We have not written a fourth equation in the longitudinal sector,
which can be shown to be a consequence of the three others.

Using the equations of motion, the bulk action \eqref{Srmusect} can be
rewritten as a boundary term:
\begin{align}
\mathcal{S}_\mathrm{reg}=-\int_{z=\epsilon} d^4x\left[\frac{F}{2z}R^t_\mu
  {R^t_{\mu}}'\right.&
+\frac{F^2}{4z^3}\sinh^22\eta(\alpha+\frac{1}{2}e^\phi C_0-\sqrt{6}r)(\alpha'+\frac{1}{2}e^{\phi}C_0')
 \notag\\
&\qquad\qquad\qquad\left.+\frac{F^2}{4z^3}e^{2\phi}\cosh^2\eta C_0C_0' \right]\ .
\label{Sregvector}
\end{align}
Note that since we are at the boundary, we should keep in mind that both
background fields $\eta$ and $\phi$ are small there,
$\eta \sim z$ at most and $\phi \sim z^4$ since we set its constant
value to zero.

We now consider in turn the cases of   explicit and spontaneous breaking
of $R$-symmetry.

\subsubsection{Explicit breaking of R symmetry: $\eta_0\neq0$}
In the case of explicit R symmetry breaking, the near boundary expansion at $z=0$ becomes, after solving the equations of motion,
\begin{subequations}
\begin{align}
&R_{\mu}^t= R_{0 \mu}^t+\frac{1}{2}(-\Box+12\eta_0^2)R_{0 \mu}^t z^2\log z +\tilde{R}_{2 \mu}^tz^2+\mathcal{O}(z^4)\ , \\
&C_0=c_0+\frac{1}{4}\Box c_0z^2-\left(\frac{\Box^2}{16}+\frac{\eta_0^2\Box}{6}\right)c_0z^4\log z +\tilde{c}_4z^4+\mathcal{O}(z^6)\ ,\\
&r=r_0-\sqrt{6}\eta_0^2\left(\alpha_0+\frac{1}{2}c_0-\sqrt{6}r_0\right)z^2\log z
+2 \sqrt{6}{\eta}_{0}^2\left(\frac{1}{8}c_0+\frac{1}{\Box}\tilde{\alpha}_2\right)z^{2}+\mathcal{O}(z^4)\ ,\\
&\alpha=\alpha_0-\frac{\Box}{2}\left({\alpha_0}+\frac{1}{2}c_0-\sqrt{6}r_0\right)z^2\log z+\tilde{\alpha}_2z^{2}+\mathcal{O}(z^4)\ ,
\end{align}
\end{subequations}

Note that here we have expressed the subleading term $\tilde{r}_2$ of the longitudinal vector fluctuation $r$ in terms of the subleading term $\tilde{\alpha}_2$ of $\alpha$. Equivalently one can do the other way around. Obviously this feature comes from a Ward identity which relates the corresponding operators, as it will become clear when we will write the renormalized action. 

The boundary action \eqref{Sregvector} has divergent terms  that need to be cancelled by the following counter terms
\begin{align}
\mathcal{S}_\mathrm{ct}=\int_{z=\epsilon} d^4x&\sqrt{\gamma}\left\{\frac{1}{4}\log z\ {\cal R}^{\mu\nu}{\cal R}_{\mu\nu}
+6\log z\ \eta^2 R^\mu R_\mu +
\frac{1}{8}C_0\Box C_0\right.\notag\\
&\qquad-\log z\ C_0\left(\frac{\Box^2}{16}+\frac{5\eta^2\Box}{12}\right)C_0+\log z\ \sqrt{6}\eta^2\partial^{\mu}R_{\mu}C_0\notag\\
&\qquad\left.-\log z\ \eta^2\left(\alpha\Box\alpha+\alpha\Box C_0-2\sqrt{6}\partial^{\mu}R_{\mu}\alpha\right)\right\}\ .
\end{align}
where as usual the indices are now contracted with $\gamma_{\mu\nu}\equiv G_{\mu\nu}|_{z=\epsilon}$. 

The final renormalized action is then  
\begin{align}
\mathcal{S}_\mathrm{ren}=\int d^4k&\left\{-R_{0\mu}^t\tilde{R}_{2\mu}^t-\frac{1}{4}R_{0\mu}^tk^2R_{0 \mu}^t-3\eta_0^2 R_{0\mu}^tR_{0 \mu}^t \right.\notag\\
&\quad-c_0\tilde{c}_4+\frac{3}{64}c_0k^4c_0+\frac{1}{24}\eta_0^2c_0k^2c_0-2\eta_0^2\left(\alpha_0+\frac{1}{2}c_0-\sqrt{6}r_0\right)\tilde{\alpha}_2 \notag\\
&\quad\left. -\frac{\eta_0^2}{2}\left(\alpha_0-\sqrt{6}r_0\right)k^2\left(\alpha_0+\frac{1}{2}c_0-\sqrt{6}r_0\right)\right\}  \ .
\end{align}
We observe that the renormalized action only depends on the gauge invariant combination $\alpha_0-\sqrt{6}r_0$. Indeed, since  the bulk boundary conditions have to preserve the same gauge invariance, also $\tilde\alpha_2$ and, because of the mixing, $\tilde c_4$, have to depend on the same combination.

Finite counter terms can modifiy the coefficient of the terms above which do not involve tilded coefficients. They can be chosen according to the preferred renormalization scheme, as exemplified in the toy model. Let us start from the longitudinal sector. 

As noted above, the Ward identity for broken R symmetry is implemented by the relation
\begin{equation}
\frac{\delta \mathcal{S}_\mathrm{ren}}{\delta r_0}=-\sqrt{6}\frac{\delta \mathcal{S}_\mathrm{ren}}{\delta \alpha_0}\ .
\end{equation}
$\alpha_0$ is the source of the imaginary part of the operator whose non zero coupling breaks R symmetry explicitly, while  $c_0$ sources the imaginary part of the operator whose real part is sourced by the dilaton. In a SYM like theory, that would be proportional to $\mathrm{tr} F^{\mu\nu}\tilde F_{\mu\nu}$. We will not be interested in the details of the correlators of such operators, besides  checking that there are no tachyonic resonances in those channels. For that, it is enough to investigate the pole structure of the functions of $k^2$ that one obtains taking the variation of $\tilde c_4$ and $\tilde \alpha_2$ with respect to  $c_0$ and $\alpha_0$. 

The only form factor that we need in the longitudinal sector is $F_1$ as defined in \eqref{<jj>}:\begin{equation}
\langle k^\mu j_\mu^R(k)\,k^\nu j_\nu^R(-k)\rangle = 
-\frac{2}{3}\frac{\delta^2  \mathcal{S}_\mathrm{ren}}{\delta r_0^2}=
- \frac{1}{3}k^2\eta_0^2\,F_1(k^2)\ .
\end{equation}
From the expression for $\mathcal{S}_\mathrm{ren}$ we gather
\begin{equation}
F_1(k^2)=\frac{8\sqrt{6}}{k^2}\frac{\delta \tilde\alpha_2}{\delta r_0} + \mathrm{finite\ terms}\ .
\end{equation}
The finite terms can be used to cancel the constant part at zero momentum, i.e.~$F_1(0)$. Note that for consistency the variation of $\tilde \alpha_2$ with respect to $r_0$ must be such that the $k^2$ at the denominator cancels, since we do not expect massless poles in the longitudinal channel when R symmetry is explicitly broken.

In the transverse sector one computes the correlator
\begin{align}
&\langle j_{\mu}^{Rt}(k)j_{\nu}^{Rt}(-k)\rangle=-\frac{2}{3}\frac{\delta^2 S_{ren}}{\delta R_{0\mu}^tR_{0\nu}^t}=-P_{\mu\nu}C_{1R}(k^2)\ .
\end{align}
This is similar to what was done for spontaneously broken R symmetry.  A possible massless pole could arise but it can be cancelled by the same finite counter term that cancels the constant piece of $F_1$. Note that this cancellation is not necessary but is a choice of scheme. Indeed in the complete current correlator such a massless pole that can be subtracted by finite counter terms, automatically cancels.

\subsubsection{Spontaneous breaking of R symmetry: $\eta_0=0$}
As in the toy model, the near boundary expansion is standard for the
fields $R_\mu^t$, $r$ and $C_0$, but contains a leading term for $\alpha$ that
goes like $z^{-2}$. Solving for the equations of motion, one obtains:
\begin{subequations}
\begin{align}
&R_{\mu}^t= R_{0 \mu}^t-\frac{1}{2}\Box R_{0 \mu}^tz^2\log z+\tilde{R}_{2 \mu}^tz^2+\mathcal{O}(z^4)\ , \\
&C_0=c_0+\frac{1}{4}\Box c_0z^2-\frac{1}{16}\Box^2 c_0z^4\log z+\tilde{c}_4z^4+\mathcal{O}(z^6)\ ,\\
&r=r_0-2\sqrt{6}\tilde \eta_2^2\frac{1}{\Box}\alpha_0 z^2+\mathcal{O}(z^4)\ ,\\
&\alpha=\alpha_0\frac{1}{z^2}-\frac{1}{2}\Box \alpha_0 \log z + \tilde
  \alpha_2+\mathcal{O}(z^2)\ .
\end{align}
\end{subequations}
The counter term action is
\begin{equation}
{\cal S}_\mathrm{ct}= \int_{z=\epsilon} d^4x\sqrt{\gamma} \left\{\frac{1}{4}\log z
{\cal R}^{\mu\nu}{\cal R}_{\mu\nu}+\frac{1}{8}C_0\Box
C_0-\frac{1}{16}\log z C_0\Box^2 C_0 -2\eta^2 \alpha^2-\eta^2 \log
z\alpha\Box \alpha\right\},
\end{equation}
so that the resulting renormalised action is then 
\begin{align}
\mathcal{S}_{\text{ren}}=\int d^4k&\left\{-R_{0\mu}^t\tilde{R}_{2
  \mu}^t-\frac{1}{4}R_{0\mu}^tk^2R_{0\mu}^t-c_0\tilde{c}_4+\frac{3}{64}c_0k^4c_0\right.
\notag \\
& \qquad\left.+2\tilde\eta_2^2\alpha_0\left(-\tilde\alpha_2+\frac{1}{2}c_0-\sqrt{6}r_0\right)+\frac{1}{2}\tilde\eta_2^2\alpha_0\Box
\alpha_0\right\}\ .
\end{align} 
Similarly as what we remarked in the toy model, the renormalized
action is not invariant under the gauge symmetry that simultaneously
shifts $\tilde \alpha_2$ and $r_0$. This is related to the Schwinger
term that we expect in this situation with spontaneously broken
R symmetry. Note that because $C_0$ mixes non trivially with the axion
$\alpha$, there is also a constant Schwinger term in the correlator
between the operators sourced by $\alpha_0$ and $c_0$.
 
Despite the complication due to the presence of the scalar $C_0$, we recognize in $\mathcal{S}_{\text{ren}}$ above the same feature as in the toy model discussed previously. The correlator of the transverse part of the current will  contain a non-local part stemming from the variation of $\tilde{R}_{2  \mu}^t$ with respect to $R_{0\mu}^t$, plus local terms specified by the renormalization scheme:
\begin{align}
&\langle j_{\mu}^{Rt}(k)j_{\nu}^{Rt}(-k)\rangle=-\frac{2}{3}\frac{\delta^2 S_\mathrm{ren}}{\delta R_{0\mu}^tR_{0\nu}^t}=-P_{\mu\nu}C_{1R}(k^2)\ ,
\end{align}
referring again to the parametrization of \eqref{<jj>}.
We should recover here a massless pole corresponding to the R axion, the Goldstone boson associated to the broken R symmetry as shown in Figure \ref{spontaneous_C1R_C_2}. This comes entirely from the profile of the fluctuations in the bulk.  

As expected, $r_0$ only appears in the terms responsible for the Schwinger term, accordingly with the fact that there are no other non trivial correlators involving the longitudinal part of the R current. In particular, this means that $F_1(k^2)=0$. The correlators of the operators sourced by $\alpha_0$ and $c_0$ can be obtained in the usual way, but we will not need them explicitely. Taking the variation of $\tilde c_4$ and $\tilde \alpha_2$ with respect to $c_0$ and $\alpha_0$ is enough to extract from the non local part of such correlators the spectrum of poles in those channels. This allows us to rule out the presence of tachyons.

\subsection{The graviton sector}
We now finally consider the sector composed of the gravition $h_{MN}$ and the two ``active" scalars $\eta$ and $\phi$. 

The relevant bulk action is 
\begin{align}
\mathcal{S}_\mathrm{graviton\ sect.}=\int d^5x \sqrt{G}&\left[-\frac{R}{2}+\partial^{M}\eta\partial_{M}\eta+\frac{1}{4}\cosh^2\eta \partial^{M}\phi\partial_{N}\phi\right.\notag \\
&\quad\left. +\frac{3}{4}\left(\cosh^22\eta-4\cosh2\eta-5\right)\right]\ ,
\end{align}
to which we have to  add the Gibbons-Hawking boundary term  in order to have a well defined variational principle for gravity
\begin{equation}
\mathcal{S}_\mathrm{G.H.}=-\int d^4x\sqrt{\gamma}\mathcal{K}\ ,
\end{equation}
where $\mathcal{K}$ is the extrinsic curvature at the boundary. 

If we parametrize (and gauge fix) the metric as 
\be
ds^2= \frac{dz^2}{z^2} + G_{\mu\nu}(z,x)dx^\mu dx^\nu\ , 
\ee
then we have the explicit expression
\be
\mathcal{K}=-\frac{z}{2}G^{\mu\nu}{G}_{\mu\nu}' \ .
\ee

We now consider fluctuations of the metric and the scalars. 
The ansatz for the metric fluctuations is 
\be
G_{\mu\nu}=\frac{F(z)}{z^2}\big(\eta_{\mu\nu}+h_{\mu\nu}(z,x)\big)\ ,
\ee
where $F(z)$ is the background profile, while $h_{\mu\nu}(z,x)$ is taken to be small.
Similarly, the scalars also split into background plus small perturbation:
\be
\eta=\eta(z)+n(z,x)\ , \qquad\qquad \phi=\phi(z)+\varphi(z,x)\ . 
\ee
The functions $F(z)$, $\eta(z)$ and $\phi(z)$ satisfy the background equations of motion \eqref{Fdprime}--\eqref{eqphi} and we choose to fluctuate around these values.

We now write the metric fluctuations separating the transverse traceless part $h_{\mu\nu}^{tt}$ from the traceful parts $h,H$: 
\begin{equation}
h_{\mu\nu}=h_{\mu\nu}^{tt}+\eta_{\mu\nu}h+\frac{\partial_{\mu}\partial_{\nu}}{\Box}H\ ,  \label{hdecomp}
\end{equation}
gauge fixing to zero the vectorial part of the metric.  
Having already fixed the gauge where $G_{zz}=z^{-2}$ and $G_{\mu z}=0$, the corresponding Einstein equations will be imposed as constraints on the equations of motion. The complete set of Einstein equations reads:
\begin{subequations}
\begin{align}
&\frac{z^{5}}{F^2}\partial_z\left(\frac{F^2}{z^3}{h^{tt}_{\mu\nu}}'\right)+\frac{z^2}{F}\Box h_{\mu\nu}^{tt}=0\ ,\\
&\frac{z^{5}}{F^2}\partial_z\left(\frac{F^2}{z^3}H'\right)+2\frac{z^2}{F}\Box h=0\ ,\\
&\frac{z^{5}}{F^2}\partial_z\left(\frac{F^2}{z^3}h'\right)+\frac{4z^2}{3}\eta'n'+\frac{z^2}{3}\cosh^2\eta\phi'\varphi'+\frac{z^2}{6}\sinh2\eta\phi'^2 n+(\sinh4\eta-4\sinh2\eta)n=0\ ,\\
&\frac{z^{9}}{F^4}\partial_z\left(\frac{F^4}{z^7}h'\right)-\frac{1}{2}\frac{z^{3}}{F}\partial_z\left(\frac{F}{z}H'\right)+2\left(\sinh4\eta-4\sinh2\eta\right)n=0\ ,\\
&\partial_{\mu}h'+\frac{4}{3}\eta'\partial_{\mu}n+\frac{1}{3}\cosh^2\eta\phi'\partial_{\mu}\varphi=0\ .
\end{align}
\end{subequations}
To complete the system we  add the two scalar fluctuations:
\begin{subequations}
\begin{align}
&\frac{z^{5}}{F^2}\partial_z\left(\frac{F^2}{z^3}n'\right)+\frac{z^2}{F}\Box n+\frac{1}{2}z^2\eta'(4h'+H')-\frac{1}{4}z^2\sinh2\eta\ \phi'\varphi'\notag\\
&\qquad\qquad\qquad\qquad
-\frac{1}{4}z^2\cosh2\eta\ {\phi'}^2 n -3(\cosh 4\eta -2 \cosh 2\eta)n=0\ .\\
&\frac{z^{5}}{F^2}\partial_z\left(\frac{F^2}{z^3}\varphi'\right)+\frac{z^2}{F}\Box \varphi+\frac{1}{2}z^2\phi'(4h'+H')+2z^2\tanh\eta\ \eta'\varphi'\notag\\
&\qquad\qquad\qquad\qquad+2z^2\tanh\eta\ {\phi}'{n}'+2z^2(1-\tanh^2\eta){\eta}'{\phi}'n=0\ .
\end{align}
\end{subequations}

The resulting boundary action at the regularizing surface reads 
\begin{align}
\mathcal{S}_\mathrm{reg}&=-\int_{z=\epsilon} d^4x \frac{F^2}{z^3}\left[
\left(\frac{3}{z}-\frac{3 F'}{2F}\right)\left(1+2h+\frac{1}{2}H-\frac{1}{4}h_{\mu\nu}^{tt}h_{\mu\nu}^{tt}
+h^2+\frac{1}{2}hH-\frac{1}{8}H^2\right)\right.\notag\\
&\qquad\quad+2{\eta}'n+\frac{1}{2}\cosh^2\eta\ {\phi}'\varphi
+\left({\eta}'n +\frac{1}{4}\cosh^2\eta{\phi}'\varphi\right)\left(2h+\frac{1}{2}H\right) 
\notag \\
&\qquad\quad \left.+\frac{1}{8}h_{\mu\nu}^{tt}{h_{\mu\nu}^{tt}}'-\frac{3}{2}hh'-\frac{3}{8}(hH'+Hh')+n{n}'+\frac{1}{4}\cosh^2\eta\varphi{\varphi}'+\frac{1}{4}\sinh 2\eta {\phi}'\varphi n\right]\ .
\label{sreggrav}
\end{align}
There are terms which are of zeroth, first and second order in the fluctuating fields $h_{\mu\nu}$, $n$ and $\varphi$.  Note in particular that recalling the definition $\gamma_{\mu\nu}\equiv G_{\mu\nu}|_{z=\epsilon}=\frac{F}{z^2}(\eta_{\mu\nu}+h_{\mu\nu})$, we have that
\begin{equation}
\sqrt{\gamma}=\frac{F^2}{z^4}\left(1+2h+\frac{1}{2}H-\frac{1}{4}h_{\mu\nu}^{tt}h_{\mu\nu}^{tt}
+h^2+\frac{1}{2}hH-\frac{1}{8}H^2+\dots\right)\ ,
\end{equation}
an expression that we recognize in the first line of \eqref{sreggrav}.

Let us first derive the near boundary expansion for the different fluctuations. Consistently with the discussion of the graviphoton sector we analyze the case with or without $\eta_0$ separately. 


\subsubsection{Explicit breaking of conformal symmetry: $\eta_0\neq0$}
 Expanding the fluctuations (and the background, using \eqref{backrel}) near the boundary we get
\begin{subequations}
 \begin{align}
&h_{\mu\nu}^{tt}=h_{0 \mu\nu}^{tt}+\frac{\Box}{4}h_{0
     \mu\nu}^{tt}z^2-\left(\frac{\Box^2}{16}-\frac{\eta_{0}^2\Box}{12}\right)h_{0
     \mu\nu}^{tt}z^4\log z+\tilde{h}_{4 \mu\nu}^{tt}z^{4}+\mathcal{O}(z^6)\ ,\\
&h=h_{0}-\frac{2}{3}\eta_{0}n_{0}z^2+\eta_0 (\frac{\Box}{6}n_0+\frac{\eta_0\Box}{12}h_0-\frac{16\eta_0^2}{3}n_0)z^4\log z\notag\\
&\qquad+(-\tilde{\eta}_{2}n_0-\frac{1}{3}\tilde{\phi}_4\varphi_0-\frac{\eta_0\tilde{n}_2}{3}-\frac{\eta_0\Box}{24}n_0-\frac{\eta_0^2\Box}{48}h_0+\frac{4\eta_0^3}{9}n_0)z^4+\mathcal{O}(z^6)\ ,\\
&H=H_{0}+\frac{\Box}{2}h_0z^2+\eta_0 (\frac{\Box}{3}n_0+\frac{\eta_0\Box}{6}h_0)z^4\log z\notag\\
&\qquad+(2\tilde{\eta}_{2}n_{0}+\frac{4}{3}\tilde{\phi}_4\varphi_0-\frac{2}{3}\eta_0\tilde{n}_2+\frac{\eta_0^2\Box}{24}h_0-\frac{\eta_0\Box}{12}
   n_0+\frac{16}{9}\eta_0^3 n_0)z^4+\mathcal{O}(z^6)\ ,\\
&n=
   n_0z-(\frac{\Box}{2}n_0-8\eta_0^2n_0+\frac{\eta_0\Box}{4}h_0)z^3\log z+
   \tilde{n}_2z^3 +\mathcal{O}(z^5)\ ,\\
&\varphi=\varphi_0+\frac{\Box}{4}\varphi_0z^2-(\frac{\Box^2}{16}\varphi_0
+\frac{\eta_0^2\Box}{6}\varphi_0)z^4\log z+\tilde{\varphi}_4z^4+\mathcal{O}(z^6)\ .
\end{align}
\end{subequations}
 Note that the traceful components of the graviton have  a
 dependence from the mode $\tilde n_2$.

We are now ready to renormalize the action.  We have the following counter term action
\begin{align}
\mathcal{S}_\mathrm{ct}=\int_{z=\epsilon}
d^4x\sqrt{\gamma}&\left[3+\eta^2+\frac{8}{3}\eta^4\log z+2\eta n+\log z\ \frac{32}{3}\eta^3n\right.\notag \\
&+\frac{1}{4}R+n^2+\frac{1}{8}\varphi\Box\varphi+\log z\left(-\frac{1}{8}(R^{\mu\nu}R_{\mu\nu}-\frac{1}{3}R^2)-n\Box n\right. \notag \\
&\left. \left.-\frac{1}{16}\varphi\Box^2 \varphi+\frac{1}{6}\eta^2R+\frac{1}{3}\eta n R+16\eta^2n^2-\frac{1}{6}\eta^2\varphi\Box\varphi\right)\right] \ .
\end{align}
As for the regularized action, also the counter term action contains parts of zeroth, first and second order in the fluctuating fields. In particular, notice that $\sqrt{\gamma}$ has terms of every order.

After the renormalization procedure we are left with a finite
action 
\begin{align}
 \mathcal{S}_\mathrm{ren}=\int d^4 k&\left[-\left( \eta_0\tilde{\eta}_2+\frac{2}{3}\eta_0^4\right)
 \left(1+2h_0+\frac{H_0}{2}  -\frac{1}{4}h_{0 \mu\nu}^{tt}h_{0 \mu\nu}^{tt}  +h_0^2+\frac{h_0H_0}{2}-\frac{H_0^2}{8}  \right)\right. \notag\\
 &-4\tilde{\eta}_2n_0-2\tilde{\phi}_4\varphi_0-\frac{16}{3}\eta_0^3n_0-\frac{1}{2}h_{0 \mu\nu}^{tt}\tilde{h}_{4\mu\nu }^{tt}
 \notag\\
& -(2n_{0}+\eta_{0}h_{0})\tilde{n}_2-\varphi_0\tilde{\varphi}_4-\tilde{\phi}_4\varphi_0(2h_0+H_0)\notag\\
  &\left.-(5\tilde{\eta}_2+\frac{1}{2}\eta_0k^2+8\eta_0^3)n_0h_0-(2\tilde{\eta}_2+\frac{8}{3}\eta_0^3)n_0H_0\right.\notag\\
&\left. +\frac{3}{128}\left(h_{0 \mu\nu}^{tt}k^4h_{0 \mu\nu}^{tt}+2\varphi_0k^4\varphi_0\right)-\left(\frac{k^2}{2}+8\eta_0^2\right)n_0^2 \right.\notag\\
&\left.+\frac{\eta_0^2}{96}\left(-h_{0 \mu\nu}^{tt}k^2h_{0 \mu\nu}^{tt}+4\varphi_0k^2\varphi_0\right)\right]\ .
\end{align}
The expression above is particularly cluttered because in the most generic background the conformal symmetry is broken both by sources ($\eta_0$) and VEVs ($\tilde \eta_2$ and $\tilde \phi_4$), and also by the regularization procedure (the $z=\epsilon$ surface). Many terms however, in particular the last two lines, can be completely removed by finite counter terms.

At zeroth order the renormalized action gives the free
energy of the field theory in the specified vacuum.  This piece of the action is proportional to $\eta_0$ which is the parameter explicitly breaking supersymmetry. However, using the local finite counter term proportional to $\sqrt{\gamma}\eta^4$ the free energy can be set to zero as a scheme choice.

From the terms which are of first
order in the renormalized action we obtain all the one-point functions. 
We see that the VEVs of the operators sourced by $n_0$ and $\varphi_0$ are, respectively, given by $\langle O_\eta\rangle=-4\tilde \eta_2$ and $\langle O_\phi\rangle=-2\tilde \phi_4$. 
Recall that $n_0$ sources an operator of
dimension 3, which can be taken to be the real part of a gaugino bilinear
$O_\eta \propto \mathrm{tr}\lambda\lambda$ in a SYM like theory, while $\varphi_0$
sources a dimension 4 operator like $O_\phi\propto\mathrm{tr}F_{\mu\nu}F^{\mu\nu}$.

We also get a non zero one point function for $T$, proportional to the same combination of $\eta_0$ and $\tilde
\eta_2$ that gives the value of the free energy. It can be shifted to zero by the same scheme choice,
consistently with the fact that its value is
arbitrary since SUSY is broken explicitly. 

The most interesting terms are the ones involving both the scalar and traceful metric perturbations. It is instructive to first recall how the various modes transform under dilatations:
\begin{equation}
\delta h_0 = -2\sigma , \quad \delta H_0 =0, \quad \delta n_0 = \eta_0 \sigma, \quad \delta \tilde n_2=3\tilde \eta_2 \sigma, \quad \delta\varphi_0=0, \quad \delta \tilde\varphi_4 = 4\phi_4\sigma\ . 
\label{weyltf}
\end{equation}
The relevant gauge invariant combinations are thus $2n_0+\eta_0 h_0$, $\varphi_0$, $2\tilde n_2 +3\tilde\eta_2 h_0$ and $\tilde\varphi_4+2\tilde\phi_4 h_0$. In the explicitly broken case, we must look for the holographic realization of the Ward identities. They can be seen most neatly in the term proportional
to $\tilde n_2$, which is scheme independent. It is multiplied by the gauge invariant combination $2n_0+\eta_0h_0$, signaling that
at the operator level we have $T = \eta_0 O_\eta$. On the other hand, since $\varphi_0$ is the
coefficient of a marginal operator, the Ward identity associated to it is trivial. Hence
only $\varphi_0$ is the coefficient of $\tilde\varphi_4$. We will come back to the other terms when we discuss the spontaneously broken case.

Regarding the correlators, 
the form factors $C_2$ and $F_2$ are given by second order
variations of $\mathcal{S}_\mathrm{ren}$ with respect to $h_{0
  \mu\nu}^{tt}$ and $h_0$ respectively. 
  In particular, for the transverse part we have
  \begin{equation}
\langle
T_{\mu\nu}^{tt}(k)T_{\rho\sigma}^{tt}(-k)\rangle=-4\frac{\delta^2
  S_{ren}}{\delta
  h_{0\mu\nu}^{tt}h_{0\rho\sigma}^{tt}}=-\frac{1}{8}X_{\mu\nu\rho\sigma}C_{2}(k^2)\ ,\label{ttcorr}
\end{equation}
where $C_2$ is defined in \eqref{<TT>}, while for the trace part we have
\begin{equation}
\langle T(k) T(-k)\rangle =-4\frac{\delta^2 S_{ren}}{\delta h_{0}h_{0}}=-\frac{3}{4}\eta_0^2k^2F_2(k^2)\ .
\end{equation}

  Since $\eta_0\neq 0$, as we have already discussed, there
are many finite local counter terms that one can add, that affect
these form factors. In particular, spurious poles at $k^2=0$ can be
cancelled in this way, if one makes such a scheme choice, which must always be possible since no massless mode is
expected in the $\langle T_{\mu\nu}T_{\rho\sigma}\rangle$ correlator
in this situation.

Finally, the variations of $\tilde n_2$ and $\tilde \varphi_4$ with
respect to $n_0$ and $\varphi_0$ will inform us of the spectrum of
resonances in the
sector of the scalar operators, excluding the presence of tachyons.


\subsubsection{Spontaneous breaking of conformal symmetry: $\eta_0=0$}
The discussion of the spontaneously broken case in the gravitational sector is very straightforward.  It suffices to set $\eta_0=0$ in all the expressions obtained in the previous section. Indeed, contrary to the graviphoton sector, the expansions of the perturbations do not depend on the presence or not of $\eta_0$. 

Accordingly, the expressions simplify considerably. 
The fluctuations near the boundary are 
\begin{subequations}
\begin{align}
&h_{\mu\nu}^{tt}=h_{0 \mu\nu}^{tt}+\frac{\Box}{4}h_{0 \mu\nu}^{tt}z^2-\frac{\Box^2}{16}h_{0 \mu\nu}^{tt}z^4\log z+\tilde{h}_{4 \mu\nu}^{tt}z^{4}+\mathcal{O}(z^6)\ ,\\
&h=h_{0}-(\tilde{\eta}_{2}n_0+\frac{1}{3}\tilde{\phi}_4\varphi_0)z^4+\mathcal{O}(z^6)\ ,\\
&H=H_{0}+\frac{\Box}{2}h_0z^2+(2\tilde{\eta}_{2}n_{0}+\frac{4}{3}\tilde{\phi}_4\varphi_0)z^4+\mathcal{O}(z^6)\ ,\\
&n= n_0z-\frac{\Box}{2}n_0z^3\log z+ \tilde{n}_2z^3+\mathcal{O}(z^5)\ ,\\
&\varphi=\varphi_0+\frac{\Box}{4}\varphi_0z^2-\frac{\Box^2}{16}\varphi_0z^4\log z+\tilde{\varphi}_4z^4+\mathcal{O}(z^6)\ .
\end{align}
\end{subequations}
Let us again stress
 that this is a nice feature of our study, in which switching off a
 parameter (without changing the SUGRA theory) we can go from the case
 in which the conformal symmetry is explicitly  broken to the case where it is spontaneously  broken.

The resulting renormalised action is
\begin{align}
 \mathcal{S}_\mathrm{ren}=\int d^4 k&\left[-4\tilde{\eta}_2n_0-2\tilde{\phi}_4\varphi_0-\frac{1}{2}h_{0 \mu\nu}^{tt}\tilde{h}_{4\mu\nu }^{tt}-2n_{0}\tilde{n}_2-\varphi_0\tilde{\varphi}_4\right.
 \notag\\
& \left.-\tilde{\eta}_2n_0(5h_0+2H_0)-\tilde{\phi}_4\varphi_0(2h_0+H_0)\right.\notag\\
&\left. +\frac{3}{128}\left(h_{0 \mu\nu}^{tt}k^4h_{0 \mu\nu}^{tt}+2\varphi_0k^4\varphi_0\right)-\frac{k^2}{2}n_0^2 \right]\ .\label{srengravspont}
\end{align}
From the first order part, we
see that the operators sourced by $n_0$ and by $\varphi_0$ have the same VEVs as in the explicitly broken case. 

On the other hand, independently of the scheme, there is no term linear
in $h_0$, and thus no expectation value in the vacuum for $T$.

In the quadratic part of the renormalized action, there are no terms
proportional to $h_0^2$, consistently with the fact that the
stress-energy tensor should be traceless. Accordingly the $F_2$ form
factor trivially vanishes in the spontaneous case. Nevertheless, the
presence of $h_0$ (and $H_0$) in \eqref{srengravspont} is related to the
Schwinger terms that appear in the correlators of $T_{\mu\nu}$ with the
operators that acquire VEVs. Indeed, the terms bilinear in $n_0h_0$
and $\varphi_0h_0$ are proportional to $\tilde \eta_2$ and $\tilde \phi_4$,
respectively.

The correct Schwinger terms are determined recalling the gauge invariance of each mode \eqref{weyltf} in the spontaneous case. In particular, it means that gauge invariant boundary conditions in the bulk imply that the non-local terms will be given by $\tilde n_2 = -\frac{3}{2}\tilde\eta_2 h_0 + f(n_0,\varphi_0)$ and $\tilde\varphi_4=-2\tilde\phi_4h_0+g(n_0,\varphi_0)$. Inserting these values we get the relevant terms
 \begin{align}
 \mathcal{S}_\mathrm{ren}&\supset \int d^4 k\left[-4\tilde{\eta}_2n_0-2\tilde{\phi}_4\varphi_0-\tilde{\eta}_2n_0(2h_0+2H_0)-\tilde{\phi}_4\varphi_0H_0 \right]
 \notag\\
&=\int d^4 k\sqrt{\gamma_0}\left[-4\tilde{\eta}_2n_0-2\tilde{\phi}_4\varphi_0+6\tilde{\eta}_2n_0h_0+4\tilde{\phi}_4\varphi_0h_0 \right]\ .
\end{align}
These are precisely the correct coefficients to obtain the Schwinger terms expected in the boundary field theory, as discussed already in \cite{Bianchi:2001de}.

In order to find the form factor $C_2$, we have to take  the variation
of $\tilde{h}_{4 \mu\nu}^{tt}$ with respect to $h_{0
  \mu\nu}^{tt}$ as in \eqref{ttcorr}. In $C_2$ we should recover a massless pole related to
the presence of the dilaton of spontaneously broken conformal
invariance.

\section{From correlators to the soft spectrum}\label{GGMSpectrum}

In this section we want to make use of the holographic SUSY breaking RG flows studied before as putative hidden sectors in the framework of General Gauge Mediation \cite{Meade:2008wd}. In order to do so, we weakly gauge the unbroken $U(1)_F$ of our strongly coupled RG flows by means of Standard Model gauge degrees of freedom. The unbroken $U(1)_F$ can be then identified with the $U(1)_Y$ of the Standard Model whose coupling constant is $g_{1}$.\footnote{Larger global symmetry group which contain the full Standard Model gauge group can be achieved by considering more complicated scenarios where localised flavor branes are added in the geometry \cite{Benini:2009ff,McGuirk:2011yg}. Note also that in our simple SUGRA model the $U(1)_F$ is actually anomalous. This fact, which is related to cubic couplings in the bulk action, obviously does not affect the two point functions that we thus take as prototypes for data holographically extracted from a well defined strongly coupled theory.} 

Integrating out the hidden sector degrees of freedom one can derive the leading SUSY breaking contributions (in the $g_{1}$ perturbative expansion) to both gaugino and sfermion masses of the visible sector. These can be written as
\begin{subequations}
\begin{align}
&m^{2}_{\tilde{f}}=\frac{2Y_{\tilde{f}}^2g^{4}_1}{16\pi^2}\int dk^2 A(k^2)\ ,\label{ggmsfmass}\\
&m_{\lambda}= g^2_1 B_{1/2}(0)\ ,\label{ggmlmass}
\end{align}
\end{subequations}
where $Y_{\tilde{f}}$ is the hypercharge of the visible sector scalars\footnote{In the following we are going to ignore the $2Y_{\tilde{f}}^2$ factor for simplicity.} while the form factors $A$ and $B_{1/2}$ (defined in \eqref{GGMA} and \eqref{bhalf} respectively) encode the SUSY breaking hidden sector dynamics.
 
Instead of building up a full viable holographic model for gauge mediation, we will be interested in studying the behavior of the ratio between gaugino and sfermion masses in the different corners of the parameter space of the simple $U(1)$ model presented in Sections \ref{sugra} and \ref{allSOL}. Before presenting our results, let us spend a few words explaining the fundamental difference between the Holographic Gauge Mediation setups we present here and more standard gauge mediation models (see \cite{Giudice:1998bp} for a review on the subject). 

In gauge mediation models it is often assumed that strongly coupled effects in the hidden sector can be decoupled from the mediation dynamics. Under this assumption the  hidden sector physics can be described in terms of weakly coupled messenger fields which are charged under the Standard Model gauge group and feel the SUSY breaking via perturbative couplings to the hidden sector. These models have the advantage of being fully calculable by means of perturbative techniques and allow one to study easily how \eqref{ggmsfmass} and \eqref{ggmlmass} behave. 

\begin{figure}[h!]
\centering
\includegraphics[width= 0.5\textwidth]{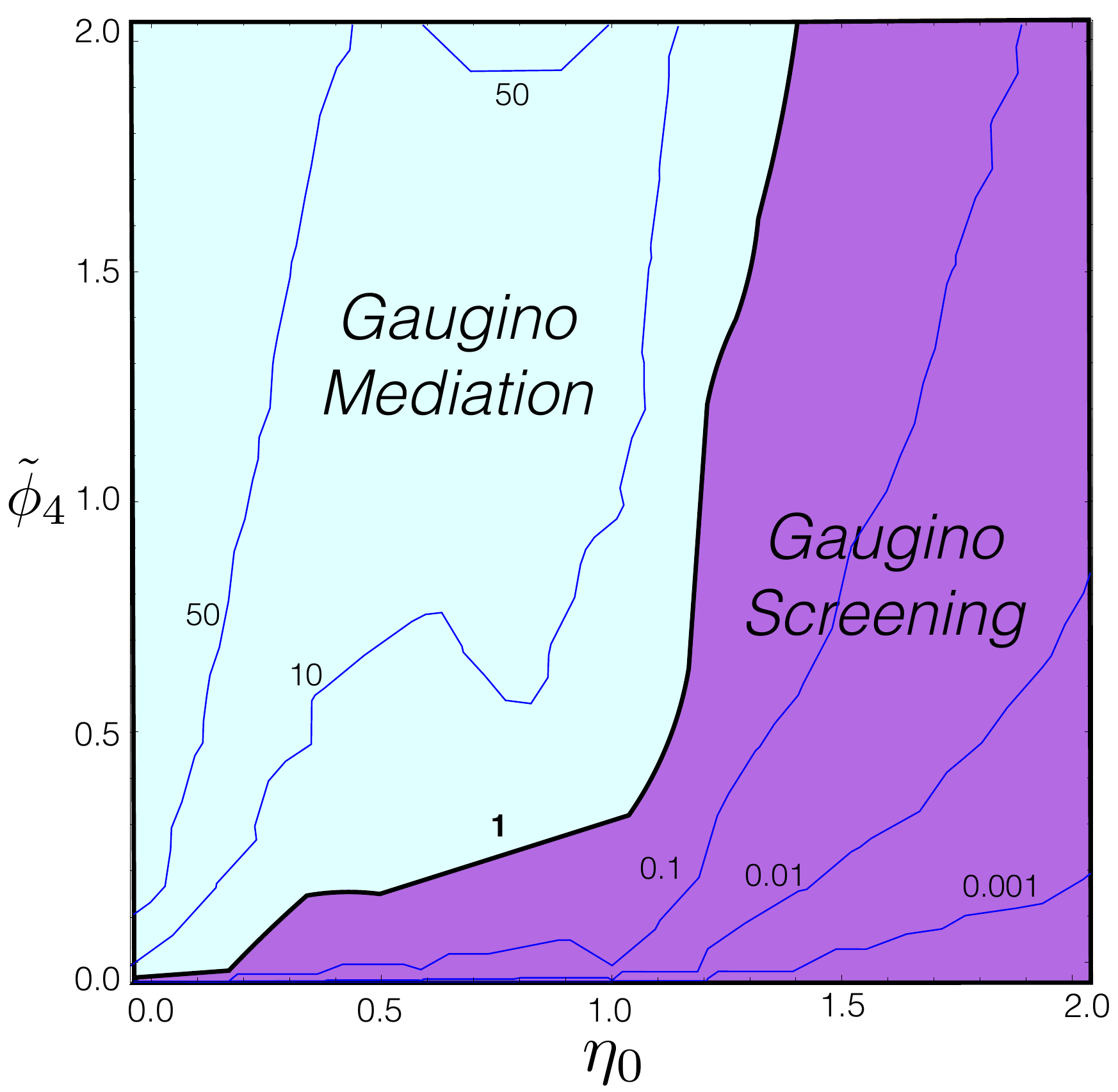}
\caption{Contours of $R$ defined in \eqref{NDAestimate} in the $(\eta_0,\tilde{\phi}_4)$ plane having fixed $\tilde{\eta}_2=1$ and $N=10$. We always consider the generic solution in the plane which corresponds to the ``$\eta$ blowing up" solutions of Section \ref{etablowingup}. In the light blue region (\emph{gaugino mediation} region) of the parameter space  $R>1$ while in the violet region (\emph{gaugino screening} region)  $R<1$.} 
\label{softmasses}
\end{figure}

If the mediation sector does not admit a weakly coupled description, very few predictions can be made about the hidden sector dynamics. A qualitative estimate of the soft spectrum behavior can be performed using Naive Dimensional Analysis (NDA) \cite{Luty:1997fk}. Applying this to holographic setups is relatively easy since the renormalized action comes already in the NDA form
\begin{equation}
\hat{S}_{\text{ren}}=\frac{N^2}{4\pi^2}S_{\text{ren}}\ .
\end{equation}
where the overall factor is fixed by uplifting the $5d$ SUGRA effective action to the full $10d$ type IIB SUGRA in the pure $AdS_{5}$ case \cite{Argurio:2012cd}. Assuming the hidden sector SUSY breaking dynamics to be well described by a single scale $\Lambda$ up to $\mathcal{O}(1)$ coefficients one gets $\int dk^2 A(k^2)\sim\frac{N^2}{4\pi^2}\Lambda$ and $B_{1/2}\sim \frac{N^2}{4\pi^2}\Lambda$ which implies  
\begin{equation}
R\equiv\frac{m_{\lambda}^2}{m^{2}_{\tilde{f}}}\sim 4 N^2\gg1\ .\label{NDAestimate}
\end{equation}
This result suggests that hidden sectors which admit a holographic description (i.e. a large $N$ expansion) deliver generically \emph{gaugino mediation} spectra where the gaugino mass is parametrically larger than the sfermion masses. In what follows, we go beyond the NDA estimate computing exactly the form factors $A$ and $B_{1/2}$ at the leading order in the large $N$ expansion by means of holographic techniques. We find out that the single scale assumption, on which the NDA estimate \eqref{NDAestimate} relies, oversimplified the physics of our model eventually yielding misleading conclusions. 


In our $\mathcal{N}=2$ SUGRA model the hidden sector dynamics is described in terms of the 3 parameters of \eqref{parspace}, which have a clear interpretation in terms of the $AdS$/CFT correspondence (see Section \ref{sugra}). This has to be contrasted with previous holographic embeddings of gauge mediation based on Randall-Sundrum (RS) setups \cite{Argurio:2012bi} (see also \cite{McGarrie:2010yk} for related works) where the features of the hidden sector dynamics were mimicked by suitable boundary conditions on the IR brane. The price to pay is the loss of an analytical handle on the form factors which we partially overcome by a detailed numerical study.

Generic solutions in the $3d$ parameter space \eqref{parspace} corresponds to the backgrounds with $\eta$ blowing up presented in sections \ref{etablowingup} and \ref{etablowingup2}. Evaluating $A$ and $B_{1/2}$ over these backgrounds we find $R\in (0,\infty)$ as it is shown in Figure \ref{softmasses}. It is remarkable that \emph{gaugino screening} scenarios with $R\ll1$ can be naturally realized in contrast with the NDA expectation \eqref{NDAestimate}. The failure of the NDA estimates is due to a non trivial dynamical feature of our holographic RG flows which could not be captured in other ways than by direct computation of the linear multiplet two point function. Increasing the value of $\eta_{0}$ until it dominates over $\tilde{\eta}_2$ we see that the gaugino mass decreases in perfect agreement with the screening mechanism of $B_{1/2}$ presented in Figure \ref{explicit_A_B}. When $\eta_{0}$ is small compared to $\tilde{\eta}_2$ the dynamics becomes more similar to the RG flows illustrated in Figure \ref{spontaneous_C1_C_2} where the SUSY breaking dynamics is triggered mostly by $\tilde{\phi}_4$ which contributes both to $A$ and to $B_{1/2}$ since $R$ symmetry is broken by a SUSY preserving VEV proportional to $\tilde{\eta}_2$. 

We now consider dilaton domain wall solutions. The latter are single scale models were Dirac gauginos are dynamically realized through the mixing of massless fermonic states from the hidden sector and the visible sector gaugino \cite{Argurio:2012cd}. An analogous mechanism to the one we find here has been advocated as a possible way of generating large gauginos masses in models of low energy SUSY breaking with a strongly coupled hidden sector at the TeV scale \cite{Gherghetta:2011na}.\footnote{We thank Mark Goodsell for pointing out this paper to us.}  

The GGM formulas (\ref{ggmsfmass}-\ref{ggmlmass}) should be modified in the presence of IR non decoupling effects by resumming the propagators of the vector multiplet fields as discussed in \cite{Buican:2009vv, Intriligator:2010be}.  After resummation is performed, the GGM formulas are
\begin{subequations}
\begin{align}
&m^{2}_{\tilde{f}}=\frac{g^{2}_1}{16\pi^2}\int d k^2\left(\frac{1}{1+g^2_1C_{0}(k^2)}-\frac{4}{1+g^2_1C_{1/2}(k^2)}+\frac{3}{1+g^2_1C_{1}(k^2)}\right)\ ,\label{fullmass1}\\[7pt]
&-m^2_{\lambda}\left(1+g^2_1 C_{1/2}(-m^2_{\lambda})\right)^2+g^4_1|B_{1/2}(-m^2_{\lambda})|^2=0\ ,\label{fullmass2}
\end{align}
\end{subequations}
where the physical mass of the gaugino is now given by the solution of the algebraic equation \eqref{fullmass2} which corresponds to the pole of the resummed propagator. If $g^2_1C_{1/2}(0)\ll1$ the pole mass is given by the solution of $m^2_{\lambda}=g^4_1|B_{1/2}(-m^2_{\lambda})|^2$ which is well approximated by \eqref{ggmlmass} if $B_{1/2}$ is varying sufficiently slowly as a function of $k^2$.

In Figure \ref{C1_C2_Gubser} we showed that $C_{1/2}$ has a pole at zero momentum in dilaton domain wall solutions. This massless pole is related to 't~Hooft fermions matching in the IR the unbroken anomalous R symmetry. The same unbroken R symmetry is forcing $B_{1/2}$ to vanish. In such a situation the residue of the massless pole in $C_{1/2}$ sets the unique SUSY breaking scale and the gaugino acquires a mass of Dirac nature 
\begin{equation}
m_\lambda^2=\frac{N^2g^2_1}{4\pi^2} f_{\pi1/2}\ .\label{Diracmass}
\end{equation}
where $f_{\pi1/2}$ is the residue of $C_{1/2}$ defined in Section \ref{dilatonblowingup2}. Since the Dirac mass arises as a tree level effect due to the mixing of the gaugino with the 't~Hooft fermions, the dependence of $m_\lambda^2$ from both $g_{1}$ and $N$ drastically changes compared to the one inferred in \eqref{NDAestimate}. 

\begin{figure}[h!]
\centering
\includegraphics[width= 0.6\textwidth]{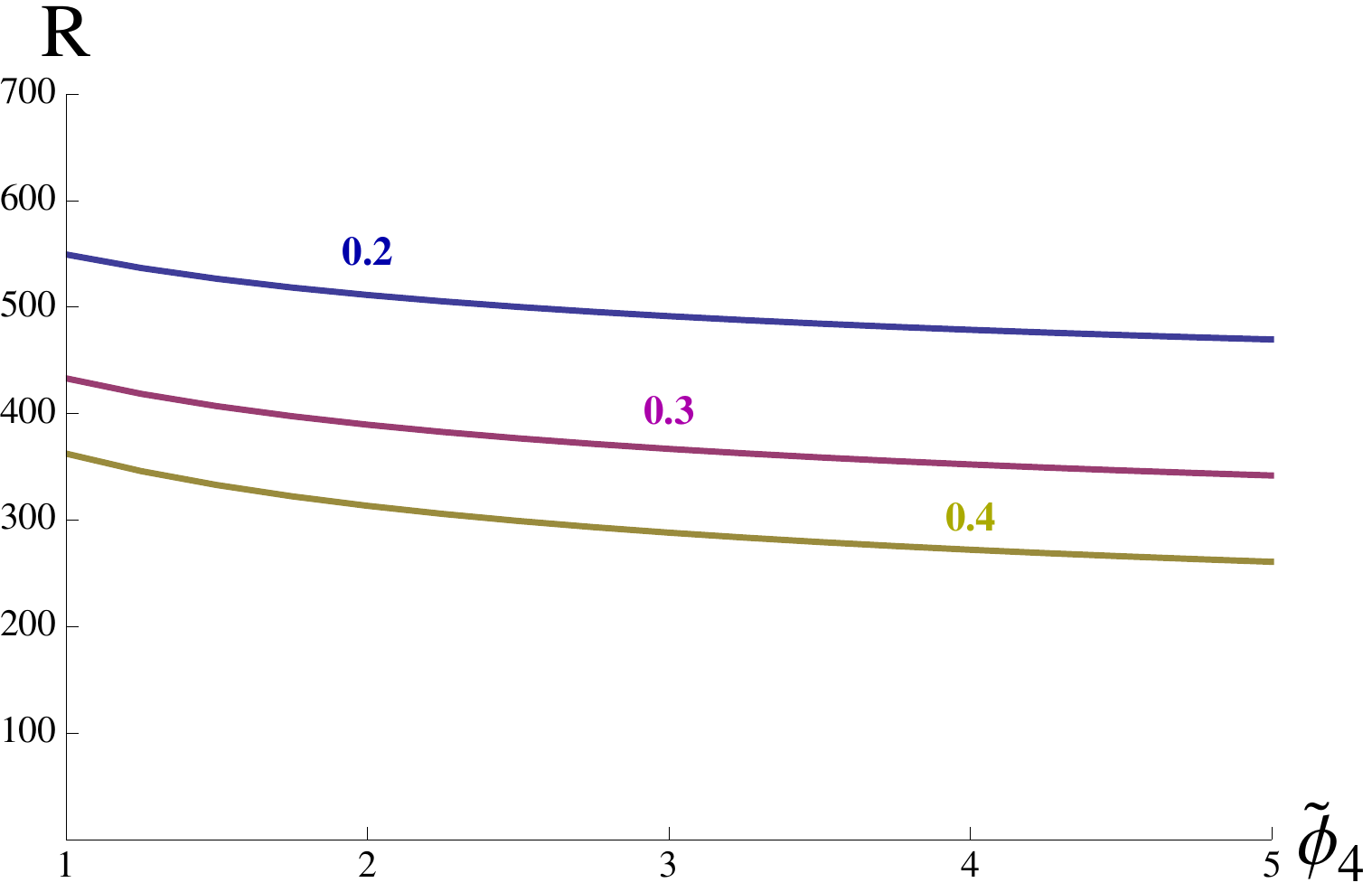}
\caption{$R$ as a function of $\tilde{\phi}_4$ in dilaton domain wall backgrounds for $N=5$. The different lines correspond to different choices of $g_{1}^2$ defined at the decoupling scale of the hidden sector. 
}
\label{softmassesGubby}
\end{figure}

The pole in $C_{1/2}$ affects also the sfermion masses and the full formula \eqref{fullmass1} should be used in order to compute them. The latter makes an NDA estimate of the scalar masses very difficult because $g_{1}$ and $N^2$ cannot be factorized in front of the mass formula. Our numerical result in Figure \ref{softmassesGubby} shows that $R\gg1$ and a \emph{Dirac gaugino mediation} scenario is realized. The hierarchy between gaugino and sfermion masses gets reduced increasing the value of the $g_{1}$ coupling. (Remember that $g_{1}$ is evauated at the scale at  which the hidden sector decouples, and thus varies according to  the MSSM RG equations for the gauge couplings \cite{Martin:1997ns}.)
From Figure \ref{softmassesGubby} we also see that the dependence of $R$ on $\tilde{\phi}_4$ (which controls $f_{\pi1/2}$) is very weak. This can be understood analytically by assuming the sfermion mass integral \eqref{fullmass1} to be dominated by the IR pole of $C_{1/2}$. Neglecting the UV contribution one gets 
\begin{equation}
m_{\tilde{f}}^2\sim \frac{g^2_1m_\lambda^2}{4\pi^2}\log\left(\frac{1}{g^2_1}\right)\ . 
\end{equation} 
Remembering \eqref{Diracmass} we find
\begin{equation}
R\sim\frac{4\pi^2}{g^2_1\log(\frac{1}{g^2_1})}\ .
\end{equation}
which is independent on $\tilde{\phi}_4$ and in qualitative agreement with the results in Figure \ref{softmassesGubby}.

Before leaving this section let us notice that in a more general situation, where $C_{1/2}(0)$ and/or $B_{1/2}(0)$
are very peaked close to the origin, one has also to use the complete expressions
\eqref{fullmass1} and \eqref{fullmass2}.  The gaugino mass will typically have both a Majorana and a
Dirac component. This indeed happens in dilaton like backgrounds where a parametrically small source of R symmetry breaking is switched on in the hidden sector. For example one can estimate from Figure \ref{Bulletto_2} that $\frac{g^2_1N^2}{4\pi^2}C_{1/2}(0)\gtrsim1$  for $N\sim\mathcal{O}(10)$ and $\eta_{w}<0.3$. Note that this situations are unlikely to be realized in weakly coupled models of gauge mediation unless the visible sector is extended beyond the MSSM field content like in \cite{Benakli:2008pg,Belanger:2009wf}. 

\section{Summary and outlook}

In this paper we develop a systematic understanding of how SUSY breaking RG flows departing from an $\mathcal{N}=1$ SCFT in the strongly coupled regime can be probed by means of two point correlators of gauge invariant operators. We assume that in the large $N$ limit the $\mathcal{N}=1$ SCFT dynamics simplifies and can be described by a handful of light operators. In the dual description this assumption allows us to consider $\mathcal{N}=2$ gauged SUGRA models which are truncations of the full $10d$ type IIB spectrum.   
 
Working within a simple model we classify all the domain wall solutions within it. Most of these have a naked singularity in the deep interior of the bulk. We propose an operational ``goodness'' criterion for singular backgrounds which is based on the absence of tachyonic modes in the dual QFT. 
Our criterion has nothing to say about possible uplifts to string theory but it allows us to show that the latter issue is not posing any obstruction to constructing effective holographic descriptions of strongly coupled SUSY breaking RG flows. 

The knowledge of two point correlators of the R symmetry current and the stress energy tensor allow us to recover Ward identities for explicitly broken R symmetry and conformal symmetry. When the latter symmetries are spontaneously broken we recover the associated Goldstone modes both in the current-current correlator and in the mixed correlator with the scalar operator taking a vacuum expectation value. The innovation of our approach with respect to previous studies is that the different dynamical features of the RG flows at the boundary can be obtained by moving into the parameter space of the bulk SUGRA truncation. A natural extension of our analysis will be to study the behavior of the supercurrent two point functions, identifying the fermionic Goldstone mode associated to spontaneously broken SUSY (i.e. the Goldstino). As already discussed in \cite{Argurio:2013uba}, the latter should appear as a Schwinger term in the supercurrent two point function very much like the Goldstone modes of spontaneously broken gauge and conformal symmetry we discussed here. A careful study of this problem is left for a forthcoming paper \cite{inpreparation}.  

The techniques we discussed so far can be seen as a first attempt to build up an effective understanding of holographic SUSY breaking RG flows at the level of two point functions. Extending our approach to higher point functions might provide further informations about the dynamics of flows which are dual to singular backgrounds in supegravity \cite{Adams:2006sv}. 

On the other hand, aiming to understand more about suggested UV completion of holographic SUSY breaking flows in full type IIB string theory (see for example \cite{Kachru:2002gs}) we should drop the requirement of asymptotically $AdS$ness. In fact there are no known examples of SUSY-breaking $AAdS_5$ SUGRA solutions whose singularity is resolved in string theory.\footnote{An $AAdS_4$ solution where SUSY gets spontaneously broken in a metastable vacuum has been recently constructed in \cite{Massai:2014wba}. It would be certainly interesting to understand more from the QFT perspective how the obstructions present in the $4d$ case can be overcome for $3d$ field theories.} From the field theory perspective this can be understood by remembering that RG flows departing from a UV fixed point with extended supersymmetry (i.e $\mathcal{N}\geq2$) are unlikely to develop SUSY breaking vacua \cite{Antoniadis:2010nj}.  Our techniques would need to be generalized to backgrounds which have asymptotically logarithmic corrections to $AdS$ (i.e.~asymptotically Klebanov-Tseytlin \cite{Klebanov:2000nc}). For a recent study of SUSY breaking solutions in this context, see \cite{inpreparation2}. Computing QFT observables over these backgrounds would hopefully shed light on their physical properties. 

From the model building perspective the examples presented in this paper illustrate a new way of constructing SUSY breaking hidden sectors with warped extra dimensions. This idea was already contained in \cite{Argurio:2012cd} but here we extend and systematize it by considering fully backreacted $AAdS$ geometries and identifying the parameter space of a given SUGRA model and gauging.  The advantage of this approach with respect to the standard Randall-Sundrum setups comes from the use of fully fledged SUGRA as a starting point. This ensures the recovering of SUSY in the UV and clarifies the interpretation of a given geometry as an RG flow interpolating between a UV $\mathcal{N}=1$ SCFT and a SUSY breaking theory with a gapped phase in the IR. In this paper, we focus our attention to a particular gauging which is providing us an unbroken $U(1)_{F}$ of the hidden sector. The flavor symmetry can then be thought of as the Standard Model $U(1)_{Y}$ in order to build toy models for gauge mediation. 

Extending the same idea, different gaugings of the same truncation can be considered. In these cases a weakly gauged flavor symmetry of the hidden sector which gets spontaneously broken along the flow can be responsible for the mediation mechanism. Such a kind of models will provide a holographic realization of the so called Higgsed gauged mediation framework \cite{Gorbatov:2008qa}.

Along the same lines one can also envisage the possibility of constructing holographic models of Higgs mediation \cite{Komargodski:2008ax}, coupling the MSSM Higgs doublets to a strongly coupled hidden sector via Yukawa like interactions. Strong coupling effects are known to  provide a possible solution of both the $\mu/ B_{\mu}$ problem and the $A/ m_{H}$ problem (see for example \cite{Roy:2007nz,Knapen:2013zla}). Using holographic techniques one can in principle go beyond the General Messenger Higgs Mediation framework \cite{Craig:2013wga} and consider both the SUSY breaking hidden sector and the messenger sector as a whole strongly coupled system. Of course this would require the extension of the model considered here to a more general one containing at least two different hypermultiplets dual to the pair of chiral operators of the hidden sector we want to couple to the MSSM Higgses. We hope to come back to these issues in the near future.

\section*{Acknowledgments}

We are grateful to M.~Bertolini, L.~Di Pietro,  M.~Goodsell, A.~Mariotti, F.~Porri and T.~Van Riet for useful discussions. D.M. would like to thank also A.~Amoretti, F.~Bigazzi, A.~Braggio, D.~Forcella, R.~K.~Gupta, N.~Maggiore, N.~Magnoli, A.~Mezzalira, H.~Raj for both useful and nice discussions.  The research of R.A. is supported in part by IISN-Belgium (conventions 4.4511.06, 4.4505.86 and 4.4514.08), by the ``Communaut\'e Fran\c{c}aise de Belgique" through the ARC program and by a ``Mandat d'Impulsion Scientifique" of the F.R.S.-FNRS. R.A. is a Senior Research Associate of the Fonds de la Recherche Scientifique--F.N.R.S. (Belgium). The research of D.R. is supported by the ERC Higgs\@ LHC. D.M. and D.R. would like to thank both ULB and the Solvay Institutes for the warm hospitality during part of this project.


\bibliographystyle{JHEP}
\bibliography{biblio}

\end{document}